\documentclass[11pt]{article}%
\usepackage[round]{natbib}
\usepackage{amssymb}
\usepackage{mathtools}
\usepackage{amsfonts}
\usepackage{color, colortbl}
\usepackage{amsmath}
\usepackage{eurosym}
\usepackage{leftidx}
\usepackage{ragged2e}
\usepackage[nohead]{geometry}
\usepackage[singlespacing]{setspace}
\usepackage[bottom, multiple]{footmisc}
\usepackage{indentfirst}
\usepackage{endnotes}
\usepackage{geometry}
\usepackage{rotating}
\usepackage{times}
\usepackage{booktabs}
\usepackage{multirow}
\usepackage{pifont}
\usepackage{amsmath}
\usepackage{graphicx}
\usepackage{setspace}
\usepackage[dvips]{epsfig}
\usepackage{enumerate}
\usepackage{tabularx}
\usepackage{hyperref}
\usepackage[center]{caption}
\usepackage{subcaption}
\usepackage{longtable}
\usepackage{lscape}
\usepackage{epstopdf}
\usepackage[disable]{todonotes}
\usepackage{graphicx}
\renewcommand{\thetable}{\Roman{table}}

\newcolumntype{C}[1]{>{\centering\arraybackslash}p{#1}}
\definecolor{Gray}{gray}{0.9}
\definecolor{LightCyan}{rgb}{0.88,1,1}
\newtheorem{theorem}{Theorem}

\makeatletter
\def\@biblabel#1{\hspace*{-\labelsep}}
\makeatother
\begin{document}
\hypersetup{pageanchor=false}

\title{\vspace{-5ex} \Huge \bf Marginals Versus Copulas: Which Account For More Model Risk In Multivariate Risk Forecasting? \\}

\newcounter{savecntr}
\newcounter{restorecntr}

\author{
\begin{large} Simon Fritzsch\end{large}\thanks{Universit\"{a}t Leipzig, Wirtschaftswissenschaftliche Fakult\"{a}t, e-mail: \textit{fritzsch@wifa.uni-leipzig.de}.} \\
\begin{large} \textit{Universit\"{a}t Leipzig}\end{large}
  \and
\begin{large} Maike Timphus\thanks{Universit\"{a}t Leipzig, Wirtschaftswissenschaftliche Fakult\"{a}t, e-mail: \textit{timphus@wifa.uni-leipzig.de}.}\end{large}\\
\begin{large} \textit{Universit\"{a}t Leipzig}\end{large}
  \and
  \begin{large} Gregor Wei{\ss}\end{large}\thanks{Corresponding author: Universit\"{a}t Leipzig, Wirtschaftswissenschaftliche Fakult\"{a}t, e-mail: \textit{weiss@wifa.uni-leipzig.de}. We thank Barbara Rogo, Thorsten Schmidt, Andreas Tsanakas, Ximeng Wang, and participants at the 24th International Congress on Insurance: Mathematics and Economics for their helpful comments. Simon Fritzsch gratefully acknowledges financial support through a PhD scholarship from LBBW Asset Management Investmentgesellschaft mbH, Stuttgart.} \\
	\begin{large} \textit{Universit\"{a}t Leipzig}\end{large} \\
}

\maketitle

\sloppy%

\vspace{-6mm}
\onehalfspacing
\begin{center}
\textbf{ABSTRACT}
\end{center}
\noindent Copulas. We study the model risk of multivariate risk models in a comprehensive empirical study on Copula-GARCH models used for forecasting Value-at-Risk and Expected Shortfall. To determine whether model risk inherent in the forecasting of portfolio risk is caused by the candidate marginal or copula models, we analyze different groups of models in which we fix either the marginals, the copula, or neither. Model risk is economically significant, is especially high during periods of crisis, and is almost completely due to the choice of the copula. We then propose the use of the model confidence set procedure to narrow down the set of available models and reduce model risk for Copula-GARCH risk models. Our proposed approach leads to a significant improvement in the mean absolute deviation of one day ahead forecasts by our various candidate risk models.

\normalsize
\vspace{0mm}

\strut

\renewcommand{\thefootnote}{\arabic{footnote}}

\thispagestyle{empty}

\pagebreak%

\onehalfspacing
\normalsize

\hypersetup{pageanchor=true}

\setcounter{page}{1}

\section{Introduction}
Financial institutions employ quantitative models in almost all aspects of their risk management (e.g., for the pricing of derivatives, the modeling of credit risk portfolios, or the forecasting of market risk measures). The increase in the multiplicity and the complexity of these risk models is driven by both the occurrence of tail risks after the Great Financial Crisis as well as the incentives set by the Basel II/III and Solvency II regulations for institutions to develop and use internal quantitative models. However, any risk measurement that does not rely on a standard approach prescribed by regulators will require the risk manager to select a candidate risk model thus introducing potential \textit{model risk}. In this paper, we study such model risk for a class of prominent models for multivariate risk forecasting: Copula-GARCH models. As our main result, we find that copulas account for considerably more model risk than marginals in multivariate models. 

We study the model risk of multivariate risk models in a comprehensive empirical study using Copula-GARCH models. Our ultimate goals are to quantify model risk in multivariate forecasts of portfolio Value-at-Risk (VaR) and Expected Shortfall (ES), and to propose ways how to reduce model uncertainty. To achieve these goals, we forecast the VaR and ES for a large number of multivariate portfolios using a variety of Copula-GARCH models. The first question which we want to answer is whether the model risk inherent in the forecasting of portfolio risk is caused by the candidate marginal or copula models. For this, we analyze different groups of models in which we fix either the marginals, the copula, or neither. We then propose the use of the model confidence set procedure proposed by \citet{Hansen.2011} to narrow down the set of available models and reduce model risk for Copula-GARCH risk models.

As our first main result, we find that model risk is economically significant for the set of candidate multivariate models that we consider. For a portfolio with a value of \$100,000 and a holding period of 10 days, model risk can account for an absolute deviation in VaR (ES) of up to \$2,678 (\$2,264). Interestingly, these high levels of model risk are almost completely due to the choice of the copula with the choice of marginal model having only a small effect on overall model uncertainty. Finally, and not surprisingly, periods of high market volatility lead to a surge in model risk. We then propose the use of the model confidence set procedure to narrow down the set of available models and reduce model risk for Copula-GARCH risk models. Our proposed approach leads to a significant improvement in the mean absolute deviation of one day ahead forecasts by our various candidate risk models.

Our paper paper contributes to several strands in the literature on both model risk and multivariate risk forecasting. First, our paper complements several studies on the importance of moral risk in financial risk forecasting. In this field, early studies focused on the choice of models for pricing option contract. For example, \citet{Green.1999} and later \citet{JohnHull.} show that model risk in the form of inaccurate volatility forecasts and modeling errors in the implied volatility function can lead to significant risk exposure for option writers. In a similar setting, \citet{Cont.} proposes a simple framework for quantifying model uncertainty in derivatives pricing. While our paper relies on similar definitions of model risk, we do no restrict our study to derivatives pricing but instead focus on the more general problem of studying model risk in the context of market risk forecasting.\footnote{It is interesting to see that the shortcomings of VaR-models were identified even as early as the mid 90s with \citet{Hendricks.1996} pointing our that even though ``\textit{[...]virtually all of the approaches produce accurate 95th percentile risk measures.}'' the models considered in his study did not sufficiently capture extreme events.} After the Great Financial Crisis, interest in the study of model risk surged again as the failure of market and credit risk models to adequately capture tail risks was seen as a major driver of the global crisis. In this strand of the literature, \citet{Alexander.2012} and \citet{Glasserman.2014} both propose new methodologies for quantifying model risk with the former concentrating on Value-at-Risk models and the latter studying credit and counterparty risk. Similarly, \citet{Danielsson.2016} study model risk of models used for forecasting systemic risk. In contrast to these related studies, we focus on \textit{multivariate} risk models and disentangle the parts of model risk that are due to the separate modeling of the marginal behavior and the dependence structure in copula models.\footnote{In this respect, our paper is related to the studies of \citet{BERNARD2015166} and \citet{Bernard.2020} who develop a framework to allocate model risk to the different assumptions inherent in a risk model and try to incorporate information on the dependence of risks into the computation of risk bounds.}  

Moreover, our study is also related to a large strand of literature on copula modeling in quantitative risk management.\footnote{Other fields in which copulas have been applied include (among others) decision trees \citep[see][]{Wang}, reliability modeling \citep[see][]{Wu}, and systemic risk modeling \citep[see, e.g.,][]{Jayech,Calabrese,Calabrese2}.} In this field of research, the majority of studies have been concerned with the model risk caused by the need of selecting the best parametric copula family in a multivariate risk model \citep[see, e.g.,][]{KOLE20072405,SavuTrede,GENEST2009199,DIMANN201352}. In addition, several recent papers propose new copula models that are suitable for modeling dependence structures in high-dimensional financial data (e.g., large market risk portfolios) \citep[see, e.g.,][]{AAS2009182,Brechmann.2013,OhPatton:2017,OhPatton:2018,Bassetti}. Finally, several papers have looked at the superiority of copula models over competing one-dimensional or correlation-based risk models \citep[see, e.g.,][]{JONDEAU2006827,Grundke,Weiss:2008}. Complementing these studies, our paper analyzes the economic significance and the source of model risk in copula risk models. As such, it is the first to quantify the extent to which model risk stemming from the choice of a parametric copula can lead to significant additional risk exposure for risk managers and investors.  

The rest of the paper is structured as follows. In Section \ref{sec:2}, we present the models and backtests employed in our empirical study as well as a description of the Model Confidence Set. In Section \ref{sec:data}, we shortly discuss the financial market data used in our study. Sections \ref{sec:empiricalAnalysis} and \ref{sec:analysisMCS} present our main results of the analysis of model risk and the proposed use of the Model Confidence set, respectively. Section \ref{sec:conclusion} concludes.

\section{Market risk models and model risk}
\label{sec:2}
This section provides details on the estimation of market risk via copula GARCH models, backtesting of the risk estimates as well as the calculation of model risk. 
In addition, we introduce the model confidence set (MCS) procedure due to \cite{Hansen.2011} yielding a set of models that contains the best model with a certain probability.

\subsection{Multivariate estimation of market risk}
\label{sec:estimation}


As a consequence of the theorem by \cite{Sklar.1959}, one can separate the modeling of the marginals and the dependence of multivariate return series. This is usually done by using GARCH-type models to filter the univariate time series while copulas are subsequently applied to model the dependence structure between different assets in a given portfolio. We will now present the corresponding two step approach to multivariate time series modeling in more detail.

We start with modeling the marginals as ARMA-GARCH-type processes with various error distributions. We  consider not only the standard GARCH model by \cite{Bollerslev.1986}, but also employ the EGARCH model by \cite{Nelson.1991}, GJR-GARCH model by \cite{Glosten.1993}, T-GARCH model by \cite{Zakoian.1994}, and aPARCH model by \cite{Ding.1993}. These models are all nested within the fGARCH model by \cite{Hentschel.1995}. We provide the specifications for an ARMA(p,q)-GARCH(r,s) process in Section \ref{sec:appendixArmaGarch} in more detail as this is a very popular representative of this class of models. For the remaining models we refer to the reference guide by \cite{Bollerslev.2010} as well as to the original papers.

We consider innovations following a normal distribution, a Student-t distribution, a skewed Student-t distribution, and a generalized error distribution.\footnote{For details on the skewed Student-t distribution and the generalized error distribution we refer to \cite{Fernandez.1998} and \cite{Nadarajah.2005}.} Apart from the normal distribution, these distributions are able to account for skewness and/or fat tails in the data.  Combining these four distributions with the five GARCH-type models yields in total 20 different (univariate) specifications that we use to fit ARMA(1,1)-GARCH(1,1)-type processes to the return series.\footnote{Estimation is performed based on the R-package \emph{rugarch} by \cite{Ghalanos.2020}.}\todo{\tiny später: ggf. order(1,1) begründen}

In a second step, we apply copula dependence models to the GARCH filtered data from the first step. 
Copulas are multivariate distributions with all marginals being uniformly distributed in the interval $[0,1]$. For a $d$-dimensional distribution with distribution function $F$ and marginals $F_1, \dots, F_d$, the copula associated with $F$ is a function $C: [0,1]^d \to [0,1]$ satisfying 
$$F(\mathbf{x})=C\big(F_1(x_1), \dots, F_d(x_d)\big)$$ for $x=(x_1, \dots, x_d) \in \mathbb{R}^d.$  
On the basis of the two step approach lies the fact that copulas enable us to model the dependence structure of multivariate distributions separately from the marginal distributions. This is a consequence of the theorem by \cite{Sklar.1959}, see Section \ref{sec:sklarsTheorem} for more details.

We consider various copula functions. We include the Gaussian and Student-t copulas because they are widely used in financial applications \citep[cf. ][]{Cherubini.2004, McNeil.2005}. We additionally employ the Archimedean copulas Clayton, Gumbel, Frank, and Joe copula. This gives us further possibilities to model multivariate dependence as for example the Clayton copula allows for modeling a positive lower tail dependence. For more details on copulas in general and the copulas used in this paper we refer to the comprehensive books by by \cite{Joe.1997} and \cite{Nelsen.2006}.

While the Gaussian and Student-t copulas can only model symmetric dependencies using correlation matrices, Archimedean copulas typically have only one (like the above mentioned copulas) or two parameters, which is very restrictive. So-called pair copula constructions (also referred to as vine copulas) offer a very convenient possibility for a highly flexible modeling of the dependence structure. Originally introduced by \cite{Joe.1996}, further significant contributions have been made by \cite{Bedford.2001, Bedford.2002}, and \cite{Kurowicka.2006}.
Additionally, we include the Gaussian mixture copula due to \cite{Tewari.2011} that can capture multi-modal dependencies as well as asymmetric and tail dependencies. Another very popular model is the Dynamic Conditional Correlation (DCC) model due to \cite{Engle.2001} and \cite{Engle.2002}. 
Though not a copula model, the DCC model fits perfectly into our framework. It allows for a two stage estimation of the model parameters where in a first stage GARCH-type models are fitted to each of the univariate return series. The conditional correlation matrix is derived in a second step to model the multivariate dependence between the return series. In the following, we will therefore subsume all models under the term copula GARCH models.\footnote{Estimation for the Gaussian, Student-t, Clayton, Gumbel, Frank, and Joe copula is performed based on the \emph{copula} R-package  by \cite{Hofert.2020}. Inference for the vine copula is performed using the \emph{VineCopula} R-package by \cite{Nagler.2019} considering the Gaussian, Student-t, Frank, Clayton, Gumbel, and Joe copula (along with rotated and survival versions of the latter three copulas) as bivariate building blocks of a regular vine copula. Selection of a particular bivariate copula is performed based on the AIC value. We estimate a Gaussian mixture copula with three components (two for capturing potentially fat tails, one for ``regular'' returns) employing the \emph{GMCM} R-package by \cite{Bilgrau.2016}. Inference for the DCC model is based on the \emph{rmgarch} R-package by \cite{Ghalanos.2019} assuming a multivariate Student-t distribution.}\todo{\tiny später: ggf. noch ausführlicher werden in Bezug auf das konkrete Schätzverfahren (z.B. itau, mle) und weiteres (siehe Notizen im Quelltext.)}

For each combination of the considered copula and ARMA-GARCH-type models we now proceed as follows to obtain one day ahead VaR and ES forecasts for a given portfolio of $K$ assets: First, we simulate 10,000 $K$-dimensional vectors of standardized residuals from the fitted copula. Based on the parameters and ex-ante mean and variance forecasts from the ARMA and GARCH-type models that have already been fitted to the univariate return series, we transform the simulated standardized residuals into 10,000 return forecasts for each of the portfolio constituents. We finally derive 10,000 portfolio returns and calculate the VaR as the sample quantile and the ES as the conditional mean of the returns falling below this quantile (both values are subsequently multiplied by $(-1)$). This procedure is repeated on a daily basis based on a moving window of data. For more details we refer to \cite{Brechmann.2013}, see also \cite{Aas.2009}, \cite{Ausin.2010}, and \cite{Nikoloulopoulos.2012}. One important advantage of this approach is that copulas and marginals only have to be fitted once to consider various portfolio weights. This is because the weights enter into the procedure only at the end when calculating the portfolio returns.\footnote{We make use of this fact to include 100 portfolios with randomly generated portfolio weights to ensure robustness of our results.}

\subsection{Backtests}
\label{sec:backtests}

Our aim is to analyze the risk associated with choosing an appropriate model from a variety of valid candidate models, not to identify the \textit{optimal} model for forecasting VaR and ES estimates.
Nonetheless, we backtest our risk forecasts before measuring the model risk itself for the following reasons:
First, in order to determine a set of \textit{valid} candidate models, we need to prevent our results from being biased by erroneous risk forecasts due to misspecified models.
Second, the approach of calculating model risk after having applied backtests is also favorable from a more practical perspective. The Basel III regulation requires banks to backtest their (internal) market risk models. For banks, uncertainty on the model choice is, essentially, uncertainty on the choice of models that have not
been rejected by backtests. This is in line with our calculation of model risk.

We measure the quality of the VaR estimates using the duration-based test of \cite{Christoffersen.2004}. The test follows a more general approach of independence rather than focussing on the independence of VaR violations only. The test assumes that in the case of independent VaR violations, the time between violations must be independent of the time to a previous violation. In simple terms, this means that the probability that a VaR violation will occur in the next 10 days must be independent of whether the last one occurred in the last 10 or 100 days.\footnote{See \cite{Campbell.}.} \cite{Christoffersen.2004} shows that his duration-based test tends to reveal VaR methods that violate the independence property in realistic situations more often and hence has better power properties than former tests. For more details on the duration-based test we refer to \ref{sec:appendixVaRbacktest} as well as to the original paper.\footnote{In addition, we implement the dynamic quantile test of \cite{Engle.2004}, which as a conditional coverage test examines not only the number and independence of VaR hits but also the independence of the estimators. The results can be found in Section \ref{sec:analysisBacktesting}.} 

Since the VaR has been replaced by the ES as the leading market risk measure in regulatory requirements, 
the debate on suitable backtesting for the ES has also been increasing in the literature. In view of the regulatory requirements and because the VaR does not have to be issued by financial institutions, ideally a backtest should only determine the quality of the ES model based on real data and the forecasts. In practice, however, many backtests require additional input variables or assumptions.\footnote{See \cite{Bayer.2020}.}
We evaluate ES estimates using a comparatively new test by \cite{Nolde.2017}, which is based on the concept of conditional calibration (CC). For a brief description of the test, we refer to \ref{sec:appendixESbacktest} and the original paper. The CC test requires in its simple version both VaR and ES forecasts.\footnote{In the general version, volatility is also taken into account.} We find the use of a joint backtest of VaR and ES predictions sensible for the following reasons:
First, in line with \cite{Bayer.2020}, we find that the ES is by definition closely related to the VaR and thus suitable ES forecasts are based on adequate VaR estimations. Secondly, we apply the MCS procedure at a later point of our analysis. This method is again based on functions that require both VaR and ES forecasts. Consequently, the use of a joint VaR and ES backtest is appropriate.\footnote{As a supplement, we use the exceedance residual test of \cite{McNeil.2000}, which is based on the ES-specified residuals that exceed the VaR. A comprehensive description of the test can also be found in \cite{Bayer.2020}. The exceedance residual test is a joint backtest as well, which is why the same reasons for use as for the conditional calibration test of \cite{Nolde.2017} apply. The results can be found in Section \ref{sec:analysisBacktesting}.}

By using a joint backtest for the ES, it should be noted that this is a possible reason why more models fail the backtest than in the case of the VaR. For this reason and because different confidence levels are considered in the baseline case, we separate the analysis of the VaR and the ES in terms of model risk.

\subsection{Model risk}
\label{sec:modelRisk}
Financial risk cannot be measured directly, but only be estimated using statistical models. However, it is well known that distinct models may differ vastly in their risk predictions \citep[see, e.g., ][]{Danielsson.2016}. In the literature one can find various approaches towards defining and measuring model risk. A large part of research focuses on the factors that control model risk within models such as the misspecification of the underlying theoretical models \citep{Green.1999} or assumptions made about unknown (or unobservable) parameters, distributions, or other model specifications \citep[e.g., ][]{JohnHull., Alexander.2012, Glasserman.2014, Boucher.2014}. 
However, we focus on a more general problem: With a variety of standard VaR and ES models within the industry, uncertainty about the choice of a particular model creates model risk per se. Especially, our aim is not to identify an \emph{optimal} model for forecasting the VaR or ES, but to analyze the risk that emerges from the presence of a large set of valid candidate models.
Our notion of model risk as uncertainty on the model choice itself in the presence of many possible alternative models is most closely related to \cite{Cont.} and \cite{Danielsson.2016}. 

We consider all VaR and ES models  that are not rejected in the respective backtest as viable candidates for measuring market risk. This leaves us with a large number of models and corresponding risk forecasts for the same quantity. As a measure of model risk, we quantify the level of disagreement between the individual estimates based on the mean absolute deviation (mad), the standard deviation (sd), and the interquartile range (iqr). The mean absolute and the standard deviation are intuitive and plausible measures since both take into account how much the models deviate from the average forecast. The iqr of an observation variable is the difference of its 75\% and 25\% percentiles and consequently a measure of the maximum disagreement within the 50\% of observations around the median. We choose the mad as our main measure of model risk for the following reasons: First, the mad is more robust against outliers than the sd. Second, it can easily be interpreted in absolute terms relative to a given portfolio value and reflects the deviation of a forecast by a randomly chosen model from the average risk forecast. Finally, since our main analysis relies on forming different groups of models, we need a measure of model risk that is as independent as possible from the number of models within a particular group. For these reasons, we focus on the  mad to measure model risk and include the 
measures sd and iqr for robustness.
\footnote{Another measure to determine model risk is the \emph{risk ratio} by \cite{Danielsson.2016} that is defined as the ratio of the highest to the lowest risk forecast within a set of candidate models. Disagreement between models is therefore captured by a risk ratio greater than 1. We do not include the risk ratio into our analysis for the following reasons: First, we want to focus on the \textit{average} deviations of risk forecasts whereas the risk ratio tends to capture the extreme, \emph{maximum} possible deviations within a set of candidate models. Moreover, the mad is more suitable in our context than the risk ratio, because risk estimates by the latter can depend on the number of models. For example, let us assume we have a group of models $A$ with maximum $\max_A$ and minimum $\min_A$. For an arbitrary subgroup $B$ of group $A$, we have $\min_B\geq \min_A$ and $\max_B\leq \max_A$. As a consequence, the risk ratio of group $B$ is smaller than (or equal to) the risk ratio of the larger group of models $A$. As our approach of studying the importance of modeling the marginals and the multivariate dependence structure heavily relies on building subgroups, this property could potentially bias our results.}

\subsection{Model confidence set}
\label{sec:mcsTheoreticalBackground}
Researchers, practitioners, and regulators are often confronted with situations where a variety of models for computing a specific estimate, e.g., for VaR or ES forecasts, exist. Optimally, one would like to know which of the many available models  is the \emph{best}. However, in many situations this question cannot be answered, especially when the set of competing methods is large and the data are not sufficiently informative. Yet, one can try to reduce the set of available models to a smaller set of alternatives. This can be done by the model confidence set (MCS) procedure by \cite{Hansen.2011}.

The MCS procedure yields a set of models (the \emph{model confidence set}) that contains the \emph{best} model with a certain \emph{probability}. That is, the procedure does in general not identify a best model nor does it assume that a particular model represents the true data generating process. Instead, the MCS can be seen as an analogue to a confidence interval that contains a parameter of interest with a specified probability.
An important advantage of the MCS procedure over methods that choose a single model is that it accounts for the informativeness of the data at hand. When data are very informative one may obtain a MCS that consists of only the best model. Less informative data, on the other hand, may lead to a MCS containing several models as the data make it hard to distinguish between models. Additionally, the MCS procedure allows for valid statements about significance that are not hampered by multiple pairwise comparisons \citep{Hansen.2011}. These attractive features make it interesting to apply the MCS procedure to our set of VaR and ES models. The MCS procedure might help in further narrowing down the set of valid candidate models (after applying the backtests) such that the remaining models exhibit a lower model risk.   

The construction of the MCS procedure relies on an \emph{equivalence test}, $\delta_\mathcal{M}$, and an \emph{elimination rule}, $e_\mathcal{M}$.\footnote{This short introduction to the MCS procedure is based on the original paper by \cite{Hansen.2011} and we refer to it for more details and proofs of the results.} First, the equivalence test is applied to the set of candidate models $\mathcal{M}^0$. If equivalence is rejected at a given confidence level $\alpha$, this implies that the candidate models are not equally ``good''. Thus, the elimination rule is applied to remove a poorly performing model from the set $\mathcal{M}^0$. These two steps are repeated until the equivalence test is not rejected for the first time. The remaining elements of $\mathcal{M}^0$ are then considered as the model confidence set $\hat{\mathcal{M}}_{1-\alpha}$.  As the same confidence level $\alpha$ is used in each iteration for the equivalence test, the procedure guarantees that $\lim_{n\to\infty} P(\mathcal{M}^*\subseteq \hat{\mathcal{M}}^*_{1-\alpha})\geq 1-\alpha$, where $\mathcal{M}^*$ denotes the true set of best models and $n$ is the number of observations per model. Additionally, the MCS procedure provides p-values for each model that can be interpreted as the probability that the respective model is among the best alternatives in $\mathcal{M}^0$. For more details on the procedure we refer to Section \ref{sec:MCS_internetAppendix} and to the original paper.

In the MCS approach, models are evaluated based on a user-defined loss function. For evaluating the forecast accuracy of risk models it is natural to compare the forecasts to the realized financial losses over a period of time. \cite{Nolde.2017} highlight that for comparing different models' risk forecasts elicitability of the risk measure is a desirable property. Generally speaking, a risk measure is elicitable if it minimizes the expected value of a \emph{scoring function}.\footnote{See \cite{Gneiting.2011} for a comprehensive literature review on elicitability as well as \cite{Frongillo.2015, Fissler.2016, Ziegel2016} for more recent advances in the field.} Elicitability is a property that has been proven to be useful for forecast ranking, comparative backtesting, and for model selection \citep{Nolde.2017}.

 As the VaR represents a quantile of a probability distribution (multiplied by -1) it is well known to be elicitable \citep[see, e.g., ][]{Koenker.1978}. The associated scoring function, the so-called check-function, is given by 
$$L_{VaR}(r_t, VaR_\alpha^t,\alpha):= \big(r_t-(-VaR_\alpha^t)\big)\cdot\big(\alpha-\mathbb{I}_{(-\infty, 0)}(r_t-(-VaR_\alpha^t))\big),$$
where $r_t$ and $VaR_\alpha^t$ denote the realized return and the VaR forecast with coverage level $\alpha$ (and confidence level $1-\alpha$) at day $t$ and $\mathbb{I}_{(-\infty, 0)}$ denotes the characteristic function of the open interval $(-\infty, 0)$. \todo{\tiny Falls wir VaR doch als negative Zahl auffassen, dann Vorzeichen beim VaR wieder rumdrehen.} We choose this scoring function as the loss function for the VaR in the MCS framework. 

As opposed to the VaR, the ES alone is not elicitable. Instead, \cite{Fissler.2016} show that ES and VaR are \emph{jointly} elicitable, see also \cite{Acerbi.2014}. There is a growing body of literature building on this result, see, e.g., \cite{Fissler.2016b, Nolde.2017} for forecast comparisons and \cite{Patton.2019, Barendse.2019, Bayer.2020} for applications in a regression procedure. As the ES is only jointly elicitable with the VaR, we choose a loss function for the ES that is based on both VaR and ES forecasts as input into the MCS procedure. This is consistent with the conditional calibration backtest introduced in Section \ref{sec:backtests} that is used to determine the set of candidate models for the ES.  As \cite{Nolde.2017} provide a joint scoring function for the VaR and the ES only in a general form, we adopt the 0-homogeneous version introduced in \cite{Patton.2019}
$$L_{ES}(VaR_\alpha^t, ES_\alpha^t, r_t,\alpha):=\frac{1}{\alpha ES_\alpha^t}\cdot \mathbb{I}_{(-\infty,0)}(r_t+VaR_\alpha^t)\cdot(-VaR_\alpha^t-r_t)+\frac{VaR_\alpha^t}{ES_\alpha^t}+\log(ES_\alpha^t)-1,$$
where the notation is as above.\footnote{Note that throughout the paper we regard VaR and ES estimates as positive values.} For an implementation of the MCS procedure we rely on the R-package \emph{MCS} by \cite{Catania.2017}. For more details we refer to Section \ref{sec:analysisMCS}.

\section{Data}
\label{sec:data}
We form well diversified portfolios consisting of equity indices (developed and emerging markets), bond indices (governement, corporate, and high-yield bonds) as well as commodity and real estate indices. Therefore, we retrieve the total return indices (in US\$) of the following set of indices from Datastream:
Stoxx Europe 600, Dow Jones Industrial Average, FTSE Developed Asia Pacific Index, MSCI Emerging Markets Index, S\&P U.S. Treasury Bond Index,  S\&P 500 Investment Grade Corporate Bond Index, S\&P U.S. High Yield Corporate Bond Index, S\&P Pan-Europe Developed Sovereign Bond Index, S\&P GSCI, and Developed Markets Datastream Real Estate Index.
The sample period is January 2001 to December 2018. Next, we calculate geometric returns that allow us to easily derive portfolio returns. For our main analysis we focus on an equally weighted portfolio.  This corresponds to a portfolio consisting of 40\% stocks, 40\% bonds, 10\% commodities, and 10\% real estate. For robustness we also consider portfolios based on random portfolio weights that were drawn from a unit-simplex. Summary statistics on the equally weighted portfolio returns as well as on the individual index returns can be found in Table \ref{tab:PF_stats}.

\begin{center}
	-- Insert Table \ref{tab:PF_stats} about here. --
\end{center}

We calculate daily VaR and ES estimates based on 180 different model specifications of copula GARCH models, see Section \ref{sec:estimation} for details. Additionally, we derive risk estimates from univariate GARCH-type models applied to the portfolio return series. Estimations are performed based on various confidence levels (99.9\%, 99\%, 97.5\%, and 95\%)\footnote{In line with the Basel II and III market risk regulations we focus on the 99\% VaR and the 97.5\% ES.} using a moving window of 500 days corresponding to approximately two years of daily observations\footnote{For robustness we also consider a moving window of 1000 days.}
and a forecast horizon of one day. We clean the risk estimates from outliers that are due to convergence errors in fitting the copula GARCH models.\footnote{We identify outliers based on the daily absolute changes of the risk forecasts. Therefore, we calculate z-scores based on per model standard deviation and mean calculated over the first 500 risk estimates. We then replace observations with a z-score above 25 with the value from the previous day. This affects on average 0.17\% of all risk estimates. Note that by this procedure we do not introduce a look-ahead bias into our analysis.}




Subsequently, we perform VaR or ES backtests to determine the set of candidate models that enter into the calculation of model risk on a daily basis. The backtests are based on a confidence level of 99\% in line with \cite{BaselCommitteeonBankingSupervision.2019} and a moving window of 500 days. 
Note that by employing a moving window for the backtests, we avoid introducing a look-ahead bias into the selection of the set of candidate models. 
In our main analysis, we rely on the duration-based VaR backtest by \cite{Christoffersen.2004} and the conditional calibration ES backtest by \cite{Nolde.2017}, see Section \ref{sec:backtests} for details.\footnote{We use the  \textit{rugarch} R-package by \cite{Ghalanos.2020} for performing the duration-based backtest and the  \emph{esback} R-package by \cite{SebastianBayer.2020} for the conditional calibration backtest (simple version one-sided using Hommel's correction). For comparison, we also run the backtests using a moving window of 1000 days as well as a fixed window over the entire period. Additionally, we perform the dynamic quantile test by \cite{Engle.2004} with the \textit{GAS} R-package by \cite{DavidArdia.2019} and the exceedance residual backtest by \cite{McNeil.2000} (one-sided) with 1000 bootstrap iterations implemented in the \textit{esback} R-package by \cite{SebastianBayer.2020}. For more details see Section \ref{sec:analysisBacktesting}.}

Afterwards, we calculate model risk on a daily basis for the risk models that have passed the respective backtest. Note that by using moving windows the composition of the set of candidate models varies over time. 
Since VaR and ES estimations as well as backtests are performed based on a 500 day moving window in our baseline analysis, we obtain daily model risk estimates from day 1001 onwards. This corresponds to the time period from November 4, 2004 until December 31, 2018. We obtain model risk estimates for both VaR and ES forecasts for various confidence levels and portfolio weights. In our main analysis we focus on the model risk of  risk forecasts for an equally weighted portfolio and the 99\% VaR and 97.5\% ES in line with the Basel II and Basel III market risk regulations.\footnote{In the following, VaR will refer to the 99\% VaR and ES will refer to the 97.5\% ES unless specified differently.} Our main measure of model risk is the mean absolute deviation (mad) of risk forecasts.

\section{Analysis of model risk}
\label{sec:empiricalAnalysis}
\subsection{All multivariate models}
\label{sec:analysisModelRisk:AllMultivariateModels}

\begin{figure}[htbp]
	\caption{Daily model risk for all multivariate models}
	\vspace{0.2cm} \noindent {\justifying \scriptsize This figure shows the model risk associated with one day ahead 99\% VaR and 97.5\% ES forecasts for a well diversified portfolio. Model risk is measured in terms of the mean absolute deviation (\emph{mad}) of one day ahead forecasts by various risk models. Values are calculated on a daily basis between November 4, 2004 until December 31, 2018 in percent of the portfolio value based on all multivariate models that passed the respective backtest, see Section \ref{sec:backtests} for details. 
	}\\
	
	\begin{center}
	\includegraphics[width=0.9\linewidth]{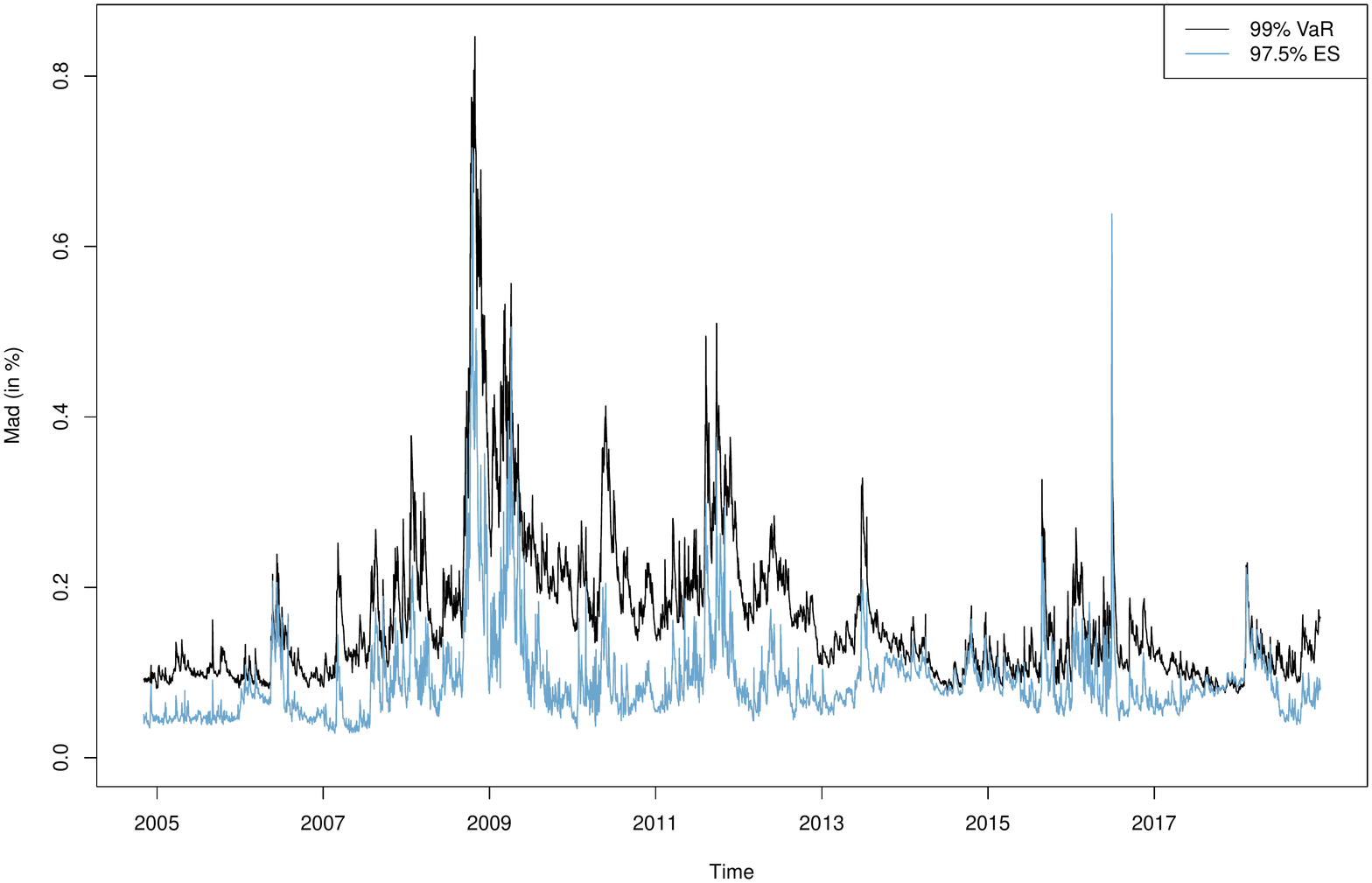}
	\label{fig:allMultivariateModelsMad}
	\end{center}
\end{figure}

We start with an analysis of the model risk over time and different market conditions of all multivariate VaR and ES models that passed the respective backtest.\footnote{We start with 180 different multivariate model specifications. After applying the backtests we are left with on average 174 99\% VaR 121 97.5\% ES models.} Figure \ref{fig:allMultivariateModelsMad} presents the daily model risk associated with one day ahead forecasts of the 99\% VaR and the 97.5\% ES in terms of mad between November 4, 2004 and December 31, 2018. The figure reveals that model risk is normally quite moderate but increases significantly during and after the global financial crisis. Summary statistics are provided in Table \ref{tab:allMultivariateModelsMad}. Over the entire time period, model risk is on average about 0.165\% of the portfolio value for VaR forecasts and about  0.092\% of the portfolio value for ES forecasts. Model risk is quite volatile over time ranging from 0.075\% to 0.847\% for the VaR and from 0.029\% to 0.716\% for the ES with a standard deviation of daily model risk estimates of more than half of the average model risk.

\begin{center}
	-- Insert Figure \ref{fig:allMultivariateModelsMad} and Table \ref{tab:allMultivariateModelsMad} about here. --
\end{center}

Model risk is especially pronounced during times of financial turmoil. During the years 2008-2009 (in the following referred to as the \emph{crisis period}) the average model risk is 0.286\% of the portfolio value for the VaR and 0.145\% for the ES. That is, the average model risk more than doubles compared to the period before 2008 (the \emph{pre-crisis period}) with an average VaR of 0.119\% and ES of 0.063~\%.\footnote{These differences between pre-crisis and crisis period are statistically significant at the 1\% level where statistical significance throughout the paper is determined based on t-tests with standard errors corrected for serial correlation and heteroskedasticity according to \cite{Newey.1987} with the automatic bandwidth selection procedure described in \cite{Newey.1994}. Note that although for robustness we include 100 randomly generated portfolio weights into the study, the portfolio bootstrap procedure according to \cite{Danielsson.2016} is not applicable in our setting. This is, because the portfolio weights enter into the risk forecast after the GARCH models and copula functions are fit.} The maximum model risk values are realized in Q4 2008 in the follow-up of the Lehman Brothers bankruptcy on the peak of the financial crisis. The extraordinary impact of the financial crisis on model risk is further highlighted by the fact that nearly all 
VaR model risk values above the 99\% quantile occurred in Q4 2008. The same is true for the majority 
of the ES model risk values.\footnote{Further high model risk values occurred in particular in Q2 2009, Q3 2011, and Q3 2015. The highest model risk value for the ES was realized on October 17, 2008, just two days after the Dow Jones Industrial Average Index experienced its largest drop in relative terms since 1987. The highest model risk value for the VaR was realized on October 29, 2008. Two days earlier, the Nikkei 225 Index lost more than 6.4\% while the Hang Seng Index decreased by 12.7\% while the consecutive day world wide stock markets saw a huge rally in anticipation of rate cuts by central banks.}
\todo{\tiny Testen, dass die Unterschiede statistisch signifikant sind ... siehe Danielsson2016 section 3.3, Argumentieren, dass die auch ökonomisch signifikant sind.}

These results can only partly be explained by an increase in volatility. On the one hand we calculate VaR and ES forecasts based on conditional volatility estimates derived from GARCH-type models. Consequently, whenever volatility is high, VaR and ES forecasts will on average also show increased levels resulting in a higher model risk. On the other hand, the average model risk in 2008-2009 is 140\% higher than in the pre-crisis period for the VaR while the average level of VaR estimates is only 113\% higher. Similarly, the model risk for the ES in 2008-2009 is increased by 130\% while the average level of ES estimates is 119\% higher. 
This disproportionately high increase of model risk in periods of crisis might be due to the fact that all models treat history and shocks quite differently such that a change in statistical regimes can be expected to lead to higher disagreements between risk forecasts \citep{Danielsson.2016}.
Following the great financial crisis, model risk does not decrease to the pre-crisis\todo{\tiny ggf. später Erklärung bzw. Vermutung warum das so ist? Steht dazu was bei Danielsson?} level (0.119\% for the VaR and 0.063\% for the ES) but remains elevated at 0.155\% (VaR) and 0.090\% (ES).\footnote{These differences between the pre-crisis and the after-crisis period are statistically significant at the 1\% level.} The findings are robust to considering randomly generated portfolio weights, see Table \ref{tab:averageWeightsAllMultModels}. Summary statistics for other measures of model risk (standard deviation, interquartile range) are provided in Table \ref{tab:otherModelRiskMeasuresAllMultModels}.

\begin{center}
	-- Insert Tables \ref{tab:averageWeightsAllMultModels} and \ref{tab:otherModelRiskMeasuresAllMultModels} about here. --
\end{center}

\begin{figure}[htbp]
	\caption{Potential portfolio value under financial distress}
	\vspace{0.2cm} \noindent {\justifying \scriptsize This figure illustrates the economic significance of model risk arising from the disparity between different VaR models. Here, we focus on the 99\% VaR for a well diversified portfolio (\$100,000) and a 10 day holding period. We provide the portfolio value minus the 5th and the 95th percentile of VaR forecasts from all multivariate models that passed the duration-based backtest by \cite{Christoffersen.2004} on a daily basis. This corresponds to the potential portfolio value under financial distress according to the more (95th percentile) or less (5th percentile) conservative VaR models. The sample period is November 4, 2004 until December 31, 2018.
	}\\
	
	\begin{center}
	\includegraphics[width=0.9\linewidth]{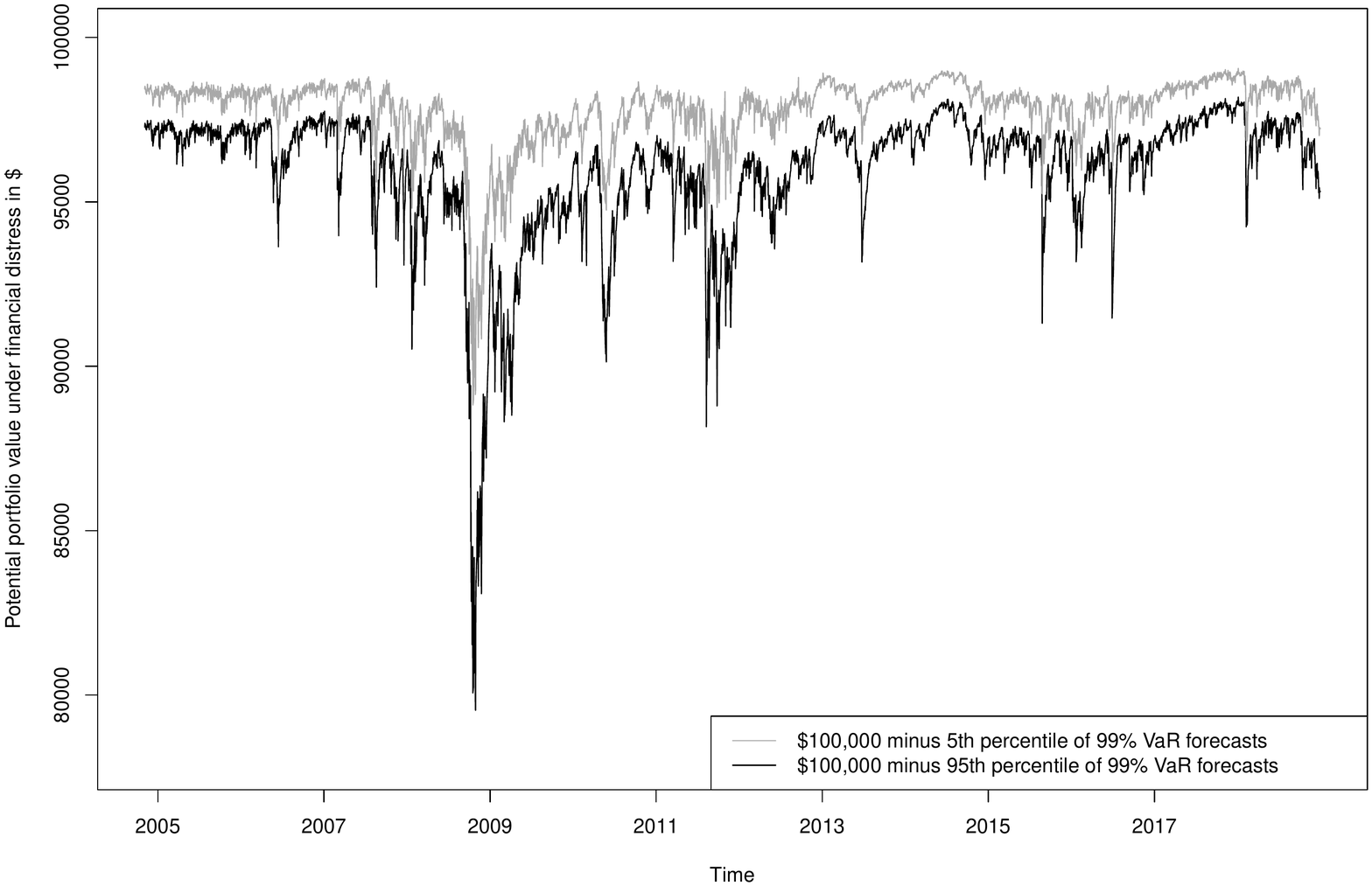}
	\label{fig:minimumPortfolioValueVaR}
	\end{center}
\end{figure}

Risk models are embedded within the Basel accords and play a central role in the regulatory process to determine bank capital. That is, expensive decisions such as the amount of capital held or portfolio allocations depend on the outputs of risk forecasting models as input. We highlight this point by providing model risk estimates in absolute terms in Table \ref{tab:allMultivariateModelsMad}. These values can be interpreted as average deviations in regulatory capital to be held according to different models (at a particular day). We calculate model risk estimates in \$ by assuming a portfolio value of \$100,000 and a holding period of 10 days. Therefore, we multiply the mad values (as percentage of the portfolio value) with \$100,000 and $\sqrt{10}$.\footnote{The square-root-of-time rule for scaling daily VaR forecasts is the industry standard although this approach might lead to underestimation \citep{Danielsson.2006, Wang.2011} or overestimation of the 10-day VaR \citep{Diebold.1997}, see \cite{Kole.2017}. A detailed analysis on the effects of different choices of temporal aggregation can be found ibid. For simplicity, we also rely on the square-root-of-time rule for scaling ES forecasts.} We thereby emphasize that differences in model risk are not only statistically but also economically significant. This is also highlighted by Figure \ref{fig:minimumPortfolioValueVaR}\todo{\tiny ggf. Figure \ref{fig:minimumPortfolioValueES} aus dem IA entfernen?!} illustrating the extent of disparity of VaR models with regard to the portfolio value. The average model risk (mad) in absolute terms over the entire time period is \$522 for the VaR and \$291 for the ES with maximum values of \$2,678  (VaR) and \$2,264 (ES). 

\begin{center}
	-- Insert Figure \ref{fig:minimumPortfolioValueVaR} about here. --
\end{center}

Note that model risk of VaR and ES forecasts cannot directly be compared to each other. This is due to the fact that risk forecasts entering into the calculation of model risk are determined based on different backtests for the VaR and the ES. As a result, on average 174 different model specifications enter into the model risk calculation for the VaR and only 121 specifications into the model risk calculation for the ES. 
The effect of choosing different backtests or no backtest at all are discussed in Section \ref{sec:analysisBacktesting} in more detail.

 When considering several confidence levels for VaR and ES forecasts we observe that model risk for the 99.9\% VaR (0.306\%) is approximately twice as high as for the 99\% VaR (0.165\%) which again is twice as high as for the 95\% VaR (0.085\%). For the 99.9\% ES (0.309\%), 99\% ES (0.142\%), 97.5\% ES (0.092\%), and 95\% ES (0.067\%) the proportions are similar, see Table \ref{tab:analysisI_robust_lev} for more details.\footnote{All differences in pairwise comparisons of the same risk measure at different confidence levels are statistically significant at the 1\% level. Note that the backtests are performed separately for each confidence level. However, when focusing on the model risk of either VaR or ES, differences in the average  percentage of models that passed the respective backtest are quite low. The average percentage of models that passed the backtest is 99.3\%, 97.6\%, 96.7\%, and 98.9\% for the VaR and 60.7\%, 67.2\%, 75.5\%, and 68.6\% for the ES at the 95\%, 97.5\%, 99\%, and 99.9\% confidence level, respectively.} This rise in model risk when increasing the underlying confidence level can, however, only partially be explained by an increase in the absolute level of the risk forecasts as a consequence of the higher confidence level. That is, even when relating the average model risk to the average level of risk forecasts, model risk is more pronounced for higher confidence levels.  This can be seen when looking on the ratio of average model risk divided by the average level of risk forecasts which is 0.208, 0.163, 0.142 and 0.126 for the VaR and 0.201, 0.111, 0.086 and 0.075 for the ES at the 99.9\%, 99\%, 97.5\% and 95\% confidence level, respectively.\todo{\tiny Quotienten sind OK, wenn wir VaR/ES als positive Zahlen auffassen.} These results are not surprising as they mainly reflect that the disagreements between different models in modeling the tails of the return distributions increase when considering more extreme quantiles.

\subsection{Analysis of the subgroups}
\label{sec:analyisModelRisk:AnalysisOfSubgroups}
The VaR and ES estimates are obtained via copula GARCH models in a two step procedure. Therefore,  we turn to the question if a greater portion of model risk is attributable to the statistical modeling of the univariate marginals (via GARCH-type models) or to the estimation of the multivariate dependence structure (via copulas).

\begin{figure}[htbp]
	\caption{Average model risk for all groups}
	\scriptsize
    This figure shows the average model risk associated with one day ahead 99\% VaR (first panel) and 97.5\% ES (second panel) forecasts for a well diversified portfolio per group. \textit{Group 1} (G1) includes all model sets in which a copula function is fixed while varying the marginal distribution. \textit{Group 2} (G2) contains analogously the model sets with fixed marginal distribution and varying copula. \textit{Group 3} (G3) consists of all multivariate and \textit{Group 4} (G4) of all univariate models. Model risk is measured in terms of the mean absolute deviation (mad) of one day ahead forecasts by various risk models within a model set. Values are calculated on a daily basis between November 4, 2004 until December 31, 2018 in percent of the portfolio value based on all models that passed the respective backtest, see Section \ref{sec:backtests} for details. 
	\begin{center}
		\begin{subfigure}[t]{\linewidth}
			\centering
			\includegraphics[width=0.8\linewidth]{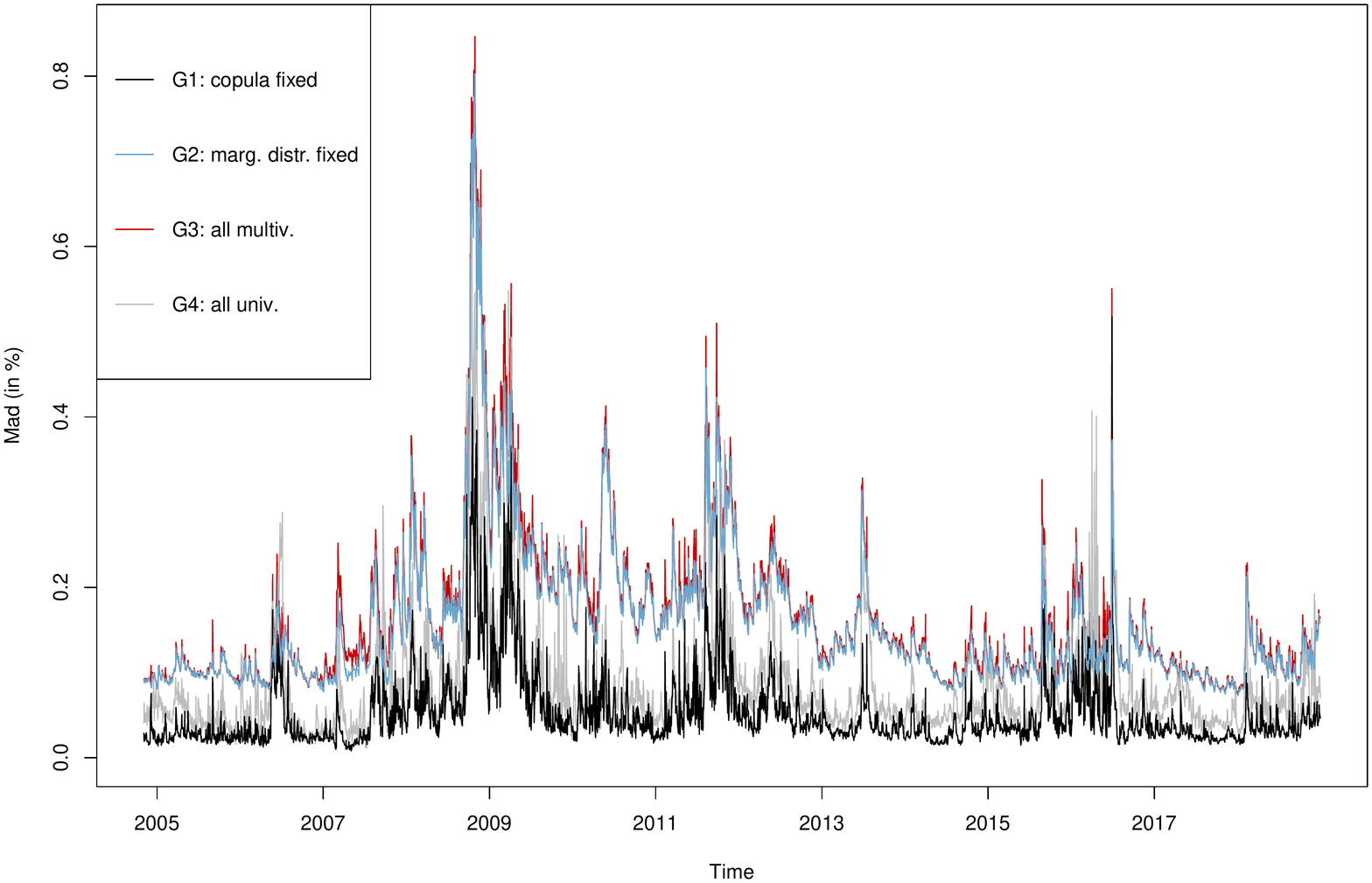}
			\caption{Average model risk (99\% VaR)}\label{fig:Rplot_AnalysisI_1}
		\end{subfigure}
		\begin{subfigure}[t]{\linewidth}
			\centering
			\includegraphics[width=0.8\linewidth]{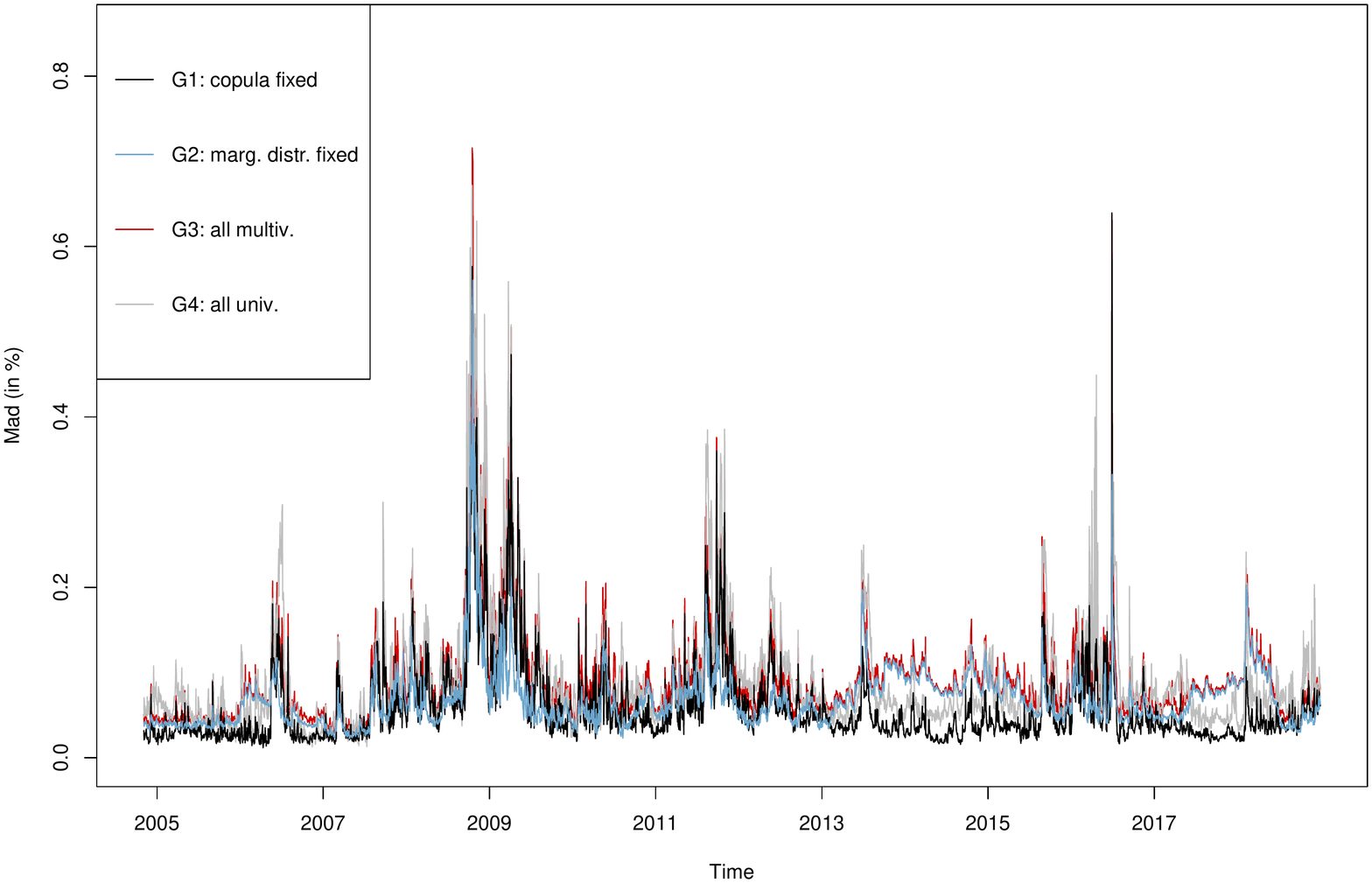}
			\caption{Average model risk (97.5\% ES)}\label{fig:Rplot_AnalysisI_2}
		\end{subfigure}
	\end{center}
	\label{fig:Rplot_AnalysisI}
\end{figure}

Figure \ref{fig:Rplot_AnalysisI} provides the average model risk for the 99\% VaR and the 97.5\% ES for on an equally weighted portfolio  from November 4, 2004 to December 31, 2018 for four different groups: Group 1 covers the average model risk across the sets of models in which a copula is fixed while the marginal distribution is varied. Group 1 thus measures the impact of choosing a specific GARCH-type model for the marginals. Analogously, Group 2 captures the average model risk among sets of models with fixed marginal distributions and varying copulas. Group 3 measures the average model risk of all multivariate and Group 4 of all univariate models. Table \ref{tab:analysisI} presents the corresponding descriptive 
statistics.
\todo{\tiny Wieso wir überhaupt Gruppen bilden. Erklären!}

\begin{center}
	-- Insert Figure \ref{fig:Rplot_AnalysisI} and Table \ref{tab:analysisI} about here. --
\end{center}

Considering the 99\% VaR, average model risk for model sets with fixed copulas is 0.052\% of the portfolio value. For sets with fixed marginal distributions, on the other hand, model risk increases significantly and is three times higher with an average of 0.157\% of the portfolio value.
Consequently, model risk is higher when choosing a copula function from a group of feasible models compared to choosing the marginal distribution. Higher model risk in the choice of a copula means that risk forecasts in groups of models with fixed marginal distributions and varying copula function differ more from each other and are thus more dissimilar than risk forecasts of groups in which the marginal distribution varies. In absolute terms, this translates into an average difference of \$164 to \$498 for a portfolio with \$100,000 in value and a holding period of ten days. Looking at the median of the averaged sets, the model risk is 3.5 times higher with a fixed marginal distribution.

The result of significant higher model risk due to the choice of a copula function is robust when considering model risk of VaR forecasts with respect to randomly generated portfolio weights. Following Table \ref{tab:analysisI_robust_weights}, the average model risk of such model sets with varying copulas is 0.142\% of the portfolio in contrast to 0.057\% when using fixed copulas.

\begin{center}
	-- Insert Table \ref{tab:analysisI_robust_weights} about here. --
\end{center}

Besides, the result is robust with respect to both the choice of a model risk measure (see Table \ref{tab:analysisIV} and Figure \ref{fig:Rplot_AnalysisIV}) and the confidence level (see Table \ref{tab:analysisI_robust_lev} and Figure \ref{fig:Rplot_AnalysisI_robust_lev1}).
Figure \ref{fig:Rplot_AnalysisI_robust_lev1} shows, in addition to the significant rise in model risk due to the choice of a copula, that increasing the confidence level from 95\% to 99.9\% triples the model risk, in case of model sets with fixed copula from 0.034\% to 0.108\% and for fixed marginal distributions from 0.078\% to 0.293\%.

\begin{center}
	-- Insert Tables \ref{tab:analysisIV} and \ref{tab:analysisI_robust_lev} about here. --
\end{center}

As we stated before, we do not compare VaR and ES results due to risk measure specific confidence levels as well as backtests and consequently different model sets in terms of constellation and number of models see Section \ref{sec:analysisBacktesting}.
Also for the 97.5\% ES, the results show that the model risk increases significantly with the choice of a copula function. With an average of 0.068\% in comparison to 0.058\% of the portfolio value, the average model risk is higher with varying compared to fixed copulas see Table \ref{tab:analysisI}. For a portfolio of \$100,000, this represents \$216 as opposed to \$184 in risk.  

Also for the ES, the result of an significant increase in model risk when choosing a copula function is robust with respect to 
the portfolio weighting (see Table \ref{tab:analysisI_robust_weights}), the model risk measure (see Table \ref{tab:analysisIV} and Figure \ref{fig:Rplot_AnalysisIV}) and the confidence level (see Table \ref{tab:analysisI_robust_lev} and Figure \ref{fig:Rplot_AnalysisI_robust_lev2}). Only for a confidence level of 95\% the average model risk shows identical values of 0.045\% of the portfolio value for fixed and varying copulas. However, when considering the median a model set with fixed marginal distributions shows a higher model risk in value (0.041\%) than a model set with fixed copulas (0.033\%). Again, increasing the confidence level from 95\% to 99.9\% results in a higher model risk. Here, the model risk triples for sets with fixed copula (from 0.045\% to 0.136\%), while the model risk increases sixfold for fixed marginal distributions (from 0.045\% to 0.270\%), see Table \ref{tab:analysisI_robust_lev}.

\begin{figure}[htbp]
	\caption{Potential portfolio value under financial distress based on the 99\% VaR for fixed copula functions}
	\scriptsize
	This figure illustrates the economic significance of model risk arising from the choice of the GARCH-type model and marginal distribution. Here, we focus on the 99\% VaR for model sets with fixed copula functions and varying univariate marginal distributions for a well diversified portfolio (\$100,000) and a 10 day holding period. We provide the portfolio value minus the 20th and the 80th percentile of VaR forecasts from all models within each model set that passed the duration-based backtest by \cite{Christoffersen.2004} on a daily basis. Values are averaged over various copula specifications. This corresponds to the potential portfolio value under financial distress according to the more (80th percentile) or less (20th percentile) conservative VaR models. The sample period is November 4, 2004 until December 31, 2018.
	
	\begin{center}
		\includegraphics[width=0.9\linewidth]{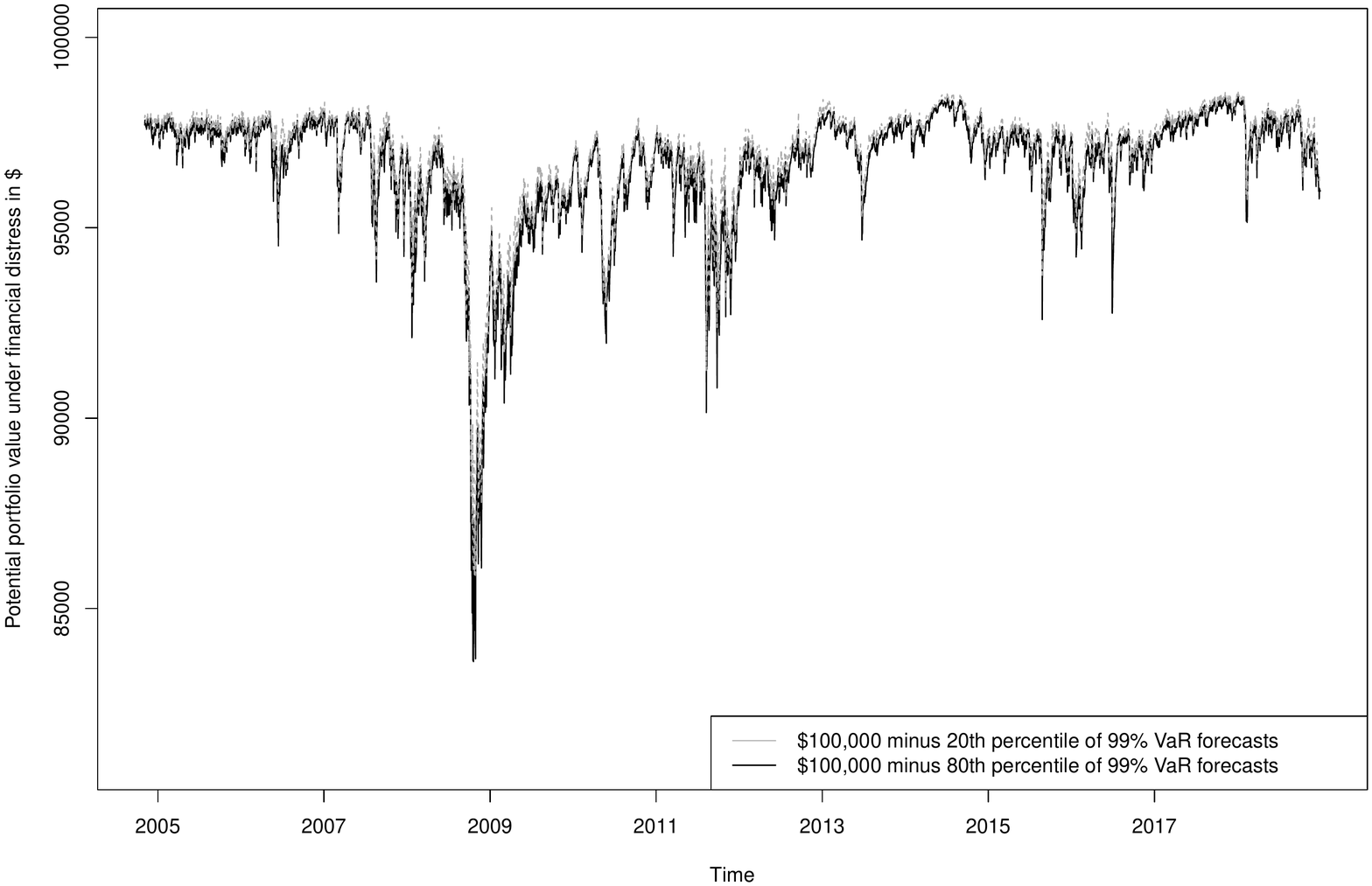}
	\end{center}
	\label{fig:minimumPortfolioValueVaRCopulasFixed.pdf}
\end{figure}

\begin{figure}[htbp]
	\caption{Potential portfolio value under financial distress based on the 99\% VaR for fixed marginal distributions}
	\vspace{0.2cm} \noindent {\justifying \scriptsize
	This figure illustrates the economic significance of model risk arising from the choice of the copula function. Here, we focus on the 99\% VaR for model sets with fixed univariate marginal distribution and varying copula functions for a well diversified portfolio (\$100,000) and a 10 day holding period. We provide the portfolio value minus the 20th and the 80th percentile of VaR forecasts from all models within each model set that passed the duration-based backtest by \cite{Christoffersen.2004} on a daily basis. Values are averaged over various marginal distributions. This corresponds to the potential portfolio value under financial distress according to the more (80th percentile) or less (20th percentile) conservative VaR models. The sample period is November 4, 2004 until December 31, 2018.}\\
	
	\begin{center}
		\includegraphics[width=0.9\linewidth]{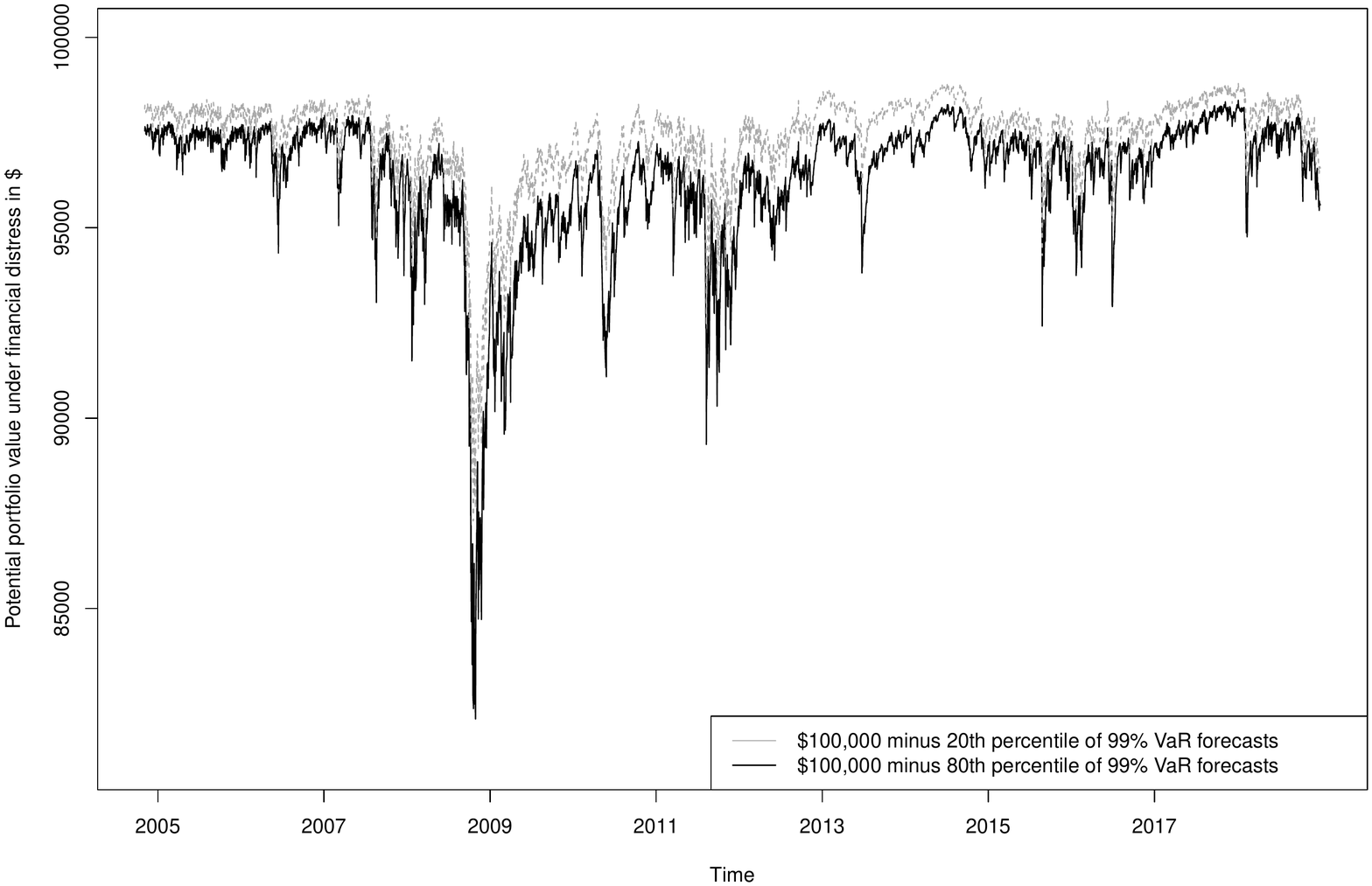}
	\end{center}
	\label{fig:minimumPortfolioValueVaRMarginalsFixed.pdf}
\end{figure}

We extend the abstract representation of model risk as the  mean absolute deviation of various risk forecasts by illustrating the impact of model choice regarding the multivariate dependence structure as well as univariate marginals on a portfolio with value \$100,000. This is illustrated in Figure \ref{fig:minimumPortfolioValueVaRCopulasFixed.pdf} for the 99\% VaR and model sets with fixed copulas. 
In this case the model risk is induced by the choice of a GARCH model and the corresponding marginal distribution.
The 20th and 80th percentile of the risk estimates are subtracted from the portfolio value to represent the range of the impact of the model choice over time. The resulting model risk range per day is averaged over the different copulas.
Analogously, we illustrate the impact of the choice of a copula model in Figure \ref{fig:minimumPortfolioValueVaRMarginalsFixed.pdf} by fixing the marginal distribution. 
Again, the model risk is illustrated over a range in terms of the 20th and 80th percentile of the ten-day-ahead VaR estimates at the 99\% confidence level and a portfolio value of \$100,000. Figure \ref{fig:minimumPortfolioValueVaRMarginalsFixed.pdf} shows a higher range of portfolio value over time, illustrating clearly the result that choosing a copula function generates higher model risk than choosing a GARCH-type model and a marginal distribution.
 
 \begin{center}
 	-- Insert Figures \ref{fig:minimumPortfolioValueVaRCopulasFixed.pdf} and \ref{fig:minimumPortfolioValueVaRMarginalsFixed.pdf} about here. --
 \end{center}

\section{Model risk for models in the model confidence set}
\label{sec:analysisMCS}
The MCS procedure by \cite{Hansen.2011} yields a set of models that contains the best model with a given confidence. 
That is, the MCS procedure does not assume a particular model to be the true or best one. Instead, it yields a set of models that can be seen as an analogue to a confidence interval for parameters, see Section \ref{sec:mcsTheoreticalBackground} for details. As in this paper we study the model risk of risk models that are valid ex-ante and but provide differing forecasts, the MCS procedure fits perfectly into our study. We apply the MCS procedure to the set of models that passed the (daily) backtests to further narrow the set of candidate models.  
\footnote{To employ the MCS procedure outlined in Section \ref{sec:mcsTheoreticalBackground} we use the \emph{MCS} R-package by \cite{Catania.2017}. We mainly rely on the default parameters. In particular, we adopt the choice of 15\% for the confidence level $\alpha$. We use the test statistic $T_R$, see Section \ref{sec:MCS_internetAppendix} for more details. The MCS procedure is computationally very expensive, especially for our large set of up to 180 models. We therefore employ the MCS procedure only every 20 days to the set of models that has not been rejected by the respective backtest on that day. Computations are performed for all confidence levels (95\%, 97.5\%, 99\%, and 99.9\%) based on a moving window of 500 days and 1000 bootstrapped samples. We additionally employ the MCS procedure on a daily basis for the 99\% VaR and the 97.5\% ES estimates based on 100 bootstrapped samples. The results remain qualitatively unchanged. Furthermore, we apply the MCS procedure not only to the equally weighted portfolio but also to 10 portfolios with randomly generated portfolio weights (again on a 20 day basis). The results stay qualitatively the same. All computations are performed on the Big-Data-Cluster Galaxy provided by the University Computing Center at Leipzig University.} 
We then determine the model risk corresponding to the models in the MCS to analyze if model risk can be reduced by these means.\footnote{Note that opposed to \cite{Santos.2013} our aim is not to determine the \emph{best} VaR or ES model nor to rank models by their forecasting accuracy. Instead we quantify the extent of non-conformity of the risk forecasts.} 

\begin{figure}[htbp]
	\caption{Model risk of the VaR before and after applying the MCS procedure}
	\vspace{0.2cm} \noindent {\justifying \scriptsize This figure compares the model risk associated with one day ahead 99\% VaR forecasts before and after applying the MCS procedure by \cite{Hansen.2011}. Model risk is measured in terms of the mean absolute deviation (\emph{mad}) of risk forecasts by all multivariate VaR models that passed the duration-based test by \cite{Christoffersen.2004} (\emph{no MCS}). Subsequently, we apply the MCS procedure to those models and recalculate model risk (\emph{MCS}). Details on the method and its implementation can be found in Sections \ref{sec:mcsTheoreticalBackground} and \ref{sec:analysisMCS}, respectively. Values are given in percent of the portfolio value between November 4, 2004 and December 31, 2018.
	}\\
	
	\begin{center}
	\includegraphics[width=0.9\linewidth]{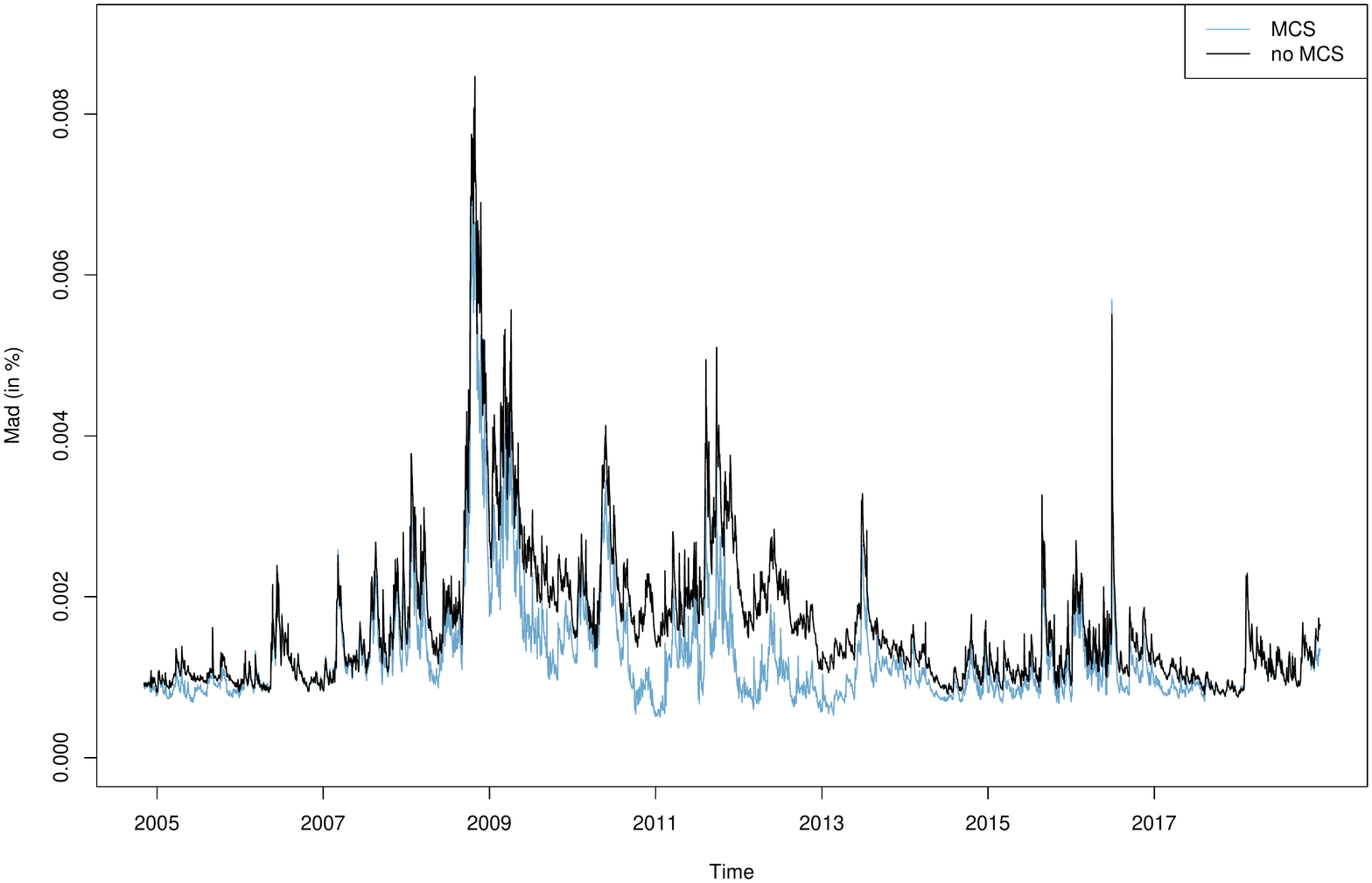}
	\label{fig:modelRiskVaRBeforeAndAfterMCS}
	\end{center}
\end{figure}

The main results are summarized in Table \ref{tab:MCSvsNoMCS} which compares model risk values before and after applying the MCS procedure in various periods of time. For the whole period, model risk before applying the MCS procedure is on average 0.165\% for the VaR and 0.092\% for the ES. These values are statistically significant reduced by 23\% to 0.127\% for the VaR and by 3\% to 0.089\% for the ES when considering only the models in the MCS.\footnote{When considering other measures of model risk (\emph{sd
} and \emph{iqr}) results are simlar and reductions range between 20\% and 30\% for VaR and 1\% and 3\% for ES forecasts.} For both VaR and ES this approach reduces model risk not only in the whole period but also in all sub-periods (\emph{pre-crisis, crisis}, and \emph{post-crisis}).\footnote{All reductions are statistically significant at the 1\% level.} The reduction of model risk for ES is, however, very small with declines ranging between 2\% and 5\%. Opposed to this, we can achieve substantial reductions in VaR model risk (8\% in the pre-crisis, 25\% in the crisis, and even 27\% in the post-crisis period). A graphical representation of the model risk of VaR forecasts before and after applying the MCS procedure can be found in Figure \ref{fig:modelRiskVaRBeforeAndAfterMCS}. 

\begin{center}
	-- Insert Figure \ref{fig:modelRiskVaRBeforeAndAfterMCS} and Table \ref{tab:MCSvsNoMCS} about here. --
\end{center}

\begin{figure}[htbp]
	\caption{Number of VaR models before and after applying the MCS procedure}
	\vspace{0.2cm} \noindent {\justifying \scriptsize 
		This figure shows the number of 99\% VaR models before and after applying the MCS procedure by \cite{Hansen.2011}, see Section \ref{sec:mcsTheoreticalBackground}. In both cases, only those models (out of 180) that were not rejected by the duration-based VaR backtest by \cite{Christoffersen.2004} enter into the MCS procedure. The sample period is November 4, 2004 until December 31, 2018.	
	}\\
	
	\begin{center}
	\includegraphics[width=0.9\linewidth]{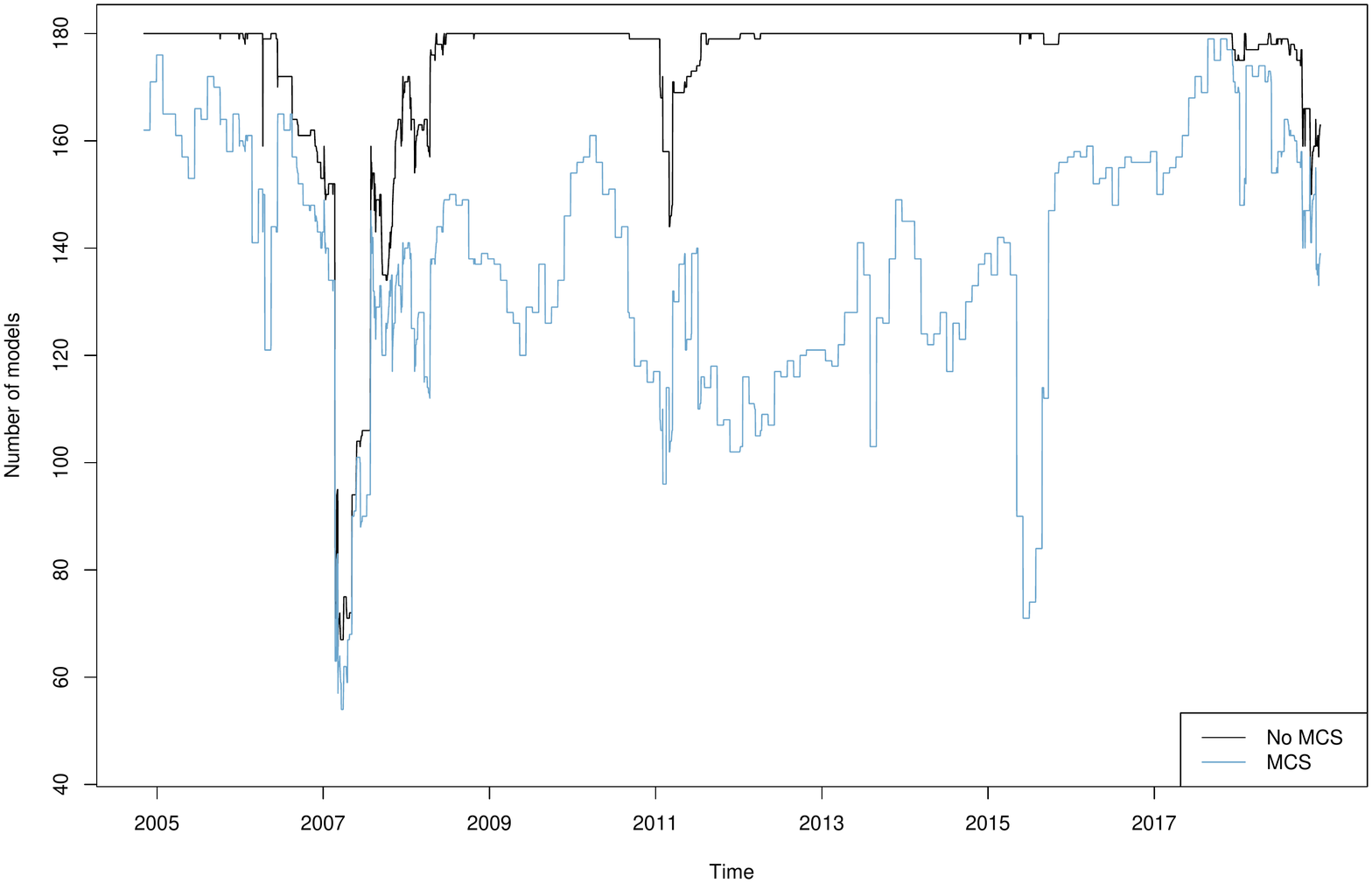}
	\label{fig:numberOfVaRModelsBeforeAndAfterMCS}
	\end{center}
\end{figure}

The reduction of model risk that can be achieved corresponds largely to the percentage of models that is excluded by the MCS procedure additionally to the models that have already been removed from the set of candidate models due to the backtests.\footnote{Note that the set of models that passed the backtests varies over time as the backtests are performed based on a moving window.} On average, only 8\% of ES models that have passed the respective backtest are excluded by the MCS procedure while the same is true for 21\% of the VaR models.\footnote{The difference in the percentages of VaR and ES models that are excluded via the MCS procedure can partially be explained by the fact that the loss function for the ES in the MCS framework is closely related to the conditional calibration backtest by \cite{Nolde.2017} while the same is not true in case of the VaR, see Sections  \ref{sec:backtests} and \ref{sec:mcsTheoreticalBackground} for more details.} However, after applying the MCS procedure on average less ES models (62\%) than VaR models (77\%) are left over. This is due to the fact that only those models that have not been rejected in the respective backtest enter into the MCS procedure and a lower fraction of ES models (67\%) has not been rejected (97\% for the VaR). While the percentage of models excluded is relatively stable over time for ES models, it varies for VaR models. In the pre-crisis period about 11\% of VaR models are excluded (additionally to the ones rejected by the backtest) while in the crisis and post-crisis period 24\% and 23\% are excluded, respectively. 
A graphical representation of the number of models before and after applying the MCS procedure can be found in Figure \ref{fig:numberOfVaRModelsBeforeAndAfterMCS}. By excluding those models that are inferior\footnote{Comparisons between different models are based on loss functions, see Section \ref{sec:mcsTheoreticalBackground} for details.} to the ones remaining in the MCS, model risk in the post-crisis period can be  reduced substantially to the level of the pre-crisis period. 

\begin{center}
	-- Insert Figure \ref{fig:numberOfVaRModelsBeforeAndAfterMCS} about here. --
\end{center}

\section{Model risk and backtesting}
\label{sec:analysisBacktesting}
In line with our notion of model risk as uncertainty on the model choice itself when having to choose among many possible alternative models \citep[cf. ][]{Cont., Danielsson.2016}, we include many different risk forecasts into the calculation of our model risk measure. We consider various GARCH-type models for the marginals and copula functions for the dependence structure yielding 180 different multivariate model specifications. However, to prevent our results from being biased by erroneous risk forecasts due to misspecified models, we first perform backtests to determine a set of valid candidate models on a daily basis. 

\begin{figure}[htbp]
	\caption{Number of market risk models passing the backtest}
	\scriptsize
    This figure shows the daily number of multivariate market risk models passing the respective backtest, see Section \ref{sec:backtests} for details. We consider risk forecasts associated with the one day ahead 99\% VaR and 97.5\% ES for a well diversified portfolio from November 4, 2004 to December 31, 2018.
    
	\begin{center}
		\includegraphics[width=0.9\linewidth]{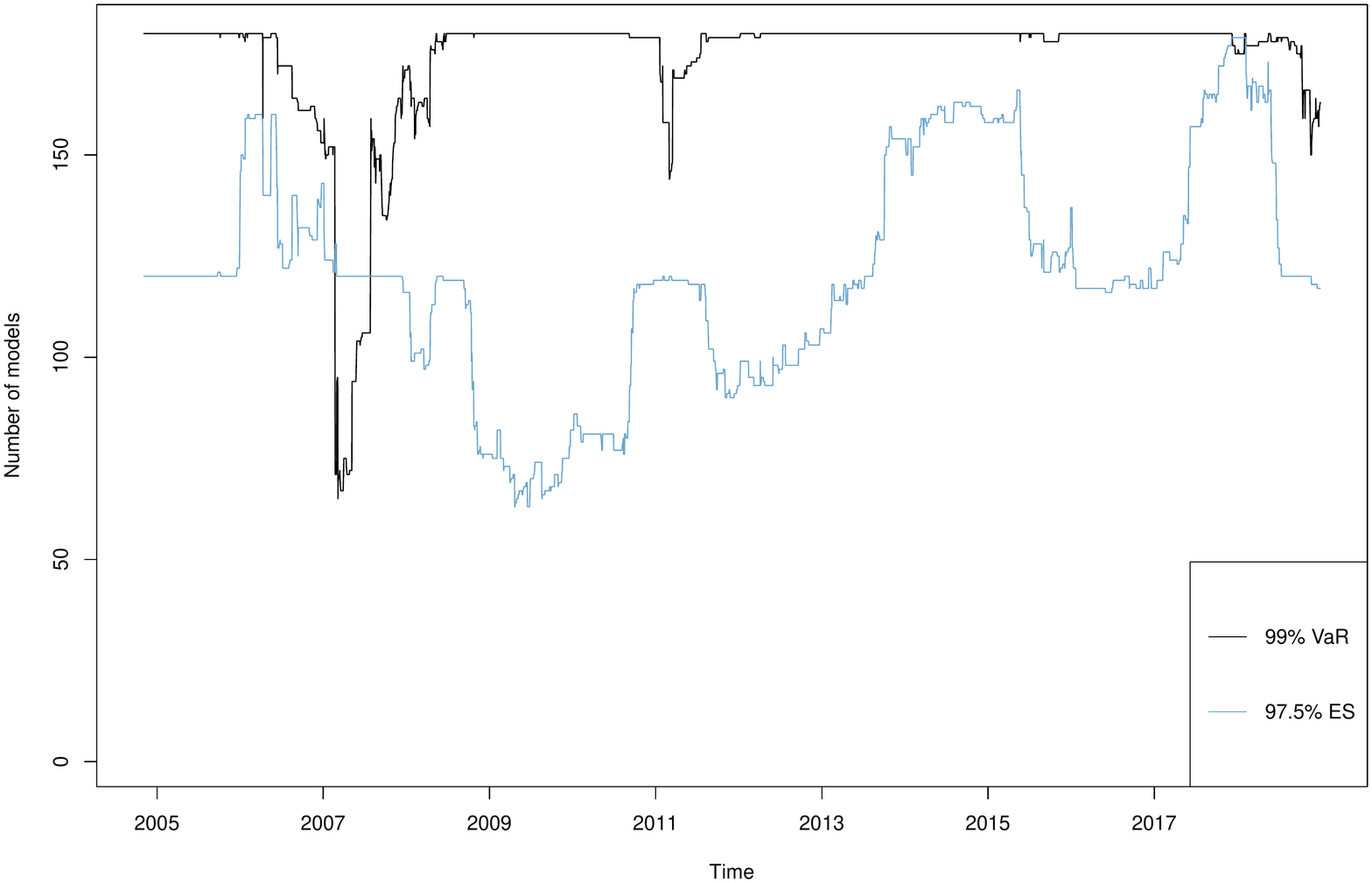} 
	\end{center}
	\label{fig:Rplot_Backtesting_AnalysisI}
\end{figure}

In our main analysis, we rely on the duration-based VaR backtest by \cite{Christoffersen.2004} and the conditional calibration ES backtest by \cite{Nolde.2017}. Table \ref{tab:backtesting_analysisI} provides summary statistics on the number of models passing  these backtests from November 4, 2004 until December 31, 2018.\footnote{The backtests are performed based on a moving window of 500 days and a confidence level of 99\%. For VaR and ES estimates themselves, we consider the confidence levels 95\%, 97.5\%, 99\%, and 99.9\%.} After having applied the backtest for the 99\%~VaR, there remain  on average 174 models while only 121 97.5\% ES models pass the ES backtest. 
This difference might be explained by the fact that the conditional calibration test is a joint VaR and ES backtest while the duration-based VaR backtest is not (see Section \ref{sec:backtests} for details). Figure \ref{fig:Rplot_Backtesting_AnalysisI} illustrates the number of VaR and ES models that are not rejected by the backtests over time.

\begin{center}
	-- Insert Figure \ref{fig:Rplot_Backtesting_AnalysisI} and Table \ref{tab:backtesting_analysisI} about here. --
\end{center}

For robustness, we provide summary statistics on the number of models that enter into the calculation of model risk when considering alternative backtests or when using different specifications in Table \ref{tab:backtesting_analysisIII}.

\begin{center}
	-- Insert Table \ref{tab:backtesting_analysisIII} about here. --
\end{center}

 When relying on the dynamic quantile test by \cite{Engle.2004}, on average 71 VaR models are not rejected while the same is true for 127 ES models in the exceedance residual test by \cite{McNeil.2000}. 
Using a moving window of 1000 days instead of 500 days reduces the average number of models to 145 for the VaR and 82 for the ES while a fixed window spanning the entire sample period\footnote{Note that the usage of a fixed window is not feasible in practice as it introduces a look-ahead-bias into the evaluation of risk models. However, we include results to provide a more complete picture.} leads on average to 113 VaR and 17 ES models. 
In the following, we focus on the impact of different backtesting specifications on our model risk estimates.

\subsection{All multivariate models}
Summary statistics for our main measure of model risk (mad) are presented in Table \ref{tab:robustnessBacktestsAllMultModels}. There, we compare the results of our main analysis to model risk estimates obtained by using no backtests or by employing alternative backtests. For robustness, we also add results when using our main backtests with a fixed estimation window and a moving window of 1000 days (instead of 500 days), respectively. 

\begin{center}
	-- Insert Table \ref{tab:robustnessBacktestsAllMultModels} about here. --
\end{center}

Over the entire sample period 
model risk (mad) for the 99\% VaR is on average 0.165\% when determining the set of candidate models via the duration-based backtest.\footnote{In our main analysis we rely on a moving estimation window of 500 days. When instead using a moving window of 1000 days we obtain an average model risk of 0.202\% and when building on a fixed estimation window an average model risk of 0.129\%. Note that although employing a fixed estimation window substantially decreases average model risk, doing so introduces a look-ahead bias to our model risk estimates and is therefore not feasible in practice.} When calculating model risk based on all 180 risk forecasts (i.e., without employing a backtest), model risk for the VaR is on average slightly lower (0.156\%). When instead relying on the dynamic quantile test by \cite{Engle.2004}, average model risk is substantially lower (0.079\%).\footnote{This can be explained by the fact that the dynamic quantile test on average rejects a larger fraction of models (60.6\%) than the duration-based backtest (3.3\%). However, there are periods of time (in total 14.0\% of all days corresponding to roughly 2 years of our sample period) with none of the 180 models passing the dynamic quantile test, which is the main reason for not using this backtest in our main analysis. Further details can be found in Table \ref{tab:backtesting_analysisIII}.} \todo{\tiny Unterschiede statistisch signifikant?} Average model risk for the 97.5\% ES over the entire sample period is 0.092\% in our main analysis\footnote{When using a moving window of 1000 days (instead of 500 days in our main analysis), we obtain an average model risk of 0.091\%. For a fixed estimation window average model risk is 0.078\%.} (conditional calibration backtest) and 0.163\% when not using any backtest. For the exceedance residual test by \cite{McNeil.2000} average model risk (0.114\%) is higher than model risk for the conditional calibration backtest, but still much lower than in the case of not using any backtest.

\subsection{The subgroups}
\begin{figure}[htbp]
	\caption{Average model risk without applying backtests (model sets with fixed and varying copula only)}
	\scriptsize
    This figure shows the average model risk associated with one day ahead 99\% VaR (first panel) and 97.5\% ES (second panel) forecasts for a well diversified portfolio per group. \textit{Group 1} (G1) includes all model sets in which a copula function is fixed while varying the marginal distribution.
    \textit{Group 2} (G2) contains analogously the model sets with fixed marginal distribution and varying copula. Model risk is measured in terms of the mean absolute deviation (mad) of one day ahead forecasts by various risk models within a model set. Values are calculated on a daily basis between November 4, 2004 until December 31, 2018 in percent of the portfolio value. In contrast to our baseline analysis, the models did not have to pass a backtest.
	\begin{center}
		\begin{subfigure}[t]{\linewidth}
			\centering
			\includegraphics[width=0.8\linewidth]{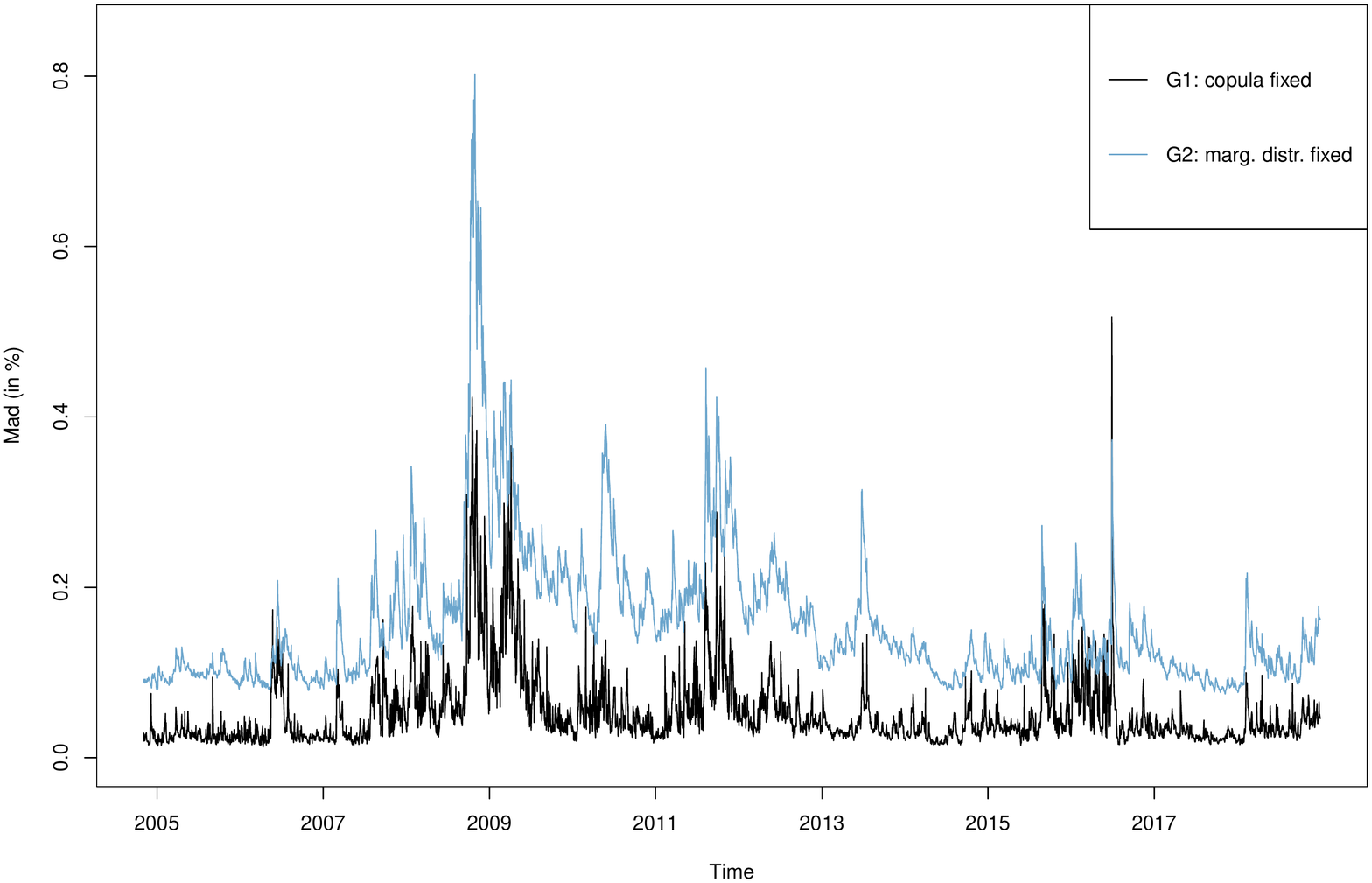}
			\caption{Average model risk (99\% VaR)}\label{fig:Rplot_AnalysisIII_1}
			\end{subfigure}
			\begin{subfigure}[t]{\linewidth}
			\centering
			\includegraphics[width=0.8\linewidth]{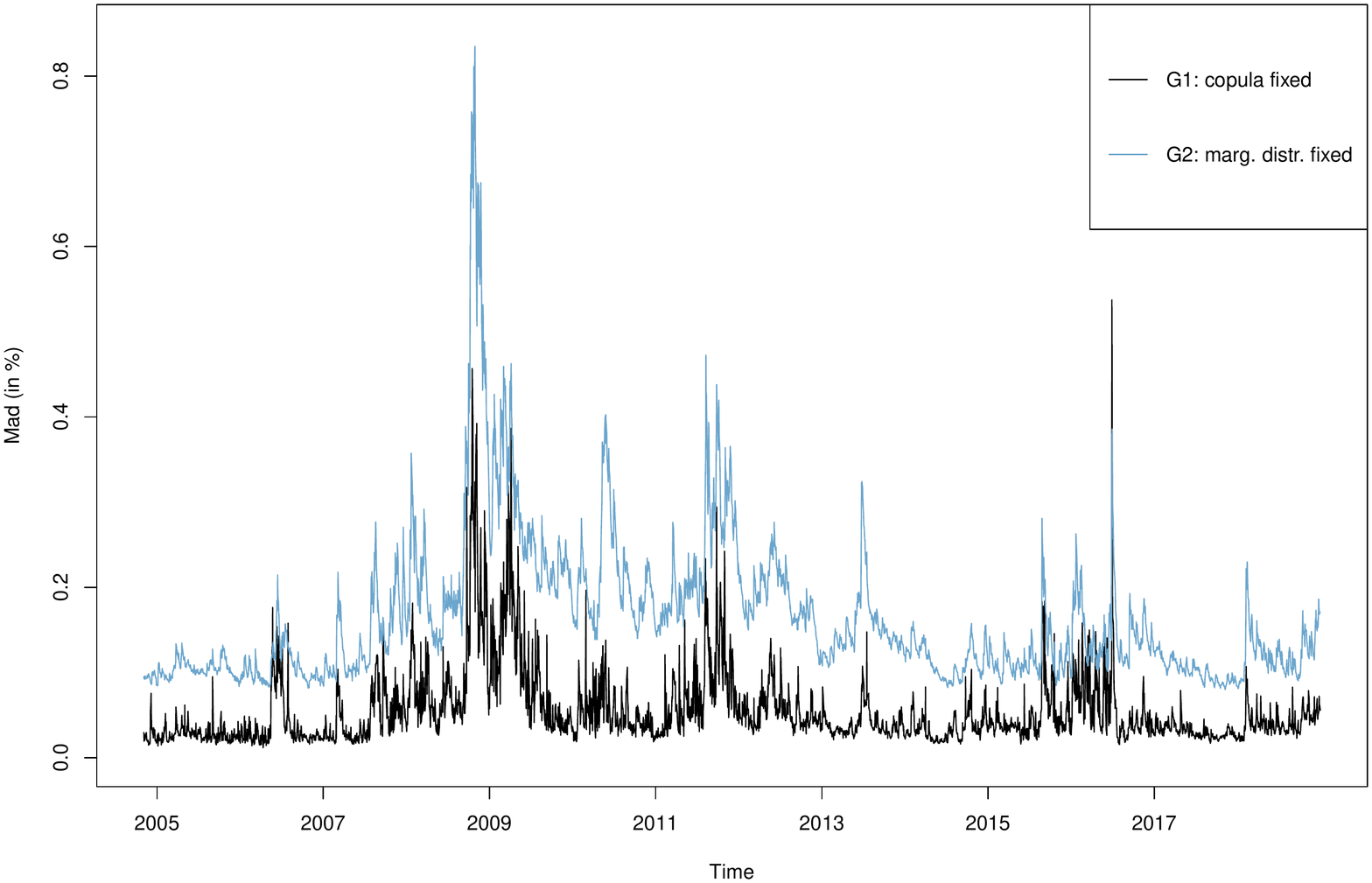}
			\caption{Average model risk (97.5\% ES)}\label{fig:Rplot_AnalysisIII_2}
		\end{subfigure}
	\end{center}
\label{fig:Rplot_AnalysisIII}
\end{figure}

Table \ref{tab:analysisIII} presents summary statistics of average model risk when using alternative backtests or no backtest at all. Again, the focus is on the comparison of model sets with fixed and model sets with varying copula function. For the 99\% VaR we find that the average model risk remains identical when we do not backtest. This is true for model sets with fixed (0.052\%) as well as for model sets with varying (0.157\%) copula function and can be explained by the fact that the duration-based backtest by \cite{Christoffersen.2004} rejects on average only about 3.3\% of VaR market risk models.
In contrast, the conditional calibration backtest by \cite{Nolde.2017} rejects on average about 33\% of models forecasting ES estimates. 
For model sets with fixed copula function, the model risk is almost similar (deterioration of 0.003 percentage points) if no backtest is performed. On the other hand, for model sets with varying copula and consequently fixed marginal distribution, the average model risk is reduced from 0.164\% to 0.068\%. In addition, Figure \ref{fig:Rplot_AnalysisIII} shows that our main result of increasing model risk by choosing a copula function is robust to no prior backtesting.

\begin{center}
	-- Insert Figure \ref{fig:Rplot_AnalysisIII} and Table \ref{tab:analysisIII} about here. --
\end{center}

For the 99\% VaR using the dynamic quantile test by \cite{Engle.2004}, the average model risk is almost not reduced for groups with fixed copula (from 0.052\% to 0.051\%) and more than three times reduced for groups with varying copula (from 0.157\% to 0.042\%).\footnote{When using a moving window of 1000 days (instead of 500 days in our main analysis), we obtain an average model risk of 0.053\% (0.159\%) for model sets with fixed (varying) copula. For a fixed estimation window average model risk is 0.048\% (0.126\%).} In contrast to our baseline backtest, 60.6\% of the market risk models are discarded on average. 
When using the exceedance residual ES backtest by \cite{McNeil.2000}, average model risk for sets with fixed copulas is robust to our baseline analysis (0.058\%). The alternative ES backtest rejects 29.7\% of market risk models on average. For the group of model sets with fixed marginal distributions, the alternative backtest leads to a slightly higher average model risk (0.077\%) than our baseline (0.068\%). Again, the average model risk is reduced in comparison to no prior backtest (0.164\%).\footnote{When using a moving window of 1000 days (instead of 500 days in our main analysis), we obtain an average model risk of 0.065\% (0.047\%) for model sets with fixed (varying) copula. For a fixed estimation window average model risk is 0.039\% (0.034\%).}  
These results highlight that model risk depends on the choice of the backtests, some of which are able to reduce average model risk (compared in particular to using no backtest). As a consequence, employing backtests can be seen as additional means for reducing model risk. A further analysis of the choice of backtests with regard to model risk is, however, beyond the scope of this paper.

\section{Conclusion}
\label{sec:conclusion}

In this paper, we study the model risk inherent in Copula-GARCH models used for forecasting financial risk. More precisely, we forecast the VaR and ES for a large number of multivariate portfolios using a variety of Copula-GARCH models. We then analyze different groups of models in which we fix either the marginals, the copula, or neither in a comprehensive empirical study to identify the main source of model risk in multivariate risk forecasting. As our first main result, we find that Copula-GARCH models come with considerable model risk that is economically significant. Interestingly, and as our second main result, we find that copulas account for considerably more model risk than marginals in multivariate models with the choice of marginal model having only a small effect on overall model uncertainty. We then propose the use of the model confidence set procedure to narrow down the set of available models and reduce model risk for Copula-GARCH risk models using ready-to-use backtests for VaR and ES, respectively. Our proposed approach leads to a significant improvement in the mean absolute deviation of one day ahead forecasts by our various candidate risk models.

The findings of our analysis stress the importance of an adequate modeling of the dependence structure inherent in financial portfolios. While the choice of marginal models is not negligible, it is however of lesser importance than selecting the right parametric copula model. In this respect, our findings are reassuring as the majority of previous papers in this field have solely concentrated on copula modeling and have relied on standard GARCH(1,1)-models for the marginals. Our quantification of the degree of model risk caused by the large set of candidate parametric copula families, however, shows that multivariate models include an economically significant amount of model risk. Using the model confidence set approach seems to alleviate this danger to some degree. Finally, our findings are of high relevance for supervisors in the banking and insurance sector as we illustrate the need for carefully checking the adequacy of a multivariate copula-based risk model.

\clearpage
\singlespacing
{\small
\bibliography{Literature}

\begin{thebibliography}{87}
\newcommand{\enquote}[1]{``#1''}
\expandafter\ifx\csname natexlab\endcsname\relax\def\natexlab#1{#1}\fi

\bibitem[\protect\citeauthoryear{Aas and Berg}{Aas and Berg}{2009}]{Aas.2009}
\textsc{Aas, K. and D.~Berg} (2009): \enquote{Models for construction of
  multivariate dependence -- a comparison study,} \emph{The European Journal of
  Finance}, 15, 639--659.

\bibitem[\protect\citeauthoryear{Aas, Czado, Frigessi, and Bakken}{Aas
  et~al.}{2009}]{AAS2009182}
\textsc{Aas, K., C.~Czado, A.~Frigessi, and H.~Bakken} (2009):
  \enquote{Pair-copula constructions of multiple dependence,} \emph{Insurance:
  Mathematics and Economics}, 44, 182--198.

\bibitem[\protect\citeauthoryear{Acerbi and Sz{\'e}kely}{Acerbi and
  Sz{\'e}kely}{2014}]{Acerbi.2014}
\textsc{Acerbi, C. and B.~Sz{\'e}kely} (2014): \enquote{Backtesting expected
  shortfall,} \emph{Risk}, 76--81.

\bibitem[\protect\citeauthoryear{Alexander and Sarabia}{Alexander and
  Sarabia}{2012}]{Alexander.2012}
\textsc{Alexander, C. and J.~M. Sarabia} (2012): \enquote{Quantile uncertainty
  and value-at-risk model risk,} \emph{Risk Analysis}, 32, 1293--1308.

\bibitem[\protect\citeauthoryear{Ausin and Lopes}{Ausin and
  Lopes}{2010}]{Ausin.2010}
\textsc{Ausin, M.~C. and H.~F. Lopes} (2010): \enquote{Time-varying joint
  distribution through copulas,} \emph{Computational Statistics {\&} Data
  Analysis}, 54, 2383--2399.

\bibitem[\protect\citeauthoryear{Barendse, Kole, and {van Dijk}}{Barendse
  et~al.}{2019}]{Barendse.2019}
\textsc{Barendse, S., E.~Kole, and D.~J. {van Dijk}} (2019):
  \enquote{Backtesting Value-at-Risk and Expected Shortfall in the Presence of
  Estimation Error,} \emph{SSRN Electronic Journal}.

\bibitem[\protect\citeauthoryear{{Basel Committee on Banking
  Supervision}}{{Basel Committee on Banking
  Supervision}}{2019}]{BaselCommitteeonBankingSupervision.2019}
\textsc{{Basel Committee on Banking Supervision}} (2019): \emph{Minimum capital
  requirements for market risk}, Basel: {Basel Committee on Banking
  Supervision}.

\bibitem[\protect\citeauthoryear{Bassetti, {De Giuli}, Nicolino, and
  Tarantola}{Bassetti et~al.}{2018}]{Bassetti}
\textsc{Bassetti, F., M.~E. {De Giuli}, E.~Nicolino, and C.~Tarantola} (2018):
  \enquote{Multivariate dependence analysis via tree copula models: An
  application to one-year forward energy contracts,} \emph{European Journal of
  Operational Research}, 269, 1107--1121.

\bibitem[\protect\citeauthoryear{Bayer and Dimitriadis}{Bayer and
  Dimitriadis}{2020{\natexlab{a}}}]{SebastianBayer.2020}
\textsc{Bayer, S. and T.~Dimitriadis} (2020{\natexlab{a}}): \enquote{esback:
  Expected Shortfall Backtesting,} .

\bibitem[\protect\citeauthoryear{Bayer and Dimitriadis}{Bayer and
  Dimitriadis}{2020{\natexlab{b}}}]{Bayer.2020}
---\hspace{-.1pt}---\hspace{-.1pt}--- (2020{\natexlab{b}}):
  \enquote{Regression-Based Expected Shortfall Backtesting,} \emph{Journal of
  Financial Econometrics}.

\bibitem[\protect\citeauthoryear{Bedford and Cooke}{Bedford and
  Cooke}{2001}]{Bedford.2001}
\textsc{Bedford, T. and R.~M. Cooke} (2001): \enquote{Probability Density
  Decomposition for Conditionally Dependent Random Variables Modeled by Vines,}
  \emph{Annals of Mathematics and Artificial Intelligence}, 32, 245--268.

\bibitem[\protect\citeauthoryear{Bedford and Cooke}{Bedford and
  Cooke}{2002}]{Bedford.2002}
---\hspace{-.1pt}---\hspace{-.1pt}--- (2002): \enquote{Vines--a new graphical
  model for dependent random variables,} \emph{The Annals of Statistics}, 30,
  1031--1068.

\bibitem[\protect\citeauthoryear{Bernard, Kazzi, and Vanduffel}{Bernard
  et~al.}{2020}]{Bernard.2020}
\textsc{Bernard, C., R.~Kazzi, and S.~Vanduffel} (2020): \enquote{A Practical
  Approach to Quantitative Model Risk Assessment,} \emph{Working Paper}.

\bibitem[\protect\citeauthoryear{Bernard and Vanduffel}{Bernard and
  Vanduffel}{2015}]{BERNARD2015166}
\textsc{Bernard, C. and S.~Vanduffel} (2015): \enquote{A new approach to
  assessing model risk in high dimensions,} \emph{Journal of Banking \&\
  Finance}, 58, 166--178.

\bibitem[\protect\citeauthoryear{Bilgrau, Eriksen, Rasmussen, Johnsen, Dybkaer,
  and Boegsted}{Bilgrau et~al.}{2016}]{Bilgrau.2016}
\textsc{Bilgrau, A.~E., P.~S. Eriksen, J.~G. Rasmussen, H.~E. Johnsen,
  K.~Dybkaer, and M.~Boegsted} (2016): \enquote{GMCM: Unsupervised Clustering
  and Meta-Analysis Using Gaussian Mixture Copula Models,} \emph{Journal of
  Statistical Software}, 70, 1--23.

\bibitem[\protect\citeauthoryear{Bollerslev}{Bollerslev}{1986}]{Bollerslev.1986}
\textsc{Bollerslev, T.} (1986): \enquote{Generalized autoregressive conditional
  heteroskedasticity,} \emph{Journal of Econometrics}, 31, 307--327.

\bibitem[\protect\citeauthoryear{Bollerslev}{Bollerslev}{2010}]{Bollerslev.2010}
---\hspace{-.1pt}---\hspace{-.1pt}--- (2010): \enquote{Glossary to ARCH (GARCH
  * ),} in \emph{Volatility and time series econometrics}, ed. by R.~F. Engle,
  M.~W. Watson, T.~Bollerslev, and J.~R. Russell, Oxford: {Oxford University
  Press}, Advanced texts in econometrics.

\bibitem[\protect\citeauthoryear{Boucher, Danielsson, Kouontchou, and
  Maillet}{Boucher et~al.}{2014}]{Boucher.2014}
\textsc{Boucher, C.~M., J.~Danielsson, P.~S. Kouontchou, and B.~B. Maillet}
  (2014): \enquote{Risk models-at-risk,} \emph{Journal of Banking {\&}
  Finance}, 44, 72--92.

\bibitem[\protect\citeauthoryear{Brechmann and Czado}{Brechmann and
  Czado}{2013}]{Brechmann.2013}
\textsc{Brechmann, E.~C. and C.~Czado} (2013): \enquote{Risk management with
  high-dimensional vine copulas: An analysis of the Euro Stoxx 50,}
  \emph{Statistics {\&} Risk Modeling}, 30.

\bibitem[\protect\citeauthoryear{Calabrese, Degl’Innocenti, and
  Osmetti}{Calabrese et~al.}{2017}]{Calabrese}
\textsc{Calabrese, R., M.~Degl’Innocenti, and S.~Osmetti} (2017):
  \enquote{The effectiveness of TARP-CPP on the US Banking Industry: a new
  copula-based approach,} \emph{European Journal of Operational Research}, 256,
  1029--1037.

\bibitem[\protect\citeauthoryear{Calabrese and Osmetti}{Calabrese and
  Osmetti}{2019}]{Calabrese2}
\textsc{Calabrese, R. and S.~A. Osmetti} (2019): \enquote{A new approach to
  measure systemic risk: A bivariate copula model for dependent censored data,}
  \emph{European Journal of Operational Research}, 279, 1053--1064.

\bibitem[\protect\citeauthoryear{Campbell}{Campbell}{2005}]{Campbell.}
\textsc{Campbell, S.~D.} (2005): \enquote{A Review of Backtesting and
  Backtesting Procedures,} \emph{FEDS Working Paper}, 21.

\bibitem[\protect\citeauthoryear{Catania and Bernardi}{Catania and
  Bernardi}{2017}]{Catania.2017}
\textsc{Catania, L. and M.~Bernardi} (2017): \enquote{MCS: Model Confidence Set
  Procedure,} .

\bibitem[\protect\citeauthoryear{Cherubini, Luciano, and Vecchiato}{Cherubini
  et~al.}{2004}]{Cherubini.2004}
\textsc{Cherubini, U., E.~Luciano, and W.~Vecchiato} (2004): \emph{Copula
  methods in finance}, Wiley finance series, Hoboken, NJ: {John Wiley {\&}
  Sons}.

\bibitem[\protect\citeauthoryear{Christoffersen}{Christoffersen}{2004}]{Christoffersen.2004}
\textsc{Christoffersen, P.} (2004): \enquote{Backtesting Value-at-Risk: A
  Duration-Based Approach,} \emph{Journal of Financial Econometrics}, 2,
  84--108.

\bibitem[\protect\citeauthoryear{Cont}{Cont}{2006}]{Cont.}
\textsc{Cont, R.} (2006): \enquote{MODEL UNCERTAINTY AND ITS IMPACT ON THE
  PRICING OF DERIVATIVE INSTRUMENTS,} \emph{Mathematical Finance}, 16,
  519--547.

\bibitem[\protect\citeauthoryear{Danielsson, James, Valenzuela, and
  Zer}{Danielsson et~al.}{2016}]{Danielsson.2016}
\textsc{Danielsson, J., K.~R. James, M.~Valenzuela, and I.~Zer} (2016):
  \enquote{Model risk of risk models,} \emph{Journal of Financial Stability},
  23, 79--91.

\bibitem[\protect\citeauthoryear{Danielsson and Zigrand}{Danielsson and
  Zigrand}{2006}]{Danielsson.2006}
\textsc{Danielsson, J. and J.-P. Zigrand} (2006): \enquote{On time-scaling of
  risk and the square-root-of-time rule,} \emph{Journal of Banking {\&}
  Finance}, 30, 2701--2713.

\bibitem[\protect\citeauthoryear{{David Ardia}, {Kris Boudt}, and {Leopoldo
  Catania}}{{David Ardia} et~al.}{2019}]{DavidArdia.2019}
\textsc{{David Ardia}, {Kris Boudt}, and {Leopoldo Catania}} (2019):
  \enquote{Generalized Autoregressive Score Models in R: The GAS Package,}
  \emph{Journal of Statistical Software}, 88, 1--28.

\bibitem[\protect\citeauthoryear{Diebold, Hickman, Inoue, and
  Schuermann}{Diebold et~al.}{1997}]{Diebold.1997}
\textsc{Diebold, F., A.~Hickman, A.~Inoue, and T.~Schuermann} (1997):
  \enquote{Converting 1-Day Volatility to h-Day Volatitlity: Scaling by Root-h
  is Worse Than You Think,} \emph{Center for Financial Institutions Working
  Papers}.

\bibitem[\protect\citeauthoryear{Diebold and Mariano}{Diebold and
  Mariano}{1995}]{Diebold.1995}
\textsc{Diebold, F. and R.~Mariano} (1995): \enquote{Comparing Predictive
  Accuracy,} \emph{Journal of Business {\&} Economic Statistics}, 13, 253--263.

\bibitem[\protect\citeauthoryear{Ding, Granger, and Engle}{Ding
  et~al.}{1993}]{Ding.1993}
\textsc{Ding, Z., C.~W. Granger, and R.~F. Engle} (1993): \enquote{A long
  memory property of stock market returns and a new model,} \emph{Journal of
  Empirical Finance}, 1, 83--106.

\bibitem[\protect\citeauthoryear{Di{\ss}mann, Brechmann, Czado, and
  Kurowicka}{Di{\ss}mann et~al.}{2013}]{DIMANN201352}
\textsc{Di{\ss}mann, J., E.~Brechmann, C.~Czado, and D.~Kurowicka} (2013):
  \enquote{Selecting and estimating regular vine copulae and application to
  financial returns,} \emph{Computational Statistics {\&} Data Analysis}, 59,
  52--69.

\bibitem[\protect\citeauthoryear{Engle}{Engle}{2002}]{Engle.2002}
\textsc{Engle, R.} (2002): \enquote{Dynamic Conditional Correlation,}
  \emph{Journal of Business {\&} Economic Statistics}, 20, 339--350.

\bibitem[\protect\citeauthoryear{Engle and Sheppard}{Engle and
  Sheppard}{2001}]{Engle.2001}
\textsc{Engle, R. and K.~Sheppard} (2001): \enquote{Theoretical and Empirical
  properties of Dynamic Conditional Correlation Multivariate GARCH,} \emph{NBER
  Working Paper}.

\bibitem[\protect\citeauthoryear{Engle}{Engle}{1982}]{Engle.1982}
\textsc{Engle, R.~F.} (1982): \enquote{Autoregressive Conditional
  Heteroscedasticity with Estimates of the Variance of United Kingdom
  Inflation,} \emph{Econometrica}, 50, 987.

\bibitem[\protect\citeauthoryear{Engle and Manganelli}{Engle and
  Manganelli}{2004}]{Engle.2004}
\textsc{Engle, R.~F. and S.~Manganelli} (2004): \enquote{CAViaR,} \emph{Journal
  of Business {\&} Economic Statistics}, 22, 367--381.

\bibitem[\protect\citeauthoryear{Fernandez and Steel}{Fernandez and
  Steel}{1998}]{Fernandez.1998}
\textsc{Fernandez, C. and M.~F.~J. Steel} (1998): \enquote{On Bayesian Modeling
  of Fat Tails and Skewness,} \emph{Journal of the American Statistical
  Association}, 93, 359.

\bibitem[\protect\citeauthoryear{Fissler and Ziegel}{Fissler and
  Ziegel}{2016}]{Fissler.2016}
\textsc{Fissler, T. and J.~F. Ziegel} (2016): \enquote{HIGHER ORDER
  ELICITABILITY AND OSBAND'S PRINCIPLE,} \emph{The Annals of Statistics}, 44,
  1680--1707.

\bibitem[\protect\citeauthoryear{Fissler, Ziegel, and Gneiting}{Fissler
  et~al.}{2016}]{Fissler.2016b}
\textsc{Fissler, T., J.~F. Ziegel, and T.~Gneiting} (2016): \enquote{Expected
  Shortfall is jointly elicitable with Value at Risk - Implications for
  backtesting,} \emph{Risk Mag.}, 58--61.

\bibitem[\protect\citeauthoryear{Frongillo and {Ian A. Kash}}{Frongillo and
  {Ian A. Kash}}{2015}]{Frongillo.2015}
\textsc{Frongillo, R. and {Ian A. Kash}} (2015): \enquote{Vector-Valued
  Property Elicitation,} \emph{Conference on Learning Theory}, 710--727.

\bibitem[\protect\citeauthoryear{Genest, Rémillard, and Beaudoin}{Genest
  et~al.}{2009}]{GENEST2009199}
\textsc{Genest, C., B.~Rémillard, and D.~Beaudoin} (2009):
  \enquote{Goodness-of-fit tests for copulas: A review and a power study,}
  \emph{Insurance: Mathematics and Economics}, 44, 199--213.

\bibitem[\protect\citeauthoryear{Ghalanos}{Ghalanos}{2019}]{Ghalanos.2019}
\textsc{Ghalanos, A.} (2019): \enquote{rmgarch: Multivariate GARCH models,} .

\bibitem[\protect\citeauthoryear{Ghalanos}{Ghalanos}{2020}]{Ghalanos.2020}
---\hspace{-.1pt}---\hspace{-.1pt}--- (2020): \enquote{rugarch: Univariate
  GARCH models,} .

\bibitem[\protect\citeauthoryear{Glasserman and Xu}{Glasserman and
  Xu}{2014}]{Glasserman.2014}
\textsc{Glasserman, P. and X.~Xu} (2014): \enquote{Robust risk measurement and
  model risk,} \emph{Quantitative Finance}, 14, 29--58.

\bibitem[\protect\citeauthoryear{Glosten, Jagannathan, and Runkle}{Glosten
  et~al.}{1993}]{Glosten.1993}
\textsc{Glosten, L.~R., R.~Jagannathan, and D.~E. Runkle} (1993): \enquote{On
  the Relation between the Expected Value and the Volatility of the Nominal
  Excess Return on Stocks,} \emph{Journal of Finance}, 48, 1779--1801.

\bibitem[\protect\citeauthoryear{Gneiting}{Gneiting}{2011}]{Gneiting.2011}
\textsc{Gneiting, T.} (2011): \enquote{Making and Evaluating Point Forecasts,}
  \emph{Journal of the American Statistical Association}, 106, 746--762.

\bibitem[\protect\citeauthoryear{Green and Figlewski}{Green and
  Figlewski}{1999}]{Green.1999}
\textsc{Green, T.~C. and S.~Figlewski} (1999): \enquote{Market Risk and Model
  Risk for a Financial Institution Writing Options,} \emph{Journal of Finance},
  54, 1465--1499.

\bibitem[\protect\citeauthoryear{Grundke and Polle}{Grundke and
  Polle}{2012}]{Grundke}
\textsc{Grundke, P. and S.~Polle} (2012): \enquote{Crisis and risk
  dependencies,} \emph{European Journal of Operational Research}, 223,
  518–528.

\bibitem[\protect\citeauthoryear{Hansen, Lunde, and Nason}{Hansen
  et~al.}{2011}]{Hansen.2011}
\textsc{Hansen, P.~R., A.~Lunde, and J.~M. Nason} (2011): \enquote{The Model
  Confidence Set,} \emph{Econometrica}, 79, 453--497.

\bibitem[\protect\citeauthoryear{Hendricks}{Hendricks}{1996}]{Hendricks.1996}
\textsc{Hendricks, D.} (1996): \enquote{Evaluation of Value-at-Risk Models
  Using Historical Data,} \emph{SSRN Electronic Journal}.

\bibitem[\protect\citeauthoryear{Hentschel}{Hentschel}{1995}]{Hentschel.1995}
\textsc{Hentschel, L.} (1995): \enquote{All in the family Nesting symmetric and
  asymmetric GARCH models,} \emph{Journal of Financial Economics}, 39, 71--104.

\bibitem[\protect\citeauthoryear{Hofert, Kojadinovic, Maechler, and Yan}{Hofert
  et~al.}{2020}]{Hofert.2020}
\textsc{Hofert, M., I.~Kojadinovic, M.~Maechler, and J.~Yan} (2020):
  \enquote{copula: Multivariate Dependence with Copulas,} .

\bibitem[\protect\citeauthoryear{Hull and Suo}{Hull and Suo}{2002}]{JohnHull.}
\textsc{Hull, J. and W.~Suo} (2002): \enquote{A Methodology for Assessing Model
  Risk and Its Application to the Implied Volatility Function Model,}
  \emph{Journal of Financial and Quantitative Analysis}, 37, 297--318.

\bibitem[\protect\citeauthoryear{Jayech}{Jayech}{2016}]{Jayech}
\textsc{Jayech, S.} (2016): \enquote{The contagion channels of
  July–August-2011 stock market crash: A DAG-copula based approach,}
  \emph{European Journal of Operational Research}, 249, 631--646.

\bibitem[\protect\citeauthoryear{Joe}{Joe}{1996}]{Joe.1996}
\textsc{Joe, H.} (1996): \enquote{Families of m-Variate Distributions with
  Given Margins and m(m-1)/2 Bivariate Dependence Parameters,} \emph{Lecture
  Notes-Monograph Series}, 28, 120--141.

\bibitem[\protect\citeauthoryear{Joe}{Joe}{2001}]{Joe.1997}
---\hspace{-.1pt}---\hspace{-.1pt}--- (2001): \emph{Multivariate models and
  dependence concepts}, vol.~73 of \emph{Monographs on statistics and applied
  probability}, Boca Raton, Fla.: {Chapman {\&} Hall/CRC}, 1. crc reprint ed.

\bibitem[\protect\citeauthoryear{Jondeau and Rockinger}{Jondeau and
  Rockinger}{2006}]{JONDEAU2006827}
\textsc{Jondeau, E. and M.~Rockinger} (2006): \enquote{The Copula-GARCH model
  of conditional dependencies: An international stock market application,}
  \emph{Journal of International Money and Finance}, 25, 827--853.

\bibitem[\protect\citeauthoryear{Koenker and Bassett}{Koenker and
  Bassett}{1978}]{Koenker.1978}
\textsc{Koenker, R. and G.~Bassett} (1978): \enquote{Regression Quantiles,}
  \emph{Econometrica}, 46, 33.

\bibitem[\protect\citeauthoryear{Kole, Koedijk, and Verbeek}{Kole
  et~al.}{2007}]{KOLE20072405}
\textsc{Kole, E., K.~Koedijk, and M.~Verbeek} (2007): \enquote{Selecting
  copulas for risk management,} \emph{Journal of Banking \& Finance}, 31,
  2405--2423.

\bibitem[\protect\citeauthoryear{Kole, Markwat, Opschoor, and {van Dijk}}{Kole
  et~al.}{2017}]{Kole.2017}
\textsc{Kole, E., T.~Markwat, A.~Opschoor, and D.~{van Dijk}} (2017):
  \enquote{Forecasting Value-at-Risk under Temporal and Portfolio Aggregation,}
  \emph{Journal of Financial Econometrics}, 15, 649--677.

\bibitem[\protect\citeauthoryear{Kurowicka and Cooke}{Kurowicka and
  Cooke}{2006}]{Kurowicka.2006}
\textsc{Kurowicka, D. and R.~Cooke} (2006): \emph{Uncertainty analysis with
  high dimensional dependence modelling}, Wiley series in probability and
  statistics, Chichester: Wiley.

\bibitem[\protect\citeauthoryear{McNeil and Frey}{McNeil and
  Frey}{2000}]{McNeil.2000}
\textsc{McNeil, A.~J. and R.~Frey} (2000): \enquote{Estimation of tail-related
  risk measures for heteroscedastic financial time series: an extreme value
  approach,} \emph{Journal of Empirical Finance}.

\bibitem[\protect\citeauthoryear{McNeil, Frey, and Embrechts}{McNeil
  et~al.}{2005}]{McNeil.2005}
\textsc{McNeil, A.~J., R.~Frey, and P.~Embrechts} (2005): \emph{Quantitative
  risk management: Concepts, techniques and tools}, Princeton series in
  finance, Princeton and Oxford: {Princeton University Press}, 2015 revised ed.

\bibitem[\protect\citeauthoryear{Nadarajah}{Nadarajah}{2005}]{Nadarajah.2005}
\textsc{Nadarajah, S.} (2005): \enquote{A generalized normal distribution,}
  \emph{Journal of Applied Statistics}, 32, 685--694.

\bibitem[\protect\citeauthoryear{Nagler, Schepsmeier, Stoeber, Brechmann,
  Graeler, and Erhardt}{Nagler et~al.}{2019}]{Nagler.2019}
\textsc{Nagler, T., U.~Schepsmeier, J.~Stoeber, E.~C. Brechmann, B.~Graeler,
  and T.~Erhardt} (2019): \enquote{VineCopula: Statistical Inference of Vine
  Copulas,} .

\bibitem[\protect\citeauthoryear{Nelsen}{Nelsen}{2006}]{Nelsen.2006}
\textsc{Nelsen, R.~B.} (2006): \emph{An introduction to Copulas}, Springer
  series in statistics, New York, NY: {Springer New York}, 2. (2010) ed.

\bibitem[\protect\citeauthoryear{Nelson}{Nelson}{1991}]{Nelson.1991}
\textsc{Nelson, D.~B.} (1991): \enquote{Conditional Heteroskedasticity in Asset
  Returns: A New Approach,} \emph{Econometrica}, 59, 347.

\bibitem[\protect\citeauthoryear{Nelson and Cao}{Nelson and
  Cao}{1992}]{Nelson.1992}
\textsc{Nelson, D.~B. and C.~Q. Cao} (1992): \enquote{Inequality Constraints in
  the Univariate GARCH Model,} \emph{Journal of Business {\&} Economic
  Statistics}, 10, 229.

\bibitem[\protect\citeauthoryear{Newey and West}{Newey and
  West}{1987}]{Newey.1987}
\textsc{Newey, W.~K. and K.~D. West} (1987): \enquote{A Simple, Positive
  Semi-Definite, Heteroskedasticity and Autocorrelation Consistent Covariance
  Matrix,} \emph{Econometrica}, 55, 703.

\bibitem[\protect\citeauthoryear{Newey and West}{Newey and
  West}{1994}]{Newey.1994}
---\hspace{-.1pt}---\hspace{-.1pt}--- (1994): \enquote{Automatic Lag Selection
  in Covariance Matrix Estimation,} \emph{Review of Economic Studies}, 61,
  631--653.

\bibitem[\protect\citeauthoryear{Nikoloulopoulos, Joe, and Li}{Nikoloulopoulos
  et~al.}{2012}]{Nikoloulopoulos.2012}
\textsc{Nikoloulopoulos, A.~K., H.~Joe, and H.~Li} (2012): \enquote{Vine
  copulas with asymmetric tail dependence and applications to financial return
  data,} \emph{Computational Statistics {\&} Data Analysis}, 56, 3659--3673.

\bibitem[\protect\citeauthoryear{Nolde and Ziegel}{Nolde and
  Ziegel}{2017}]{Nolde.2017}
\textsc{Nolde, N. and J.~F. Ziegel} (2017): \enquote{Elicitability and
  backtesting: Perspectives for banking regulation,} \emph{The Annals of
  Applied Statistics}, 11, 1833--1874.

\bibitem[\protect\citeauthoryear{Oh and Patton}{Oh and
  Patton}{2017}]{OhPatton:2017}
\textsc{Oh, D.~H. and A.~J. Patton} (2017): \enquote{Modeling Dependence in
  High Dimensions With Factor Copulas,} \emph{Journal of Business \& Economic
  Statistics}, 35, 139--154.

\bibitem[\protect\citeauthoryear{Oh and Patton}{Oh and
  Patton}{2018}]{OhPatton:2018}
---\hspace{-.1pt}---\hspace{-.1pt}--- (2018): \enquote{Time-Varying Systemic
  Risk: Evidence From a Dynamic Copula Model of CDS Spreads,} \emph{Journal of
  Business \& Economic Statistics}, 36, 181--195.

\bibitem[\protect\citeauthoryear{Patton, Ziegel, and Chen}{Patton
  et~al.}{2019}]{Patton.2019}
\textsc{Patton, A.~J., J.~F. Ziegel, and R.~Chen} (2019): \enquote{Dynamic
  semiparametric models for expected shortfall (and Value-at-Risk),}
  \emph{Journal of Econometrics}, 211, 388--413.

\bibitem[\protect\citeauthoryear{Santos, Nogales, and Ruiz}{Santos
  et~al.}{2013}]{Santos.2013}
\textsc{Santos, A. A.~P., F.~J. Nogales, and E.~Ruiz} (2013):
  \enquote{Comparing Univariate and Multivariate Models to Forecast Portfolio
  Value-at-Risk,} \emph{Journal of Financial Econometrics}, 11, 400--441.

\bibitem[\protect\citeauthoryear{Savu and Trede}{Savu and
  Trede}{2008}]{SavuTrede}
\textsc{Savu, C. and M.~Trede} (2008): \enquote{Goodness-of-fit tests for
  parametric families of Archimedean copulas,} \emph{Quantitative Finance}, 8,
  109--116.

\bibitem[\protect\citeauthoryear{Sklar}{Sklar}{1959}]{Sklar.1959}
\textsc{Sklar, A.} (1959): \enquote{Fonctions de r{\'e}partition {\`a} n
  dimensions et leurs marges,} \emph{Publications de l'Institut de Statistique
  de l'Universit{\'e} de Paris}, 8, 229--231.

\bibitem[\protect\citeauthoryear{Tewari, Giering, and Raghunathan}{Tewari
  et~al.}{2011}]{Tewari.2011}
\textsc{Tewari, A., M.~J. Giering, and A.~Raghunathan} (2011):
  \enquote{Parametric Characterization of Multimodal Distributions with
  Non-gaussian Modes,} in \emph{2011 IEEE 11th International Conference on Data
  Mining workshops (ICDMW 2011)}, ed. by M.~Spiliopoulou, Piscataway, NJ: IEEE,
  286--292.

\bibitem[\protect\citeauthoryear{Wang, Yeh, and Cheng}{Wang
  et~al.}{2011}]{Wang.2011}
\textsc{Wang, J.-N., J.-H. Yeh, and N.~Y.-P. Cheng} (2011): \enquote{How
  accurate is the square-root-of-time rule in scaling tail risk: A global
  study,} \emph{Journal of Banking {\&} Finance}, 35, 1158--1169.

\bibitem[\protect\citeauthoryear{Wang and Dyer}{Wang and Dyer}{2012}]{Wang}
\textsc{Wang, T. and J.~S. Dyer} (2012): \enquote{A Copulas-Based Approach to
  Modeling Dependence in Decision Trees,} \emph{Operations Research}, 60,
  225--242.

\bibitem[\protect\citeauthoryear{Wei\ss}{Wei\ss}{2013}]{Weiss:2008}
\textsc{Wei\ss, G.} (2013): \enquote{Copula-GARCH versus dynamic conditional
  correlation: an empirical study on VaR and ES forecasting accuracy,}
  \emph{Review of Quantitative Finance and Accounting}, 41, 179--202.

\bibitem[\protect\citeauthoryear{West}{West}{1996}]{West.1996}
\textsc{West, K.~D.} (1996): \enquote{Asymptotic Inference about Predictive
  Ability,} \emph{Econometrica}, 64, 1067.

\bibitem[\protect\citeauthoryear{Wu}{Wu}{2014}]{Wu}
\textsc{Wu, S.} (2014): \enquote{Construction of asymmetric copulas and its
  application in two-dimensional reliability modelling,} \emph{European Journal
  of Operational Research}, 238, 476--485.

\bibitem[\protect\citeauthoryear{Zakoian}{Zakoian}{1994}]{Zakoian.1994}
\textsc{Zakoian, J.-M.} (1994): \enquote{Threshold heteroskedastic models,}
  \emph{Journal of Economic Dynamics and Control}, 18, 931--955.

\bibitem[\protect\citeauthoryear{Ziegel}{Ziegel}{2016}]{Ziegel2016}
\textsc{Ziegel, J.~F.} (2016): \enquote{Coherence and elicitability,}
  \emph{Mathematical Finance}, 26, 901--918.

\end{thebibliography}
\bibliographystyle{ecta}
}

\newpage

\newpage

\clearpage
\newpage
\begin{sidewaystable}[htbp]
\caption{Summary statistics of index and portfolio returns}
\scriptsize
The table provides summary statistics for the returns of the indices  that form the basis of our well-diversified portfolio as well as for the resulting portfolio returns with equal weighting.
We include equity indices (developed and emerging markets), bond indices (governement, corporate, and high-yield bonds), a commodity index, and a real estate index. We retrieve the total return indices (in \$) from \textit{Datastream} for a period from January 2001 to December 2018. We first calculate geometric returns and derive equally weighted portfolio returns next. This corresponds to a portfolio consisting of 40\% stocks, 40\% bonds, 10\% commodities and 10\% real estate. We provide minimum (Min), median, mean, maximum (Max), and standard deviation (SD) of daily returns in percent as well as the kurtosis and skewness of the time series.
\begin{center}
	\footnotesize
    \begin{tabular}{rrrrrrrr}
          &       &       &       &       &       &       &  \\
\midrule
& \multicolumn{7}{c}{} \\
& \multicolumn{1}{c}{Min} & \multicolumn{1}{c}{Median} & \multicolumn{1}{c}{Mean} & \multicolumn{1}{c}{Max} & \multicolumn{1}{c}{SD} & \multicolumn{1}{c}{Kurtosis} & \multicolumn{1}{c}{Skewness} \\
& \multicolumn{1}{c}{(in \%)} & \multicolumn{1}{c}{(in \%)} & \multicolumn{1}{c}{(in \%)} & \multicolumn{1}{c}{(in \%)} & \multicolumn{1}{c}{(in \%)} &       &  \\
\cmidrule{2-8}    \multicolumn{1}{l}{\textbf{Indices}} &       &       &       &       &       &       &  \\
\multicolumn{1}{l}{Stoxx Europe 600} & \multicolumn{1}{c}{-9.691} & \multicolumn{1}{c}{0.042} & \multicolumn{1}{c}{0.024} & \multicolumn{1}{c}{11.284} & \multicolumn{1}{c}{1.341} & \multicolumn{1}{c}{10.642} & \multicolumn{1}{c}{-0.004} \\
\multicolumn{1}{l}{Dow Jones Industrial Average} & \multicolumn{1}{c}{-7.873} & \multicolumn{1}{c}{0.033} & \multicolumn{1}{c}{0.032} & \multicolumn{1}{c}{11.080} & \multicolumn{1}{c}{1.107} & \multicolumn{1}{c}{12.686} & \multicolumn{1}{c}{0.100} \\
\multicolumn{1}{l}{FTSE Developed Asia
	Pacific Index} & \multicolumn{1}{c}{-8.524} & \multicolumn{1}{c}{0.047} & \multicolumn{1}{c}{0.024} & \multicolumn{1}{c}{9.928} & \multicolumn{1}{c}{1.185} & \multicolumn{1}{c}{8.097} & \multicolumn{1}{c}{-0.262} \\
\multicolumn{1}{l}{MSCI Emerging Markets Index} & \multicolumn{1}{c}{-9.484} & \multicolumn{1}{c}{0.095} & \multicolumn{1}{c}{0.040} & \multicolumn{1}{c}{10.598} & \multicolumn{1}{c}{1.178} & \multicolumn{1}{c}{11.372} & \multicolumn{1}{c}{-0.306} \\
\multicolumn{1}{l}{S\&P U.S. Treasury Bond Index} & \multicolumn{1}{c}{-1.658} & \multicolumn{1}{c}{0.012} & \multicolumn{1}{c}{0.014} & \multicolumn{1}{c}{1.758} & \multicolumn{1}{c}{0.232} & \multicolumn{1}{c}{5.995} & \multicolumn{1}{c}{-0.120} \\
\multicolumn{1}{l}{S\&P 500 Investment Grade Corporate Bond Index} & \multicolumn{1}{c}{-1.668} & \multicolumn{1}{c}{0.027} & \multicolumn{1}{c}{0.020} & \multicolumn{1}{c}{1.845} & \multicolumn{1}{c}{0.284} & \multicolumn{1}{c}{5.028} & \multicolumn{1}{c}{-0.232} \\
\multicolumn{1}{l}{S\&P U.S. High Yield Corporate Bond Index} & \multicolumn{1}{c}{-3.715} & \multicolumn{1}{c}{0.042} & \multicolumn{1}{c}{0.027} & \multicolumn{1}{c}{2.226} & \multicolumn{1}{c}{0.246} & \multicolumn{1}{c}{32.336} & \multicolumn{1}{c}{-2.377} \\
\multicolumn{1}{l}{S\&P Pan-Europe Developed Sovereign Bond Index} & \multicolumn{1}{c}{-3.705} & \multicolumn{1}{c}{0.020} & \multicolumn{1}{c}{0.021} & \multicolumn{1}{c}{5.028} & \multicolumn{1}{c}{0.617} & \multicolumn{1}{c}{5.897} & \multicolumn{1}{c}{0.138} \\
\multicolumn{1}{l}{S\&P GSCI } & \multicolumn{1}{c}{-8.762} & \multicolumn{1}{c}{0.000} & \multicolumn{1}{c}{-0.003} & \multicolumn{1}{c}{7.483} & \multicolumn{1}{c}{1.434} & \multicolumn{1}{c}{5.787} & \multicolumn{1}{c}{-0.148} \\
\multicolumn{1}{l}{Developed Markets Datastream Real Estate Index} & \multicolumn{1}{c}{-6.849} & \multicolumn{1}{c}{0.061} & \multicolumn{1}{c}{0.036} & \multicolumn{1}{c}{7.223} & \multicolumn{1}{c}{0.939} & \multicolumn{1}{c}{11.411} & \multicolumn{1}{c}{-0.370} \\
&       &       &       &       &       &       &  \\
\multicolumn{1}{l}{\textbf{Equally weighted portfolio}} & \multicolumn{1}{c}{-4.009} & \multicolumn{1}{c}{0.042} & \multicolumn{1}{c}{0.023} & \multicolumn{1}{c}{4.110} & \multicolumn{1}{c}{0.552} & \multicolumn{1}{c}{11.085} & \multicolumn{1}{c}{-0.516} \\
\midrule
&       &       &       &       &       &       &  \\
    \end{tabular}%
\end{center}
\label{tab:PF_stats}%
\end{sidewaystable}%

\begin{table}[htbp]
	\caption{Model risk and the great financial crisis}
	\vspace{0.2cm} {\justifying \scriptsize \noindent This table presents summary statistics for the time series of model risk associated with 99\% VaR and 97.5\% ES forecasts for a well diversified portfolio. Model risk is measured in terms of the mean absolute deviation (mad) of risk forecasts by various models from November 4, 2004 until December 31, 2018. With \emph{crisis} we refer to the years 2008-2009 while \emph{pre-crisis} and  \emph  {after-crisis} denote the period before and after, respectively. Model risk is calculated based on all multivariate models that passed the respective backtest, see Section \ref{sec:backtests} for details. We report model risk estimates for one day ahead risk forecasts in percent of the portfolio value (first and second column) and in absolute terms (third and fourth column) for an equally weighted portfolio. The results in absolute terms are based on a portfolio value of \$100,000 and a 10 day forecast horizon obtained by applying the square-root-of-time rule. We provide minimum (Min), median, mean, maximum (Max), and standard deviation (SD) of the daily model risk estimates. Further results for randomly generated portfolio weights as well as alternative measures of model risk can be found in Tables \ref{tab:averageWeightsAllMultModels} and \ref{tab:otherModelRiskMeasuresAllMultModels}.
	}	\vspace{1cm}

	\centering
	\footnotesize
	
	    \begin{tabular}{clcccc}
	    	\toprule
		&       & \multicolumn{2}{c}{\textbf{Model risk (in \%)}} & \multicolumn{2}{c}{\textbf{Model risk (in \$)}} \\
		&       & \multicolumn{1}{l}{99\% VaR} & \multicolumn{1}{l}{97.5\% ES} & \multicolumn{1}{l}{99\% VaR } & \multicolumn{1}{l}{97.5\% ES} \\
		\cmidrule{2-6}    \multicolumn{1}{c}{\multirow{5}[2]{*}{\begin{sideways}\textbf{Whole period}\end{sideways}}} & Min   & 0.075 & 0.029 & 237   & 92 \\
		& Median & 0.138 & 0.080 & 436   & 253 \\
		& Mean  & 0.165 & 0.092 & 522   & 291 \\
		& Max   & 0.847 & 0.716 & 2,678 & 2,264 \\
		& SD    & 0.091 & 0.056 & 288   & 177 \\
		\cmidrule{2-6}          &       &       &       &       &  \\
		&       &       &       &       &  \\
		\cmidrule{2-6}    \multirow{5}[2]{*}{\begin{sideways}\textbf{Pre-crisis}\end{sideways}} & Min   & 0.081 & 0.029 & 256   & 92 \\
		& Median & 0.105 & 0.050 & 332   & 158 \\
		& Mean  & 0.119 & 0.063 & 376   & 199 \\
		& Max   & 0.280 & 0.207 & 885   & 655 \\
		& SD    & 0.036 & 0.030 & 114   & 95 \\
		\cmidrule{2-6}          &       &       &       &       &  \\
		\cmidrule{2-6}    \multirow{5}[2]{*}{\begin{sideways}\textbf{Crisis}\end{sideways}} & Min   & 0.116 & 0.046 & 367   & 145 \\
		& Median & 0.241 & 0.108 & 762   & 342 \\
		& Mean  & 0.286 & 0.145 & 904   & 459 \\
		& Max   & 0.847 & 0.716 & 2,678 & 2,264 \\
		& SD    & 0.139 & 0.101 & 440   & 319 \\
		\cmidrule{2-6}          &       &       &       &       &  \\
		\cmidrule{2-6}    \multirow{5}[2]{*}{\begin{sideways}\textbf{Post-crisis}\end{sideways}} & Min   & 0.075 & 0.034 & 237   & 108 \\
		& Median & 0.138 & 0.082 & 436   & 259 \\
		& Mean  & 0.155 & 0.090 & 490   & 285 \\
		& Max   & 0.551 & 0.638 & 1,742 & 2,018 \\
		& SD    & 0.064 & 0.037 & 202   & 117 \\
		\bottomrule    \end{tabular}%
	\label{tab:allMultivariateModelsMad}%
\end{table}%

\begin{table}[htbp]
	\caption{Model risk for all multivariate models averaged over 100 random portfolios}
\vspace{0.2cm} {\justifying \scriptsize \noindent This table provides the same results as Table \ref{tab:allMultivariateModelsMad} but for 100 random portfolios obtained by drawing portfolio weights from a unit-simplex. Therefore, we first calculate summary statistics for the time series of model risk for each of the portfolios. These statistics are then averaged over all 100 portfolios. The results can thus be interpreted as summary statistics for the model risk of an average portfolio.}	\vspace{1cm}

\centering
\footnotesize
       \begin{tabular}{clcccc}
    \toprule
   	&       & \multicolumn{2}{c}{\textbf{Model risk (in \%)}} & \multicolumn{2}{c}{\textbf{Model risk (in \$)}} \\
   	&       & \multicolumn{1}{l}{99\% VaR} & \multicolumn{1}{l}{97.5\% ES} & \multicolumn{1}{l}{99\% VaR } & \multicolumn{1}{l}{97.5\% ES} \\
   	\cmidrule{2-6}    \multicolumn{1}{c}{\multirow{5}[2]{*}{\begin{sideways}\textbf{Whole period}\end{sideways}}} & Min   & 0.065 & 0.034 & 206   & 106 \\
   	& Median & 0.127 & 0.085 & 401   & 268 \\
   	& Mean  & 0.153 & 0.099 & 484   & 313 \\
   	& Max   & 0.813 & 0.763 & 2570  & 2413 \\
   	& SD    & 0.086 & 0.057 & 272   & 181 \\
   	\cmidrule{2-6}          &       &       &       &       &  \\
   	&       &       &       &       &  \\
   	\cmidrule{2-6}    \multirow{5}[2]{*}{\begin{sideways}\textbf{Pre-crisis}\end{sideways}} & Min   & 0.069 & 0.037 & 218   & 118 \\
   	& Median & 0.097 & 0.064 & 307   & 203 \\
   	& Mean  & 0.111 & 0.074 & 351   & 233 \\
   	& Max   & 0.281 & 0.233 & 890   & 735 \\
   	& SD    & 0.036 & 0.029 & 113   & 91 \\
   	\cmidrule{2-6}          &       &       &       &       &  \\
   	\cmidrule{2-6}    \multirow{5}[2]{*}{\begin{sideways}\textbf{Crisis}\end{sideways}} & Min   & 0.108 & 0.046 & 342   & 146 \\
   	& Median & 0.221 & 0.114 & 697   & 361 \\
   	& Mean  & 0.267 & 0.153 & 843   & 483 \\
   	& Max   & 0.807 & 0.711 & 2551  & 2249 \\
   	& SD    & 0.133 & 0.105 & 420   & 331 \\
   	\cmidrule{2-6}          &       &       &       &       &  \\
   	\cmidrule{2-6}    \multirow{5}[2]{*}{\begin{sideways}\textbf{Post-crisis}\end{sideways}} & Min   & 0.069 & 0.039 & 217   & 123 \\
   	& Median & 0.127 & 0.087 & 401   & 276 \\
   	& Mean  & 0.143 & 0.096 & 451   & 303 \\
   	& Max   & 0.593 & 0.626 & 1876  & 1980 \\
   	& SD    & 0.060 & 0.039 & 189   & 123 \\
   	\bottomrule    \end{tabular}%
  \label{tab:averageWeightsAllMultModels}%
\end{table}%

\begin{table}[htbp]
  \caption{Alternative measures of model risk}
  \vspace{0.2cm} {\justifying \scriptsize \noindent This table provides summary statistics for different measures of model risk for 99\% VaR and 97.5\% forecasts. Our baseline measure is the mean absolute deviation (mad). We additionally include the standard deviation (sd) and interquartile range (iqr) of the risk forecasts by various models in percent of the portfolio value, see Section \ref{sec:modelRisk} for more details. Model risk is calculated for all multivariate models that passed the respective backtest. We provide minimum (Min), median, mean, maximum (Max), and standard deviation (SD) of the daily model risk estimates (according to the different model risk measures) over the period November 4, 2004 until December 31, 2018.  }	\vspace{1cm}
  
  \centering
  \footnotesize
    \begin{tabular}{clccccc}
    	\toprule
          &       & \multicolumn{5}{c}{\textbf{Model risk}} \\
          &       & \textbf{Min} & \textbf{Median} & \textbf{Mean} & \textbf{Max} & \textbf{SD} \\
\cmidrule{3-7}          & \textbf{Measure} &       &       &       &       &  \\
    \multirow{3}[0]{*}{\begin{sideways}\textbf{VaR}\end{sideways}} & mad (in \%) & 0.075 & 0.138 & 0.165 & 0.847 & 0.091 \\
          & sd (in \%) & 0.088 & 0.161 & 0.195 & 0.992 & 0.108 \\
          & iqr (in \%) & 0.130 & 0.269 & 0.325 & 1.631 & 0.180 \\
          &       &       &       &       &       &  \\
    \multirow{3}[0]{*}{\begin{sideways}\textbf{ES}\end{sideways}} & mad (in \%) & 0.029 & 0.080 & 0.092 & 0.716 & 0.056 \\
          & sd (in \%) & 0.037 & 0.099 & 0.117 & 0.867 & 0.072 \\
          & iqr (in \%) & 0.040 & 0.133 & 0.152 & 1.399 & 0.094 \\
          \bottomrule
    \end{tabular}%
  \label{tab:otherModelRiskMeasuresAllMultModels}%
\end{table}%

\begin{table}[htbp]
   \caption{Summary statistics of average model risk for all groups}
 \scriptsize
 This table presents summary statistics for the time series of average model risk associated with 99\% VaR (Panel A) and 97.5\% ES (Panel B) forecasts for a well diversified portfolio per group. Model risk is measured in terms of the mean absolute deviation (mad) of risk forecasts by various models within a model set from November 4, 2004 until December 31, 2018. All models passed the respective backtest, see Section \ref{sec:backtests} for details. 
 \textit{Group 1} (G1) includes all model sets in which a copula function is fixed while varying the marginal distribution. \textit{Group 2} (G2) contains analogously the model sets with fixed marginal distribution and varying copula. \textit{Group 3} (G3) consists of all multivariate and \textit{Group 4} (G4) of all univariate models. We report model risk estimates for one day ahead risk forecasts in percent of the portfolio value (columns 1-5) and in absolute terms (column 6) for an equally weighted portfolio. The results in absolute terms are based on a portfolio value of \$100,000 and a 10 day forecast horizon obtained by applying the square-root-of-time rule. We provide minimum (Min), median, mean, maximum (Max), and standard deviation (SD) of the daily averaged model risk estimates per group.
 Further results for randomly generated portfolio weights, various VaR and ES confidence levels as well as alternative measures of model risk (Groups 1 and 2 only) can be found in Tables \ref{tab:analysisI_robust_weights}, \ref{tab:analysisI_robust_lev}, and \ref{tab:analysisIV}. *** denotes a statistically significant increase in average model risk at the 1\% level compared to Group 1.
 \begin{center}
 	\footnotesize
    \begin{tabular}{rrrrrrrr}
          &       &       &       &       &       &       &  \\
\midrule
\multicolumn{1}{l}{\textit{Panel A: 99\% VaR}} &       &       &       &       &       &       &  \\
& \multicolumn{5}{c}{\textbf{Average model risk (in \%)}} &       & \multicolumn{1}{l}{\textbf{Average model risk (in \$)}} \\
& \multicolumn{1}{c}{Min} & \multicolumn{1}{c}{Median} & \multicolumn{1}{c}{Mean} & \multicolumn{1}{c}{Max} & \multicolumn{1}{c}{SD} &       & \multicolumn{1}{c}{Mean} \\
\cmidrule{2-6}\cmidrule{8-8}    \multicolumn{1}{c}{\textbf{Group}} &       &       &       &       &       &       &  \\
\multicolumn{1}{l}{G1: copula fixed} & \multicolumn{1}{c}{0.008} & \multicolumn{1}{c}{0.037} & \multicolumn{1}{c}{0.052} & \multicolumn{1}{c}{0.518} & \multicolumn{1}{c}{0.044} &       & \multicolumn{1}{c}{164} \\
\multicolumn{1}{l}{G2: marg. distr. fixed} & \multicolumn{1}{c}{0.073} & \multicolumn{1}{c}{0.130} & \multicolumn{1}{c}{0.157***} & \multicolumn{1}{c}{0.803} & \multicolumn{1}{c}{0.084} &       & \multicolumn{1}{c}{498} \\
\multicolumn{1}{l}{G3: all multivariate} & \multicolumn{1}{c}{0.075} & \multicolumn{1}{c}{0.138} & \multicolumn{1}{c}{0.165} & \multicolumn{1}{c}{0.847} & \multicolumn{1}{c}{0.091} &       & \multicolumn{1}{c}{523} \\
\multicolumn{1}{l}{G4: all univariate} & \multicolumn{1}{c}{0.011} & \multicolumn{1}{c}{0.065} & \multicolumn{1}{c}{0.085} & \multicolumn{1}{c}{0.651} & \multicolumn{1}{c}{0.069} &       & \multicolumn{1}{c}{269} \\
\midrule
&       &       &       &       &       &       &  \\
\multicolumn{1}{l}{\textit{Panel B: 97.5\% ES}} &       &       &       &       &       &       &  \\
& \multicolumn{5}{c}{\textbf{Average model risk (in \%)}} &       & \multicolumn{1}{l}{\textbf{Average model risk (in \$)}} \\
& \multicolumn{1}{c}{Min} & \multicolumn{1}{c}{Median} & \multicolumn{1}{c}{Mean} & \multicolumn{1}{c}{Max} & \multicolumn{1}{c}{SD} &       & \multicolumn{1}{c}{Mean} \\
\cmidrule{2-6}\cmidrule{8-8}    \multicolumn{1}{c}{\textbf{Group}} &       &       &       &       &       &       &  \\
\multicolumn{1}{l}{G1: copula fixed} & \multicolumn{1}{c}{0.012} & \multicolumn{1}{c}{0.042} & \multicolumn{1}{c}{0.058} & \multicolumn{1}{c}{0.640} & \multicolumn{1}{c}{0.049} &       & \multicolumn{1}{c}{184} \\
\multicolumn{1}{l}{G2: marg. distr. fixed} & \multicolumn{1}{c}{0.022} & \multicolumn{1}{c}{0.061} & \multicolumn{1}{c}{0.068***} & \multicolumn{1}{c}{0.561} & \multicolumn{1}{c}{0.036} &       & \multicolumn{1}{c}{216} \\
\multicolumn{1}{l}{G3: all multivariate} & \multicolumn{1}{c}{0.029} & \multicolumn{1}{c}{0.080} & \multicolumn{1}{c}{0.092} & \multicolumn{1}{c}{0.716} & \multicolumn{1}{c}{0.056} &       & \multicolumn{1}{c}{290} \\
\multicolumn{1}{l}{G4: all univariate} & \multicolumn{1}{c}{0.013} & \multicolumn{1}{c}{0.070} & \multicolumn{1}{c}{0.090} & \multicolumn{1}{c}{0.672} & \multicolumn{1}{c}{0.070} &       & \multicolumn{1}{c}{285} \\
\midrule
&       &       &       &       &       &       &  \\
    \end{tabular}%
\end{center}
  \label{tab:analysisI}%
\end{table}%

\begin{table}[htbp]
  \caption{Summary statistics of average model risk for all groups over 100 random portfolios}
\scriptsize
This table presents summary statistics for the time series of average model risk per group associated with 99\% VaR (Panel A) and 97.5\% ES (Panel B) forecasts for 100 random portfolios obtained by drawing portfolio weights from a unit-simplex. Therefore, we first calculate summary statistics for the time series of model risk for each of the portfolios. These statistics are then averaged over all 100 portfolios. The results can thus be interpreted as summary statistics for the model risk of an average portfolio. Model risk is measured in terms of the mean absolute deviation (mad) of risk forecasts by various models within a model set from November 4, 2004 until December 31, 2018. All models passed the respective backtest, see Section \ref{sec:backtests} for details. \textit{Group 1} (G1) includes all model sets in which a copula function is fixed while varying the marginal distribution. \textit{Group 2} (G2) contains analogously the model sets with fixed marginal distribution and varying copula. \textit{Group 3} (G3) consists of all multivariate and \textit{Group 4} (G4) of all univariate models. We report model risk estimates for one day ahead risk forecasts in percent of the portfolio value. We provide minimum (Min), median, mean, maximum (Max) and standard deviation (SD) of the daily averaged model risk estimates per group. *** denotes a statistically significant increase in average model risk at the 1\% level compared to Group 1.
\scriptsize
\begin{center} 
    \begin{tabular}{rrrrrr}
          &       &       &       &       &  \\
\midrule
\multicolumn{1}{l}{\textit{Panel A: 99\% VaR}} &       &       &       &       &  \\
& \multicolumn{5}{c}{\textbf{Average model risk (mad in \%)}} \\
& \multicolumn{1}{c}{Min} & \multicolumn{1}{c}{Median} & \multicolumn{1}{c}{Mean} & \multicolumn{1}{c}{Max} & \multicolumn{1}{c}{SD} \\
\cmidrule{2-6}    \multicolumn{1}{c}{\textbf{Group}} &       &       &       &       &  \\
\multicolumn{1}{l}{G1: copula fixed} & \multicolumn{1}{c}{0.013} & \multicolumn{1}{c}{0.041} & \multicolumn{1}{c}{0.057} & \multicolumn{1}{c}{0.588} & \multicolumn{1}{c}{0.048} \\
\multicolumn{1}{l}{G2: marg. distr. fixed} & \multicolumn{1}{c}{0.056} & \multicolumn{1}{c}{0.118} & \multicolumn{1}{c}{0.142***} & \multicolumn{1}{c}{0.753} & \multicolumn{1}{c}{0.078} \\
\multicolumn{1}{l}{G3: all multivariate} & \multicolumn{1}{c}{0.065} & \multicolumn{1}{c}{0.127} & \multicolumn{1}{c}{0.153} & \multicolumn{1}{c}{0.813} & \multicolumn{1}{c}{0.086} \\
\multicolumn{1}{l}{G4: all univariate} & \multicolumn{1}{c}{0.012} & \multicolumn{1}{c}{0.066} & \multicolumn{1}{c}{0.087} & \multicolumn{1}{c}{0.751} & \multicolumn{1}{c}{0.071} \\
\midrule
&       &       &       &       &  \\
\multicolumn{1}{l}{\textit{Panel B: 97.5\% ES}} &       &       &       &       &  \\
& \multicolumn{5}{c}{\textbf{Average model risk (mad in \%)}} \\
& \multicolumn{1}{c}{Min} & \multicolumn{1}{c}{Median} & \multicolumn{1}{c}{Mean} & \multicolumn{1}{c}{Max} & \multicolumn{1}{c}{SD} \\
\cmidrule{2-6}    \multicolumn{1}{c}{\textbf{Group}} &       &       &       &       &  \\
\multicolumn{1}{l}{G1: copula fixed} & \multicolumn{1}{c}{0.013} & \multicolumn{1}{c}{0.045} & \multicolumn{1}{c}{0.062} & \multicolumn{1}{c}{0.692} & \multicolumn{1}{c}{0.052} \\
\multicolumn{1}{l}{G2: marg. distr. fixed} & \multicolumn{1}{c}{0.024} & \multicolumn{1}{c}{0.069} & \multicolumn{1}{c}{0.075***} & \multicolumn{1}{c}{0.485} & \multicolumn{1}{c}{0.035} \\
\multicolumn{1}{l}{G3: all multivariate} & \multicolumn{1}{c}{0.034} & \multicolumn{1}{c}{0.085} & \multicolumn{1}{c}{0.099} & \multicolumn{1}{c}{0.763} & \multicolumn{1}{c}{0.057} \\
\multicolumn{1}{l}{G4: all univariate} & \multicolumn{1}{c}{0.015} & \multicolumn{1}{c}{0.071} & \multicolumn{1}{c}{0.093} & \multicolumn{1}{c}{0.762} & \multicolumn{1}{c}{0.072} \\
\midrule
&       &       &       &       &  \\
    \end{tabular}%
\end{center}
\label{tab:analysisI_robust_weights}%
\end{table}%

\begin{table}[htbp]
\caption{Summary statistics of average model risk for alternative model risk measures (model sets with fixed and varying copula only)}
\scriptsize
This table presents summary statistics for the time series of average model risk associated with 99\% VaR (Panel A) and 97.5\% ES (Panel B) forecasts for a well diversified portfolio per group. Model risk is captured by different measures based on risk forecasts by various models within a model set from November 4, 2004 until December 31, 2018. All models passed the respective backtest, see Section \ref{sec:backtests} for details. Our baseline measure is the mean absolute deviation (mad). We additionally include the standard deviation (sd) and interquartile range (iqr), see Section \ref{sec:modelRisk} for more details. \textit{Group 1} (G1) includes all model sets in which a copula function is fixed while varying the marginal distribution. \textit{Group 2} (G2) contains analogously the model sets with fixed marginal distribution and varying copula. We report model risk estimates for one day ahead risk forecasts in percent of the portfolio value. We provide minimum (Min), median, mean, maximum (Max) and standard deviation (SD) of the daily averaged model risk estimates per group. *** (**) denotes a statistically significant increase in average model risk at the 1\% (5\%) level compared to Group 1.
\begin{center}
	\footnotesize
    \begin{tabular}{rrrrrrr}
          &       &       &       &       &       &  \\
\midrule
\multicolumn{1}{l}{\textit{Panel A: 99\% VaR}} &       &       &       &       &       &  \\
&       & \multicolumn{5}{c}{\textbf{Average model risk (in \%)}} \\
&       & \multicolumn{1}{c}{Min} & \multicolumn{1}{c}{Median} & \multicolumn{1}{c}{Mean} & \multicolumn{1}{c}{Max} & \multicolumn{1}{c}{SD} \\
\cmidrule{3-7}    \multicolumn{1}{c}{\textbf{Group}} & \multicolumn{1}{c}{\textbf{Measure}} &       &       &       &       &  \\
\multicolumn{1}{l}{G1: Copula fixed} & \multicolumn{1}{l}{mad} & \multicolumn{1}{c}{0.008} & \multicolumn{1}{c}{0.037} & \multicolumn{1}{c}{0.052} & \multicolumn{1}{c}{0.518} & \multicolumn{1}{c}{0.044} \\
\multicolumn{1}{l}{G1: Copula fixed} & \multicolumn{1}{l}{sd} & \multicolumn{1}{c}{0.016} & \multicolumn{1}{c}{0.048} & \multicolumn{1}{c}{0.069} & \multicolumn{1}{c}{0.636} & \multicolumn{1}{c}{0.063} \\
\multicolumn{1}{l}{G1: Copula fixed} & \multicolumn{1}{l}{iqr} & \multicolumn{1}{c}{0.011} & \multicolumn{1}{c}{0.057} & \multicolumn{1}{c}{0.075} & \multicolumn{1}{c}{0.869} & \multicolumn{1}{c}{0.060} \\
&       &       &       &       &       &  \\
\multicolumn{1}{l}{G2: Marg. distr. fixed} & \multicolumn{1}{l}{mad} & \multicolumn{1}{c}{0.073} & \multicolumn{1}{c}{0.130} & \multicolumn{1}{c}{0.157***} & \multicolumn{1}{c}{0.803} & \multicolumn{1}{c}{0.084} \\
\multicolumn{1}{l}{G2: Marg. distr. fixed} & \multicolumn{1}{l}{sd} & \multicolumn{1}{c}{0.089} & \multicolumn{1}{c}{0.158} & \multicolumn{1}{c}{0.191***} & \multicolumn{1}{c}{0.975} & \multicolumn{1}{c}{0.101} \\
\multicolumn{1}{l}{G2: Marg. distr. fixed} & \multicolumn{1}{l}{iqr} & \multicolumn{1}{c}{0.103} & \multicolumn{1}{c}{0.255} & \multicolumn{1}{c}{0.311***} & \multicolumn{1}{c}{1.631} & \multicolumn{1}{c}{0.181} \\
\midrule
&       &       &       &       &       &  \\
\multicolumn{1}{l}{\textit{Panel B: 97.5\% ES}} &       &       &       &       &       &  \\
&       & \multicolumn{5}{c}{\textbf{Average model risk (in \%)}} \\
&       & \multicolumn{1}{c}{Min} & \multicolumn{1}{c}{Median} & \multicolumn{1}{c}{Mean} & \multicolumn{1}{c}{Max} & \multicolumn{1}{c}{SD} \\
\cmidrule{3-7}    \multicolumn{1}{c}{\textbf{Group}} & \multicolumn{1}{c}{\textbf{Measure }} &       &       &       &       &  \\
\multicolumn{1}{l}{G1: Copula fixed} & \multicolumn{1}{l}{mad} & \multicolumn{1}{c}{0.012} & \multicolumn{1}{c}{0.042} & \multicolumn{1}{c}{0.058} & \multicolumn{1}{c}{0.640} & \multicolumn{1}{c}{0.049} \\
\multicolumn{1}{l}{G1: Copula fixed} & \multicolumn{1}{l}{sd} & \multicolumn{1}{c}{0.015} & \multicolumn{1}{c}{0.055} & \multicolumn{1}{c}{0.077} & \multicolumn{1}{c}{0.734} & \multicolumn{1}{c}{0.067} \\
\multicolumn{1}{l}{G1: Copula fixed} & \multicolumn{1}{l}{iqr} & \multicolumn{1}{c}{0.017} & \multicolumn{1}{c}{0.065} & \multicolumn{1}{c}{0.085} & \multicolumn{1}{c}{1.261} & \multicolumn{1}{c}{0.068} \\
&       &       &       &       &       &  \\
\multicolumn{1}{l}{G2: Marg. distr. fixed} & \multicolumn{1}{l}{mad} & \multicolumn{1}{c}{0.022} & \multicolumn{1}{c}{0.061} & \multicolumn{1}{c}{0.068***} & \multicolumn{1}{c}{0.561} & \multicolumn{1}{c}{0.036} \\
\multicolumn{1}{l}{G2: Marg. distr. fixed} & \multicolumn{1}{l}{sd} & \multicolumn{1}{c}{0.031} & \multicolumn{1}{c}{0.081} & \multicolumn{1}{c}{0.090***} & \multicolumn{1}{c}{0.720} & \multicolumn{1}{c}{0.046} \\
\multicolumn{1}{l}{G2: Marg. distr. fixed} & \multicolumn{1}{l}{iqr} & \multicolumn{1}{c}{0.023} & \multicolumn{1}{c}{0.080} & \multicolumn{1}{c}{0.094**} & \multicolumn{1}{c}{0.857} & \multicolumn{1}{c}{0.057} \\
\midrule
&       &       &       &       &       &  \\
    \end{tabular}%
\end{center}
\label{tab:analysisIV}%
\end{table}%

\begin{sidewaystable}[htbp]
  \caption{Summary statistics of average model risk for all groups under various confidence levels}
  \scriptsize
  This table presents summary statistics for the time series of average model risk associated with 99.9\%, 99\%, 97.5\% and 95\% VaR (Panel A) and 99.9\%, 99\%, 97.5\% and 95\% ES (Panel B) forecasts for a well diversified portfolio per group. Model risk is measured in terms of the mean absolute deviation (mad) of risk forecasts by various models within a model set from November 4, 2004 until December 31, 2018. All models passed the respective backtest, see Section \ref{sec:backtests} for details. \textit{Group 1} (G1) includes all model sets in which a copula function is fixed while varying the marginal distribution. \textit{Group 2} (G2) contains analogously the model sets with fixed marginal distribution and varying copula. \textit{Group 3} (G3) consists of all multivariate and \textit{Group 4} (G4) of all univariate models. We report model risk estimates for one day ahead risk forecasts in percent of the portfolio value. We provide minimum (Min), median, mean, maximum (Max) and standard deviation (SD) of the daily averaged model risk estimates per group. *** denotes a statistically significant increase in average model risk at the 1\% level compared to Group 1.
  \scriptsize
  \begin{center} 
    \begin{tabular}{rclrrrrrrclrrrrrr}
              &       &       &       &       &       &       &       &       &       &       &       &       &       &  \\
    \midrule
    \multicolumn{1}{l}{\textit{\textbf{Panel A: VaR }}} &       &       &       &       &       &       &       & \multicolumn{1}{l}{\textit{\textbf{Panel B: ES}}} &       &       &       &       &       &  \\
    &       & \multicolumn{5}{c}{\textbf{Average model risk (mad in \%)}} &       &       &       & \multicolumn{5}{c}{\textbf{Average model risk (mad in \%)}} \\
    &       & \multicolumn{1}{c}{Min} & \multicolumn{1}{c}{Median} & \multicolumn{1}{c}{Mean} & \multicolumn{1}{c}{Max} & \multicolumn{1}{c}{SD} &       &       &       & \multicolumn{1}{c}{Min} & \multicolumn{1}{c}{Median} & \multicolumn{1}{c}{Mean} & \multicolumn{1}{c}{Max} & \multicolumn{1}{c}{SD} \\
    \cmidrule{3-7}\cmidrule{11-15}    \multicolumn{1}{c}{\textbf{Conf. level}} & \multicolumn{1}{c}{\textbf{Group}} &       &       &       &       &       &       & \multicolumn{1}{c}{\textbf{Conf. level}} & \multicolumn{1}{c}{\textbf{Group}} &       &       &       &       &  \\
    \multicolumn{1}{c}{\multirow{4}[0]{*}{\textbf{99.9\%}}} & \multicolumn{1}{l}{G1: copula fixed} & \multicolumn{1}{c}{0.031} & \multicolumn{1}{c}{0.088} & \multicolumn{1}{c}{0.108} & \multicolumn{1}{c}{0.747} & \multicolumn{1}{c}{0.070} &       & \multicolumn{1}{c}{\multirow{4}[0]{*}{\textbf{99.9\%}}} & \multicolumn{1}{l}{G1: copula fixed} & \multicolumn{1}{c}{0.023} & \multicolumn{1}{c}{0.119} & \multicolumn{1}{c}{0.136} & \multicolumn{1}{c}{0.875} & \multicolumn{1}{c}{0.083} \\
    & \multicolumn{1}{l}{G2: marg. distr. fixed} & \multicolumn{1}{c}{0.122} & \multicolumn{1}{c}{0.248} & \multicolumn{1}{c}{0.293***} & \multicolumn{1}{c}{1.507} & \multicolumn{1}{c}{0.156} &       &       & \multicolumn{1}{l}{G2: marg. distr. fixed} & \multicolumn{1}{c}{0.054} & \multicolumn{1}{c}{0.245} & \multicolumn{1}{c}{0.270***} & \multicolumn{1}{c}{1.435} & \multicolumn{1}{c}{0.125} \\
    & \multicolumn{1}{l}{G3: all multivariate} & \multicolumn{1}{c}{0.127} & \multicolumn{1}{c}{0.261} & \multicolumn{1}{c}{0.306} & \multicolumn{1}{c}{1.553} & \multicolumn{1}{c}{0.162} &       &       & \multicolumn{1}{l}{G3: all multivariate} & \multicolumn{1}{c}{0.068} & \multicolumn{1}{c}{0.279} & \multicolumn{1}{c}{0.309} & \multicolumn{1}{c}{1.506} & \multicolumn{1}{c}{0.139} \\
    & \multicolumn{1}{l}{G4: all univariate} & \multicolumn{1}{c}{0.031} & \multicolumn{1}{c}{0.169} & \multicolumn{1}{c}{0.197} & \multicolumn{1}{c}{1.454} & \multicolumn{1}{c}{0.124} &       &       & \multicolumn{1}{l}{G4: all univariate} & \multicolumn{1}{c}{0.000} & \multicolumn{1}{c}{0.221} & \multicolumn{1}{c}{0.244} & \multicolumn{1}{c}{1.521} & \multicolumn{1}{c}{0.161} \\
    &       &       &       &       &       &       &       &       &       &       &       &       &       &  \\
    \multicolumn{1}{c}{\multirow{4}[0]{*}{\textbf{99.0\%}}} & \multicolumn{1}{l}{G1: copula fixed} & \multicolumn{1}{c}{0.008} & \multicolumn{1}{c}{0.037} & \multicolumn{1}{c}{0.052} & \multicolumn{1}{c}{0.518} & \multicolumn{1}{c}{0.044} &       & \multicolumn{1}{c}{\multirow{4}[0]{*}{\textbf{99.0\%}}} & \multicolumn{1}{l}{G1: copula fixed} & \multicolumn{1}{c}{0.019} & \multicolumn{1}{c}{0.060} & \multicolumn{1}{c}{0.079} & \multicolumn{1}{c}{0.731} & \multicolumn{1}{c}{0.062} \\
    & \multicolumn{1}{l}{G2: marg. distr. fixed} & \multicolumn{1}{c}{0.073} & \multicolumn{1}{c}{0.130} & \multicolumn{1}{c}{0.157***} & \multicolumn{1}{c}{0.803} & \multicolumn{1}{c}{0.084} &       &       & \multicolumn{1}{l}{G2: marg. distr. fixed} & \multicolumn{1}{c}{0.047} & \multicolumn{1}{c}{0.105} & \multicolumn{1}{c}{0.117***} & \multicolumn{1}{c}{0.682} & \multicolumn{1}{c}{0.054} \\
    & \multicolumn{1}{l}{G3: all multivariate} & \multicolumn{1}{c}{0.075} & \multicolumn{1}{c}{0.138} & \multicolumn{1}{c}{0.165} & \multicolumn{1}{c}{0.847} & \multicolumn{1}{c}{0.091} &       &       & \multicolumn{1}{l}{G3: all multivariate} & \multicolumn{1}{c}{0.056} & \multicolumn{1}{c}{0.124} & \multicolumn{1}{c}{0.142} & \multicolumn{1}{c}{0.820} & \multicolumn{1}{c}{0.073} \\
    & \multicolumn{1}{l}{G4: all univariate} & \multicolumn{1}{c}{0.011} & \multicolumn{1}{c}{0.065} & \multicolumn{1}{c}{0.085} & \multicolumn{1}{c}{0.651} & \multicolumn{1}{c}{0.069} &       &       & \multicolumn{1}{l}{G4: all univariate} & \multicolumn{1}{c}{0.017} & \multicolumn{1}{c}{0.104} & \multicolumn{1}{c}{0.126} & \multicolumn{1}{c}{0.848} & \multicolumn{1}{c}{0.085} \\
    &       &       &       &       &       &       &       &       &       &       &       &       &       &  \\
    \multicolumn{1}{c}{\multirow{4}[0]{*}{\textbf{97.5\%}}} & \multicolumn{1}{l}{G1: copula fixed} & \multicolumn{1}{c}{0.007} & \multicolumn{1}{c}{0.028} & \multicolumn{1}{c}{0.040} & \multicolumn{1}{c}{0.391} & \multicolumn{1}{c}{0.036} &       & \multicolumn{1}{c}{\multirow{4}[0]{*}{\textbf{97.5\%}}} & \multicolumn{1}{l}{G1: copula fixed} & \multicolumn{1}{c}{0.012} & \multicolumn{1}{c}{0.042} & \multicolumn{1}{c}{0.058} & \multicolumn{1}{c}{0.640} & \multicolumn{1}{c}{0.049} \\
    & \multicolumn{1}{l}{G2: marg. distr. fixed} & \multicolumn{1}{c}{0.049} & \multicolumn{1}{c}{0.090} & \multicolumn{1}{c}{0.109***} & \multicolumn{1}{c}{0.581} & \multicolumn{1}{c}{0.061} &       &       & \multicolumn{1}{l}{G2: marg. distr. fixed} & \multicolumn{1}{c}{0.022} & \multicolumn{1}{c}{0.061} & \multicolumn{1}{c}{0.068***} & \multicolumn{1}{c}{0.561} & \multicolumn{1}{c}{0.036} \\
    & \multicolumn{1}{l}{G3: all multivariate} & \multicolumn{1}{c}{0.050} & \multicolumn{1}{c}{0.097} & \multicolumn{1}{c}{0.117} & \multicolumn{1}{c}{0.626} & \multicolumn{1}{c}{0.067} &       &       & \multicolumn{1}{l}{G3: all multivariate} & \multicolumn{1}{c}{0.029} & \multicolumn{1}{c}{0.080} & \multicolumn{1}{c}{0.092} & \multicolumn{1}{c}{0.716} & \multicolumn{1}{c}{0.056} \\
    & \multicolumn{1}{l}{G4: all univariate} & \multicolumn{1}{c}{0.009} & \multicolumn{1}{c}{0.045} & \multicolumn{1}{c}{0.062} & \multicolumn{1}{c}{0.506} & \multicolumn{1}{c}{0.056} &       &       & \multicolumn{1}{l}{G4: all univariate} & \multicolumn{1}{c}{0.013} & \multicolumn{1}{c}{0.070} & \multicolumn{1}{c}{0.090} & \multicolumn{1}{c}{0.672} & \multicolumn{1}{c}{0.070} \\
    &       &       &       &       &       &       &       &       &       &       &       &       &       &  \\
    \multicolumn{1}{c}{\multirow{4}[1]{*}{\textbf{95.0\%}}} & \multicolumn{1}{l}{G1: copula fixed} & \multicolumn{1}{c}{0.006} & \multicolumn{1}{c}{0.023} & \multicolumn{1}{c}{0.034} & \multicolumn{1}{c}{0.308} & \multicolumn{1}{c}{0.031} &       & \multicolumn{1}{c}{\multirow{4}[1]{*}{\textbf{95.0\%}}} & \multicolumn{1}{l}{G1: copula fixed} & \multicolumn{1}{c}{0.008} & \multicolumn{1}{c}{0.033} & \multicolumn{1}{c}{0.045} & \multicolumn{1}{c}{0.543} & \multicolumn{1}{c}{0.039} \\
    & \multicolumn{1}{l}{G2: marg. distr. fixed} & \multicolumn{1}{c}{0.032} & \multicolumn{1}{c}{0.065} & \multicolumn{1}{c}{0.078***} & \multicolumn{1}{c}{0.453} & \multicolumn{1}{c}{0.047} &       &       & \multicolumn{1}{l}{G2: marg. distr. fixed} & \multicolumn{1}{c}{0.012} & \multicolumn{1}{c}{0.041} & \multicolumn{1}{c}{0.045} & \multicolumn{1}{c}{0.369} & \multicolumn{1}{c}{0.025} \\
    & \multicolumn{1}{l}{G3: all multivariate} & \multicolumn{1}{c}{0.032} & \multicolumn{1}{c}{0.070} & \multicolumn{1}{c}{0.085} & \multicolumn{1}{c}{0.484} & \multicolumn{1}{c}{0.054} &       &       & \multicolumn{1}{l}{G3: all multivariate} & \multicolumn{1}{c}{0.018} & \multicolumn{1}{c}{0.058} & \multicolumn{1}{c}{0.067} & \multicolumn{1}{c}{0.651} & \multicolumn{1}{c}{0.044} \\
    & \multicolumn{1}{l}{G4: all univariate} & \multicolumn{1}{c}{0.006} & \multicolumn{1}{c}{0.035} & \multicolumn{1}{c}{0.050} & \multicolumn{1}{c}{0.412} & \multicolumn{1}{c}{0.046} &       &       & \multicolumn{1}{l}{G4: all univariate} & \multicolumn{1}{c}{0.009} & \multicolumn{1}{c}{0.052} & \multicolumn{1}{c}{0.069} & \multicolumn{1}{c}{0.521} & \multicolumn{1}{c}{0.059} \\
    \midrule
    &       &       &       &       &       &       &       &       &       &       &       &       &       &  \\	
	\end{tabular}%
\end{center}
  \label{tab:analysisI_robust_lev}%
\end{sidewaystable}%

\begin{table}[htbp]
	\caption{Model risk before and after applying the MCS procedure}
	\vspace{0.2cm} {\justifying \scriptsize \noindent This table compares the model risk associated with one day ahead forecasts of the 99\% VaR and the 97.5\% ES before and after applying the model confidence set (MCS) procedure by \cite{Hansen.2011}. Details on the method and its implementation can be found in Sections \ref{sec:mcsTheoreticalBackground} and \ref{sec:analysisMCS}, respectively. Model risk is measured in terms of the mean absolute deviation (mad) of risk forecasts by various models in percent of the portfolio value. Model risk is calculated on a daily basis from November 4, 2004 until December 31, 2018 for all multivariate models that passed the respective backtest. The term \emph{crisis} refers to the years 2008-2009 while \emph{pre-crisis} and   \emph{after-crisis} denote to the period before and after, respectively. We provide minimum (Min), median, mean, maximum (Max) and standard deviation (SD) of the daily model risk estimates.  *** denotes statistically significant reductions of mean model risk at the 1\% level by applying the MCS procedure.}	\vspace{1cm}

\footnotesize\centering
		        \begin{tabular}{clcccc}
		        	\toprule
		    	&       & \multicolumn{2}{c}{\textbf{No MCS}} & \multicolumn{2}{c}{\textbf{MCS}} \\
		    	\cmidrule{3-6}          &       & \multicolumn{2}{c}{Model risk (in \%)} & \multicolumn{2}{c}{Model risk (in \%)} \\
		    	&       & \multicolumn{1}{l}{99\% VaR} & \multicolumn{1}{l}{97.5\% ES} & \multicolumn{1}{l}{99\% VaR} & \multicolumn{1}{l}{97.5\% ES} \\
		    	\cmidrule{2-6}    \multicolumn{1}{c}{\multirow{5}[2]{*}{\begin{sideways}\textbf{Whole period}\end{sideways}}} & Min   & 0.075 & 0.029 & 0.051 & 0.022 \\
		    	& Median & 0.138 & 0.080 & 0.106 & 0.078 \\
		    	& Mean  & 0.165 & 0.092 & 0.127*** & 0.089*** \\
		    	& Max   & 0.847 & 0.716 & 0.696 & 0.703 \\
		    	& SD    & 0.091 & 0.056 & 0.070 & 0.054 \\
		    	\cmidrule{2-6}          &       &       &       &       &  \\
		    	&       &       &       &       &  \\
		    	\cmidrule{2-6}    \multirow{5}[2]{*}{\begin{sideways}\textbf{Pre-crisis}\end{sideways}} & Min   & 0.081 & 0.029 & 0.069 & 0.022 \\
		    	& Median & 0.105 & 0.050 & 0.098 & 0.049 \\
		    	& Mean  & 0.119 & 0.063 & 0.110*** & 0.062*** \\
		    	& Max   & 0.280 & 0.207 & 0.259 & 0.214 \\
		    	& SD    & 0.036 & 0.030 & 0.034 & 0.030 \\
		    	\cmidrule{2-6}          &       &       &       &       &  \\
		    	\cmidrule{2-6}    \multirow{5}[2]{*}{\begin{sideways}\textbf{Crisis}\end{sideways}} & Min   & 0.116 & 0.046 & 0.087 & 0.045 \\
		    	& Median & 0.241 & 0.108 & 0.172 & 0.099 \\
		    	& Mean  & 0.286 & 0.145 & 0.215*** & 0.138*** \\
		    	& Max   & 0.847 & 0.716 & 0.696 & 0.703 \\
		    	& SD    & 0.139 & 0.101 & 0.120 & 0.097 \\
		    	\cmidrule{2-6}          &       &       &       &       &  \\
		    	\cmidrule{2-6}    \multirow{5}[2]{*}{\begin{sideways}\textbf{Post-crisis}\end{sideways}} & Min   & 0.075 & 0.034 & 0.051 & 0.034 \\
		    	& Median & 0.138 & 0.082 & 0.099 & 0.081 \\
		    	& Mean  & 0.155 & 0.090 & 0.113*** & 0.088*** \\
		    	& Max   & 0.551 & 0.638 & 0.569 & 0.647 \\
		    	& SD    & 0.064 & 0.037 & 0.045 & 0.037 \\
		    	\bottomrule   \end{tabular}%
	\label{tab:MCSvsNoMCS}%
\end{table}%

\begin{table}[htbp]
  \caption{Summary statistics of market risk models passing the backtest}
\scriptsize
This table presents summary statistics on the number of multivariate market risk models passing the respective backtest, see Section \ref{sec:backtests} for details. We consider VaR (Panel A) and ES (Panel B) forecasts for a well diversified portfolio from November 4, 2004 until December 31, 2018 and confidence levels (\textit{Conf. level}) ranging from 95\% to 99.9\%.
We provide minimum (Min), median, mean, maximum (Max) and standard deviation (SD) of the daily number of models. 
\begin{center} 
    \begin{tabular}{rrrrrr}
          &       &       &       &       &  \\
\midrule
\multicolumn{1}{l}{\textit{Panel A: VaR}} &       &       &       &       &  \\
& \multicolumn{5}{c}{\textbf{Models passing backtest}} \\
& \multicolumn{1}{c}{Min} & \multicolumn{1}{c}{Median} & \multicolumn{1}{c}{Mean} & \multicolumn{1}{c}{Max} & \multicolumn{1}{c}{SD} \\
\cmidrule{2-6}    \multicolumn{1}{c}{\textbf{Conf. level}} &       &       &       &       &  \\
\multicolumn{1}{c}{99.9\%} & \multicolumn{1}{c}{137} & \multicolumn{1}{c}{180} & \multicolumn{1}{c}{178} & \multicolumn{1}{c}{180} & \multicolumn{1}{c}{5} \\
\multicolumn{1}{c}{99.0\%} & \multicolumn{1}{c}{65} & \multicolumn{1}{c}{180} & \multicolumn{1}{c}{174} & \multicolumn{1}{c}{180} & \multicolumn{1}{c}{17} \\
\multicolumn{1}{c}{97.5\%} & \multicolumn{1}{c}{89} & \multicolumn{1}{c}{180} & \multicolumn{1}{c}{176} & \multicolumn{1}{c}{180} & \multicolumn{1}{c}{11} \\
\multicolumn{1}{c}{95.0\%} & \multicolumn{1}{c}{158} & \multicolumn{1}{c}{180} & \multicolumn{1}{c}{179} & \multicolumn{1}{c}{180} & \multicolumn{1}{c}{3} \\
\midrule
&       &       &       &       &  \\
\multicolumn{1}{l}{\textit{Panel B: ES}} &       &       &       &       &  \\
& \multicolumn{5}{c}{\textbf{Models passing backtest}} \\
& \multicolumn{1}{c}{Min} & \multicolumn{1}{c}{Median} & \multicolumn{1}{c}{Mean} & \multicolumn{1}{c}{Max} & \multicolumn{1}{c}{SD} \\
\cmidrule{2-6}    \multicolumn{1}{c}{\textbf{Conf. level}} &       &       &       &       &  \\
\multicolumn{1}{c}{99.9\%} & \multicolumn{1}{c}{45} & \multicolumn{1}{c}{113} & \multicolumn{1}{c}{123} & \multicolumn{1}{c}{180} & \multicolumn{1}{c}{46} \\
\multicolumn{1}{c}{99.0\%} & \multicolumn{1}{c}{97} & \multicolumn{1}{c}{130} & \multicolumn{1}{c}{136} & \multicolumn{1}{c}{180} & \multicolumn{1}{c}{25} \\
\multicolumn{1}{c}{97.5\%} & \multicolumn{1}{c}{63} & \multicolumn{1}{c}{120} & \multicolumn{1}{c}{121} & \multicolumn{1}{c}{179} & \multicolumn{1}{c}{27} \\
\multicolumn{1}{c}{95.0\%} & \multicolumn{1}{c}{24} & \multicolumn{1}{c}{119} & \multicolumn{1}{c}{109} & \multicolumn{1}{c}{168} & \multicolumn{1}{c}{31} \\
\midrule
&       &       &       &       &  \\
    \end{tabular}%
\end{center}
\label{tab:backtesting_analysisI}%
\end{table}%

\begin{table}[htbp]
  \caption{Summary statistics of market risk models passing alternative backtests}
\scriptsize
This table presents summary statistics on the number of multivariate market risk models passing a set of backtesting alternatives.
For our main analysis, we rely on the duration-based VaR backtest by \cite{Christoffersen.2004} and the conditional calibration ES backtest by \cite{Nolde.2017} with a moving window of 500 days (\emph{Baseline}). For robustness, we also consider the dynamic quantile VaR backtest by \cite{Engle.2004} and the exceedance residual ES backtest by \cite{McNeil.2000} (\emph{Alternative backtests}), for more details we refer to Section \ref{sec:backtests}. Additionally, we provide results when not applying any backtest (\emph{No backtest}) or when using the baseline backtest with a fixed window (\emph{Fixed window}) or a moving window of 1000 days (\emph{1000 days mov. window}). We consider VaR (Panel A) and ES (Panel B) forecasts for a well diversified portfolio from November 4, 2004 until December 31, 2018 and confidence levels (\textit{Conf. level}) ranging from 95\% to 99.9\%. We provide minimum (Min), median, mean, maximum (Max) and standard deviation (SD) of the daily number of models.
\begin{center}
    \begin{tabular}{rrrrrr}
          &       &       &       &       &  \\
\midrule
\multicolumn{1}{l}{\textit{Panel A: 99\% VaR}} &       &       &       &       &  \\
& \multicolumn{5}{c}{\textbf{Models passing backtest}} \\
& \multicolumn{1}{c}{Min} & \multicolumn{1}{c}{Median} & \multicolumn{1}{c}{Mean} & \multicolumn{1}{c}{Max} & \multicolumn{1}{c}{SD} \\
\cmidrule{2-6}          &       &       &       &       &  \\
\multicolumn{1}{l}{Baseline} & \multicolumn{1}{c}{65} & \multicolumn{1}{c}{180} & \multicolumn{1}{c}{174} & \multicolumn{1}{c}{180} & \multicolumn{1}{c}{17} \\
\multicolumn{1}{l}{No backtest} & \multicolumn{1}{c}{180} & \multicolumn{1}{c}{180} & \multicolumn{1}{c}{180} & \multicolumn{1}{c}{180} & \multicolumn{1}{c}{0} \\
\multicolumn{1}{l}{Alternative backtest} & \multicolumn{1}{c}{0} & \multicolumn{1}{c}{65} & \multicolumn{1}{c}{71} & \multicolumn{1}{c}{160} & \multicolumn{1}{c}{42} \\
\multicolumn{1}{l}{Fixed window} & \multicolumn{1}{c}{113} & \multicolumn{1}{c}{113} & \multicolumn{1}{c}{113} & \multicolumn{1}{c}{113} & \multicolumn{1}{c}{0} \\
\multicolumn{1}{l}{1000 days mov. window} & \multicolumn{1}{c}{0} & \multicolumn{1}{c}{179} & \multicolumn{1}{c}{145} & \multicolumn{1}{c}{180} & \multicolumn{1}{c}{62} \\
\midrule
&       &       &       &       &  \\
\multicolumn{1}{l}{\textit{Panel B: 97.5\% ES}} &       &       &       &       &  \\
& \multicolumn{5}{c}{\textbf{Models passing backtest}} \\
& \multicolumn{1}{c}{Min} & \multicolumn{1}{c}{Median} & \multicolumn{1}{c}{Mean} & \multicolumn{1}{c}{Max} & \multicolumn{1}{c}{SD} \\
\cmidrule{2-6}          &       &       &       &       &  \\
\multicolumn{1}{l}{Baseline} & \multicolumn{1}{c}{63} & \multicolumn{1}{c}{120} & \multicolumn{1}{c}{121} & \multicolumn{1}{c}{179} & \multicolumn{1}{c}{27} \\
\multicolumn{1}{l}{No backtest} & \multicolumn{1}{c}{180} & \multicolumn{1}{c}{180} & \multicolumn{1}{c}{180} & \multicolumn{1}{c}{180} & \multicolumn{1}{c}{0} \\
\multicolumn{1}{l}{Alternative backtest} & \multicolumn{1}{c}{99} & \multicolumn{1}{c}{119} & \multicolumn{1}{c}{127} & \multicolumn{1}{c}{180} & \multicolumn{1}{c}{21} \\
\multicolumn{1}{l}{Fixed window} & \multicolumn{1}{c}{17} & \multicolumn{1}{c}{17} & \multicolumn{1}{c}{17} & \multicolumn{1}{c}{17} & \multicolumn{1}{c}{0} \\
\multicolumn{1}{l}{1000 days mov. window} & \multicolumn{1}{c}{0} & \multicolumn{1}{c}{95} & \multicolumn{1}{c}{82} & \multicolumn{1}{c}{120} & \multicolumn{1}{c}{39} \\
\midrule
&       &       &       &       &  \\
    \end{tabular}%
  \end{center}
\label{tab:backtesting_analysisIII}%
\end{table}%

\begin{table}[htbp]
  \caption{Model risk and alternative backtests (all multivariate models)}
  	\vspace{0.2cm} {\justifying \scriptsize \noindent This table provides summary statistics on the model risk associated with one day ahead forecasts of 99\% VaR and 97.5\% ES after having different (or differently specified) backtests. For our main analysis, we rely on the duration-based VaR backtest by \cite{Christoffersen.2004} and the conditional calibration ES backtest by \cite{Nolde.2017} with a moving window of 500 days (\emph{Baseline}). For robustness, we also consider the dynamic quantile VaR backtest by \cite{Engle.2004} and the exceedance residual ES backtest by \cite{McNeil.2000} (\emph{Alternative backtests}), for more details we refer to Section \ref{sec:backtests}. Additionally, we provide results when not applying any backtest (\emph{No backtest}) or when using the baseline backtest with a fixed window (\emph{Fixed window}) or a moving window of 1000 days (\emph{1000 days mov. window}).\\  		
  	Model risk is measured in terms of the mean absolute deviation (mad) of the risk forecasts that passed the respective backtest on a daily basis from November 4, 2004 until December 31, 2018. We report model risk estimates in percent of the portfolio value (first and second column) and in absolute terms (third and fourth column) for an equally weighted portfolio. The results in absolute terms are based on a portfolio value of \$100,000 and a 10 day forecast horizon. We provide minimum (Min), median, mean, maximum (Max), and standard deviation (SD) of the daily model risk estimates.}	\vspace{0.5cm}
  
  \footnotesize\centering
        \begin{tabular}{cclcccc}
        	\toprule
    	&       &       & \multicolumn{2}{c}{\textbf{Model risk (in \%)}} & \multicolumn{2}{c}{\textbf{Model risk (in \$)}} \\
    	&       &       & \multicolumn{1}{l}{99\% VaR} & \multicolumn{1}{l}{97.5\% ES} & \multicolumn{1}{l}{99\% VaR } & \multicolumn{1}{l}{97.5\% ES} \\
    	\cmidrule{3-7}          & \multicolumn{1}{c}{\multirow{5}[2]{*}{\begin{sideways}\textbf{Baseline}\end{sideways}}} & Min   & 0.075 & 0.029 & 237   & 92 \\
    	&       & Median & 0.138 & 0.080 & 436   & 253 \\
    	&       & Mean  & 0.165 & 0.092 & 522   & 291 \\
    	&       & Max   & 0.847 & 0.716 & 2,678 & 2,264 \\
    	&       & SD    & 0.091 & 0.056 & 288   & 177 \\
    	\cmidrule{3-7}          &       &       &       &       &       &  \\
    	\cmidrule{3-7}          & \multicolumn{1}{c}{\multirow{5}[2]{*}{\begin{sideways}\textbf{No backtest}\end{sideways}}} & Min   & 0.056 & 0.058 & 177   & 183 \\
    	&       & Median & 0.127 & 0.133 & 402   & 421 \\
    	&       & Mean  & 0.156 & 0.163 & 493   & 515 \\
    	&       & Max   & 0.847 & 0.876 & 2678  & 2770 \\
    	&       & SD    & 0.089 & 0.092 & 281   & 291 \\
    	\cmidrule{3-7}          &       &       &       &       &       &  \\
    	\cmidrule{3-7}    \multirow{5}[2]{*}{\begin{sideways}\textbf{Alternative}\end{sideways}} & \multicolumn{1}{c}{\multirow{5}[2]{*}{\begin{sideways}\textbf{backtests}\end{sideways}}} & Min   & 0.000 & 0.032 & 0     & 101 \\
    	&       & Median & 0.065 & 0.095 & 206   & 300 \\
    	&       & Mean  & 0.079 & 0.114 & 250   & 360 \\
    	&       & Max   & 0.672 & 0.864 & 2125  & 2732 \\
    	&       & SD    & 0.054 & 0.077 & 171   & 243 \\
    	\cmidrule{3-7}          &       &       &       &       &       &  \\
    	\cmidrule{3-7}    \multirow{5}[2]{*}{\begin{sideways}\textbf{Fixed}\end{sideways}} & \multicolumn{1}{c}{\multirow{5}[2]{*}{\begin{sideways}\textbf{window}\end{sideways}}} & Min   & 0.035 & 0.011 & 111   & 35 \\
    	&       & Median & 0.097 & 0.068 & 307   & 215 \\
    	&       & Mean  & 0.129 & 0.078 & 408   & 247 \\
    	&       & Max   & 1.081 & 0.712 & 3418  & 2252 \\
    	&       & SD    & 0.103 & 0.049 & 326   & 155 \\
    	\cmidrule{3-7}          &       &       &       &       &       &  \\
    	\cmidrule{3-7}    \multirow{5}[2]{*}{\begin{sideways}\textbf{1000 days}\end{sideways}} & \multicolumn{1}{c}{\multirow{5}[2]{*}{\begin{sideways}\textbf{mov. window}\end{sideways}}} & Min   & 0.048 & 0.025 & 152   & 79 \\
    	&       & Median & 0.170 & 0.072 & 538   & 228 \\
    	&       & Mean  & 0.202 & 0.091 & 639   & 288 \\
    	&       & Max   & 1.118 & 0.962 & 3535  & 3042 \\
    	&       & SD    & 0.113 & 0.067 & 357   & 212 \\
    	\bottomrule    \end{tabular}%
  \label{tab:robustnessBacktestsAllMultModels}%
\end{table}%

\begin{table}[htbp]
\caption{Model risk and alternative backtests (model sets with fixed and varying copula only)}
\scriptsize
This table presents summary statistics for the time series of average model risk per group associated with 99\% VaR (Panel A) and 97.5\% ES (Panel B) forecasts for a well diversified portfolio and a set of backtesting alternatives. For our main analysis, we rely on the duration-based VaR backtest by \cite{Christoffersen.2004} and the conditional calibration ES backtest by \cite{Nolde.2017} with a moving window of 500 days (\emph{Baseline}). For robustness, we also consider the dynamic quantile VaR backtest by \cite{Engle.2004} and the exceedance residual ES backtest by \cite{McNeil.2000} (\emph{Alternative backtests}), for more details we refer to Section \ref{sec:backtests}. Additionally, we provide results when not applying any backtest (\emph{No backtest}) or when using the baseline backtest with a fixed window (\emph{Fixed window}) or a moving window of 1000 days (\emph{1000 days mov. window}). Model risk is measured in terms of the mean absolute deviation (mad) of risk forecasts by various models within a model set that passed the respective backtest on a daily basis from November 4, 2004 until December 31, 2018. 
\textit{Group 1} (G1) includes all model sets in which a copula function is fixed while varying the marginal distribution. \textit{Group 2} (G2) contains analogously the model sets with fixed marginal distribution and varying copula. We report model risk estimates for one day ahead risk forecasts in percent of the portfolio value. We provide minimum (Min), median, mean, maximum (Max), and standard deviation (SD) of the daily averaged model risk estimates per group.
\begin{center} 
	\footnotesize
    \begin{tabular}{rrrrrrr}
          &       &       &       &       &       &  \\
\midrule
\multicolumn{1}{l}{\textit{Panel A: 99\% VaR}} &       &       &       &       &       &  \\
&       & \multicolumn{5}{c}{\textbf{Average model risk (in \%)}} \\
&       & \multicolumn{1}{c}{Min} & \multicolumn{1}{c}{Median} & \multicolumn{1}{c}{Mean} & \multicolumn{1}{c}{Max} & \multicolumn{1}{c}{SD} \\
\cmidrule{3-7}    \multicolumn{1}{c}{\textbf{Group}} &       &       &       &       &       &  \\
\multicolumn{1}{c}{\multirow{5}[0]{*}{G1: Copula fixed}} & \multicolumn{1}{l}{Baseline} & \multicolumn{1}{c}{0.008} & \multicolumn{1}{c}{0.037} & \multicolumn{1}{c}{0.052} & \multicolumn{1}{c}{0.518} & \multicolumn{1}{c}{0.044} \\
& \multicolumn{1}{l}{No backtest} & \multicolumn{1}{c}{0.012} & \multicolumn{1}{c}{0.037} & \multicolumn{1}{c}{0.052} & \multicolumn{1}{c}{0.518} & \multicolumn{1}{c}{0.044} \\
& \multicolumn{1}{l}{Alternative backtest} & \multicolumn{1}{c}{0.000} & \multicolumn{1}{c}{0.039} & \multicolumn{1}{c}{0.051} & \multicolumn{1}{c}{0.524} & \multicolumn{1}{c}{0.041} \\
& \multicolumn{1}{l}{Fixed window} & \multicolumn{1}{c}{0.011} & \multicolumn{1}{c}{0.033} & \multicolumn{1}{c}{0.048} & \multicolumn{1}{c}{0.516} & \multicolumn{1}{c}{0.043} \\
& \multicolumn{1}{l}{1000 days mov. window} & \multicolumn{1}{c}{0.014} & \multicolumn{1}{c}{0.040} & \multicolumn{1}{c}{0.053} & \multicolumn{1}{c}{0.518} & \multicolumn{1}{c}{0.041} \\
&       &       &       &       &       &  \\
\multicolumn{1}{c}{\multirow{5}[1]{*}{G2: Marg. distr. fixed}} & \multicolumn{1}{l}{Baseline} & \multicolumn{1}{c}{0.073} & \multicolumn{1}{c}{0.130} & \multicolumn{1}{c}{0.157} & \multicolumn{1}{c}{0.803} & \multicolumn{1}{c}{0.084} \\
& \multicolumn{1}{l}{No backtest} & \multicolumn{1}{c}{0.075} & \multicolumn{1}{c}{0.130} & \multicolumn{1}{c}{0.157} & \multicolumn{1}{c}{0.803} & \multicolumn{1}{c}{0.084} \\
& \multicolumn{1}{l}{Alternative backtest} & \multicolumn{1}{c}{0.000} & \multicolumn{1}{c}{0.038} & \multicolumn{1}{c}{0.042} & \multicolumn{1}{c}{0.439} & \multicolumn{1}{c}{0.024} \\
& \multicolumn{1}{l}{Fixed window} & \multicolumn{1}{c}{0.058} & \multicolumn{1}{c}{0.103} & \multicolumn{1}{c}{0.126} & \multicolumn{1}{c}{0.717} & \multicolumn{1}{c}{0.072} \\
& \multicolumn{1}{l}{1000 days mov. window} & \multicolumn{1}{c}{0.075} & \multicolumn{1}{c}{0.137} & \multicolumn{1}{c}{0.159} & \multicolumn{1}{c}{0.669} & \multicolumn{1}{c}{0.073} \\
\midrule
&       &       &       &       &       &  \\
\multicolumn{1}{l}{\textit{Panel B: 97.5\% ES}} &       &       &       &       &       &  \\
&       & \multicolumn{5}{c}{\textbf{Average model risk (in \%)}} \\
&       & \multicolumn{1}{c}{Min} & \multicolumn{1}{c}{Median} & \multicolumn{1}{c}{Mean} & \multicolumn{1}{c}{Max} & \multicolumn{1}{c}{SD} \\
\cmidrule{3-7}    \multicolumn{1}{c}{\textbf{Group}} &       &       &       &       &       &  \\
\multicolumn{1}{c}{\multirow{5}[0]{*}{G1: Copula fixed}} & \multicolumn{1}{l}{Baseline} & \multicolumn{1}{c}{0.012} & \multicolumn{1}{c}{0.042} & \multicolumn{1}{c}{0.058} & \multicolumn{1}{c}{0.640} & \multicolumn{1}{c}{0.049} \\
& \multicolumn{1}{l}{No backtest} & \multicolumn{1}{c}{0.012} & \multicolumn{1}{c}{0.040} & \multicolumn{1}{c}{0.055} & \multicolumn{1}{c}{0.537} & \multicolumn{1}{c}{0.045} \\
& \multicolumn{1}{l}{Alternative backtest} & \multicolumn{1}{c}{0.012} & \multicolumn{1}{c}{0.042} & \multicolumn{1}{c}{0.058} & \multicolumn{1}{c}{0.581} & \multicolumn{1}{c}{0.050} \\
& \multicolumn{1}{l}{Fixed window} & \multicolumn{1}{c}{0.006} & \multicolumn{1}{c}{0.027} & \multicolumn{1}{c}{0.039} & \multicolumn{1}{c}{0.546} & \multicolumn{1}{c}{0.039} \\
& \multicolumn{1}{l}{1000 days mov. window} & \multicolumn{1}{c}{0.017} & \multicolumn{1}{c}{0.047} & \multicolumn{1}{c}{0.065} & \multicolumn{1}{c}{0.640} & \multicolumn{1}{c}{0.053} \\
&       &       &       &       &       &  \\
\multicolumn{1}{c}{\multirow{5}[1]{*}{G2: Marg. distr. fixed}} & \multicolumn{1}{l}{Baseline} & \multicolumn{1}{c}{0.022} & \multicolumn{1}{c}{0.061} & \multicolumn{1}{c}{0.068} & \multicolumn{1}{c}{0.561} & \multicolumn{1}{c}{0.036} \\
& \multicolumn{1}{l}{No backtest} & \multicolumn{1}{c}{0.080} & \multicolumn{1}{c}{0.137} & \multicolumn{1}{c}{0.164} & \multicolumn{1}{c}{0.835} & \multicolumn{1}{c}{0.087} \\
& \multicolumn{1}{l}{Alternative backtest} & \multicolumn{1}{c}{0.029} & \multicolumn{1}{c}{0.069} & \multicolumn{1}{c}{0.077} & \multicolumn{1}{c}{0.561} & \multicolumn{1}{c}{0.045} \\
& \multicolumn{1}{l}{Fixed window} & \multicolumn{1}{c}{0.003} & \multicolumn{1}{c}{0.029} & \multicolumn{1}{c}{0.034} & \multicolumn{1}{c}{0.423} & \multicolumn{1}{c}{0.026} \\
& \multicolumn{1}{l}{1000 days mov. window} & \multicolumn{1}{c}{0.015} & \multicolumn{1}{c}{0.039} & \multicolumn{1}{c}{0.047} & \multicolumn{1}{c}{0.546} & \multicolumn{1}{c}{0.035} \\
\midrule
&       &       &       &       &       &  \\
    \end{tabular}%
\end{center}
\label{tab:analysisIII}%
\end{table}%

\clearpage
\renewcommand{\thetable}{IA.\arabic{table}}
\setcounter{table}{0}

\renewcommand{\thefigure}{IA.\arabic{figure}}
\setcounter{figure}{0}
\long\def\symbolfootnote[#1]#2{\begingroup%
\def\thefootnote{\fnsymbol{footnote}}\footnote[#1]{#2}\endgroup} 

\begin{center}
\LARGE{\bf Internet Appendix for \\
	
	``Marginals Versus Copulas: Which Account For More Model Risk In Multivariate Risk Forecasting?''}
\end{center}
\vspace{4cm}
\singlespacing
This Internet Appendix contains theoretical foundations as well as several additional tables and figures that complement the results presented in the main paper. 
\thispagestyle{empty}
\newpage
\doublespacing
\renewcommand\thesection{IA.\arabic{section}}
\setcounter{section}{0}
\setcounter{footnote}{0}
\setcounter{page}{1}

\newpage
%
%
%
%

\clearpage
\doublespacing

\section{Theoretical foundations}
 \label{sec:appendixTheoreticalFoundations}
\subsection{ARMA-GARCH process}
\label{sec:appendixArmaGarch}
In an ARMA(p,q)-GARCH(r,s) process, the conditional mean of a (univariate) time series is modeled by the ARMA part while the conditional volatility is captured by the GARCH part.
With ARMA(p,q) we denote a model with $p$ autoregressive and $q$ moving average terms. More formally, we have the following specification
$$r_t=\mu + \sum_{j=1}^p \phi_j r_{t-j} + \sum_{j=1}^q \theta_j\epsilon_{t-j} + \epsilon_t,$$
where $r_\tau, \tau = t-p, \dots, t$ are observations from the time series and $\phi_j, j=1, \dots, p$, $\theta_j, j=1, \dots, q$, and $\mu$ denote parameters. 
%
%
%
%
%
The Generalized Autoregressive Conditional Heteroskedasticity (GARCH) model by \cite{Bollerslev.1986} extends the ARCH model due to \cite{Engle.1982} by including lags of the conditional variances. More exactly, the variance equation according to the GARCH(r,s) model at time $t$ is given by 
$$\sigma_t^2=\omega + \sum_{j=1}^r\beta_j\sigma^2_{t-j} +\sum_{j=1}^s\alpha_j\epsilon^2_{t-j},$$
where $\sigma^2_\tau, \tau=t-r, \dots, t$ denotes the conditional variance, $\alpha_j, j=1, \dots, s$, $\beta_j, j=1, \dots, r$, and $\omega$ are parameters and all $\epsilon_t$ are of the form $\epsilon_t=z_t\sigma_t$ where $z_t$ is an iid process with zero mean and unit variance. 
The parameters must fulfill some conditions in order to guarantee that the GARCH conditional variance estimates are always positive, see \cite{Nelson.1992} for details. In case of the GARCH(1,1) model, forecasts can be calculated as
$$\sigma^2_{t+h|t}=\sigma^2 + (\alpha+\beta)^{h-1}(\sigma^2_{t+1}-\sigma^2),$$
where $h>2$ denotes the horizon of the forecasts and $\sigma^2$ denotes the unconditional variance given by $\sigma^2=\frac{\omega}{1-\alpha-\beta}$ \citep[see ][]{Bollerslev.2010}.\footnote{Conditional variance estimates for the GARCH(1,1) model are positive almost surely given that $\omega>0, \alpha\geq 0$ and $\beta\geq 0$. The model is covariance stationary provided that $\alpha + \beta<1$.}

\subsection{Sklar's theorem}
\label{sec:sklarsTheorem}
The popularity of copulas in multivariate dependence modeling is due to the theorem by \cite{Sklar.1959}. Loosely speaking the theorem states that modeling of the marginals and of the multivariate dependence can be separated by means of copulas functions. 

\begin{theorem}[Sklar's theorem]
	Let $\mathbf{X}=(X_1, \dots, X_d)\sim F$ be a $d$-dimensional random variable with marginal distributions $F_j, j=1, \dots, d.$ We then have $$F(x_1, \dots, x_d)=C\big(F_1(x_1), \dots, F_d(x_d)\big),$$ where $C$ denotes some appropriate $d$-dimensional copula. If the multivariate distribution function $F$ is absolutely continuous and the marginal distributions $F_1, \dots, F_d$ are strictly increasing continuous, we have 
	$$f(x_1, \dots, x_d)=\Big(\prod_{j=1}^df_j(x_j) \Big)\cdot c\big(F_1(x_1), \dots, F_d(x_d)\big)$$ with the small letters denoting the corresponding probability density functions.
\end{theorem}
The second part of the theorem highlights that the joint distribution of the random variable $\mathbf{X}$ can be modeled separately in terms of a ``marginal term'' $\prod_{j=1}^d f_j(x_j)$ and a ``dependence term'' $c\big(F_1(x_1), \dots, F_d(x_d)\big)$. The marginal term is based on information from the (univariate) marginals alone and does not contain any information about the multivariate dependence. On the other hand, the random variables $F_1(X_1), \dots, F_d(X_d)$ are all uniformly distributed in the interval $[0,1]$ and therefore do not contain any information on the marginals. For more details we refer to the books by \cite{Joe.1997} and \cite{Nelsen.2006}.

\subsection{Duration-based backtest}
\label{sec:appendixVaRbacktest}
The duration-based VaR backtest by \cite{Christoffersen.2004}, as the name implies, is based on the duration of days between VaR violations. The hit sequence of $VaR_t$ violations is defined as
$$I_t = \begin{cases}
	&1, \quad \text{if} ~ r_t < -VaR_t(p) \\
	&0, \quad \text{else}
\end{cases}$$
where $r_t$ is a time series of daily ex-post portfolio returns and $VaR_t(p)$ a time series of ex-ante VaR forecasts with a coverage rate $p$.
The time in days between two VaR violations is called the no-hit duration $D_i = t_i - t_{i-1}$ where $t_i$ denotes the day of violation number $i$.
The null hypothesis of the test is then: If the VaR model is correctly specified for coverage rate $p$, the no-hit duration or in other words the conditional expected duration between VaR violations should have no memory and a mean duration of $1/p$ days. Thus, under the null hypothesis the no-hit duration follows the exponential distribution
$$f_{exp}(D;p)= p~ exp(-pD)$$
whereas the alternative that allows for duration dependence follows a Weibull distribution
$$f_W(D;a,b)=a^b b D^{b-1}exp (-(aD)^b).$$ The tested null hypothesis of independence is then defined as $$H_{0,ind}: b=1.$$
For a detailed derivation of the no-memory  property in terms of the discrete probability distribution and its hazard function, please see \cite{Christoffersen.2004}.

\subsection{Conditional calibration backtest}
\label{sec:appendixESbacktest}
\cite{Nolde.2017} introduce a conditional calibration (CC) test and show that well-known traditional backtests can be unified within the concept of CC. The CC test comes in two versions: The simple version used in our analysis requires only risk forecasts (VaR and ES), whereas the general version additionally needs information on conditional volatility. 
Following \cite{Nolde.2017}, $\mathcal{P}_0$ defines the class of Borel-probability distributions on $\mathbb{R}$. Moreover, $\mathcal{P}_1 \subseteq \mathcal{P}_0$ denotes the class of all distributions with finite mean whereas $\mathcal{P}_V \subset \mathcal{P}_0$ describes distributions with unique quantiles. 
The CC backtests rely on so-called identification functions. We further follow the notation of \cite{Nolde.2017} and define $\Theta=(\rho_1, ..., \rho_k)$ as the identifiable functional with an identification function $V$ with respect to $\mathcal{P}$.
Let $\{r_t\}_{t \in \mathbb{N}}$ be a series of negated log-returns that are adapted to the filtration $\mathcal{F} = \{\mathcal{F}_t\}_{t \in \mathbb{N}} $. Further, we define $\{x_t\}_{t \in \mathbb{N}}$ as a sequence of predictions of $\Theta$ that are $\mathcal{F}_{t-1}$-measurable. All conditional distributions $\mathcal{L}(r_t \vert \mathcal{F}_{t-1})$ and all unconditional distributions $\mathcal{L}(r_t)$ are assumed to belong to $\mathcal{P}$ almost surely.
The identification function used for the pair $(VaR_\nu,ES_\nu)$ for the level $\nu \in (0,1)$ is defined as
$$V(x_1,x_2,r) = \left(\frac{1- \nu - \mathbb{I}_{(0, \infty)}(r- x_1)}{x_1-x_2-\frac{1}{1-\nu} \mathbb{I}_{(0, \infty)}(r- x_1) (x_1-r)}\right)$$
with respect to $\mathcal{P}_1 \cap \mathcal{P}_V$, where $\mathbb{I}_{(0,\infty)}$ denotes the characteristic function of the open interval $(0,\infty)$. The sequence of predictions $\{x_t\}_{t \in \mathbb{N}}$ is conditionally calibrated for $\Theta$ if $$\mathbb{E}(V(x_t,r_t) \vert \mathcal{F}_{t-1}) = 0 \quad \text{almost surely}, \forall t \in \mathbb{N}.$$
The null hypothesis of the traditional backtest for CC considers exactly this requirement: The sequence of predictions $\{x_t\}_{t \in \mathbb{N}}$ is conditionally calibrated for $\Theta$. Here, the requirement for the expected value is equivalent to 
$\mathbb{E}(h^{'}_t V(x_t,r_t)) = 0$ for all $\mathcal{F}_{t-1}$-measurable $\mathbb{R}^k$-valued functions $h_t$. 
\cite{Nolde.2017} consider a $\mathcal{F}$-predictable sequence $\{\mathbf{h}_t\}_{t \in \mathbb{N}}$ of $q \times k$-matrices $\mathbf{h}_t$ called test functions to construct a Wald-type test statistic. For the simple version of the CC test, $\mathbf{h}_t$ equals the identity matrix. For more information on the Wald-type test statistic as well as a complete derivation of the CC test in both versions, please refer to the original paper.

\subsection{Model confidence set procedure}
\label{sec:MCS_internetAppendix}
In this section we more formally introduce the MCS procedure outlined in Section \ref{sec:mcsTheoreticalBackground}. Let therefore $\mathcal{M}^0$ denote the set of candidate models, $\mathcal{M}^*$ the true set of best models, and $\hat{\mathcal{M}}^*_{1-\alpha}$ the model confidence set at confidence level $\alpha$. We further assume that $\mathcal{M}^0$ consists of a finite number $m_0$ of models.  Based on an evaluation criterion (the \emph{loss function}) one can calculate the losses $L_{i,t}$ that are associated with model $i$ at time $t$. In the case of a VaR forecast, e.g., one might compare the risk forecast $VaR_\alpha^{i,t}$ of model $i$ at time $t$ with the actual realized return $r_t$ by setting $L_{i,t}:=L(VaR_\alpha^{i,t}, r_t)$ where $L$ denotes an appropriate loss function. For $i,j \in \mathcal{M}^0$ one then defines the relative performance variable 
$$d_{ij,t}:=L_{i,t}-L_{j,t}$$
as well as the expected value of the performance variable 
$$\mu_{ij}:=E(d_{ij,t}).$$ 
The alternatives in $\mathcal{M}^0$ are now ranked based on their expected loss. That is, model $i$ is preferred over model $j$ if $\mu_{ij}<0.$
The MCS is now constructed based on a sequence of significance tests
$$H_{0,\mathcal{M}}: \mu_{ij}=0 \quad \text{for all } i,j \in \mathcal{M},$$
with $\mathcal{M}\subseteq \mathcal{M}^0$. If $H_{0,\mathcal{M}}$ can be rejected, the elimination rule is applied to remove a model from $\mathcal{M}$ that is inferior to the remaining ones. The model confidence set is then defined as any subset of $\mathcal{M}^0$ that contains all best models with a given probability $1-\alpha$. More shortly, the MCS procedure can be summarized in the following algorithmic form \citep[see][]{Hansen.2011}:
\begin{align*}
	\text{Step 0: } & \text{Set } \mathcal{M}:=\mathcal{M}^0.\\
	\text{Step 1: } & \text{Test the null hypothesis } H_{0,\mathcal{M}} \text{ based on the equivalence test } \delta_\mathcal{M} \text{ at the confidence level } \alpha.\\
	\text{Step 2: } & \text{If } H_{0,\mathcal{M}} \text{ is not rejected, set } \hat{\mathcal{M}}^*_{1-\alpha}:=\mathcal{M} \text{, otherwise use the elimination rule } e_\mathcal{M} \text{ to elimi-}\\ 
	& \text{nate a model from } \mathcal{M} \text{ and repeat the procedure from step 1.}
\end{align*}

\cite{Hansen.2011} provide two t-statistics for the hypothesis test in step 1. We opt for the statistic $t_{ij}$ that is also used in the test for comparing two forecasts \citep[see ][]{Diebold.1995, West.1996}. We therefore define the sample loss statistic $\bar{d}_{ij}:=\frac{1}{n}\sum_{t=1}^n d_{ij,t}$ and set
$$t_{ij}:=\frac{\bar{d}_{ij}}{\sqrt{\hat{var}(\bar{d}_{ij})}},$$
where $\hat{var}(\bar{d}_{ij})$ denotes an estimate of $var(\bar{d}_{ij})$. The final test statistic is then defined as
$$T_{R,\mathcal{M}}:=\max_{i,j\in\mathcal{M}}|t_{ij}|.$$ The asymptotic distribution of $T_{R,\mathcal{M}}$ is non-standard and derived via a bootstrapping scheme, see \cite{Hansen.2011} for details. The natural elimination rule corresponding to the test statistic $T_{R,\mathcal{M}}$ is $e_{R,\mathcal
	M}:=\arg \max_{i\in\mathcal{M}} \sup_{j\in\mathcal{M}}t_{ij}$ because the corresponding model is such that $t_{e_{R,\mathcal{M}}j}=T_{R,\mathcal{M}}$ is fulfilled for some $j \in \mathcal{M}$. Removing model $e_{R,\mathcal{M}}$ will therefore reduce (or at least not increase) the test statistic $T_{R,\mathcal{M}}$.

\clearpage

\section{Figures}
\begin{figure}[htbp]
	\caption{Potential portfolio value under financial distress according to the 97.5\% ES}
	\vspace{0.2cm} \noindent {\justifying \scriptsize This figure illustrates the economic significance of model risk arising from the disparity between different ES models. Here, we focus on the 97.5\% ES for a well diversified portfolio (\$100,000) and a 10 day holding period between November 4, 2004 and December 31, 2018. We provide the portfolio value minus the 5th and the 95th percentile of ES forecasts from all multivariate models that passed the conditional calibration backtest by \cite{Nolde.2017}  on a daily basis. This corresponds to the potential portfolio value under financial distress according to the more (95th percentile) or less (5th percentile) conservative ES models.
	}\\
	
	\begin{center}
	\includegraphics[width=0.9\linewidth]{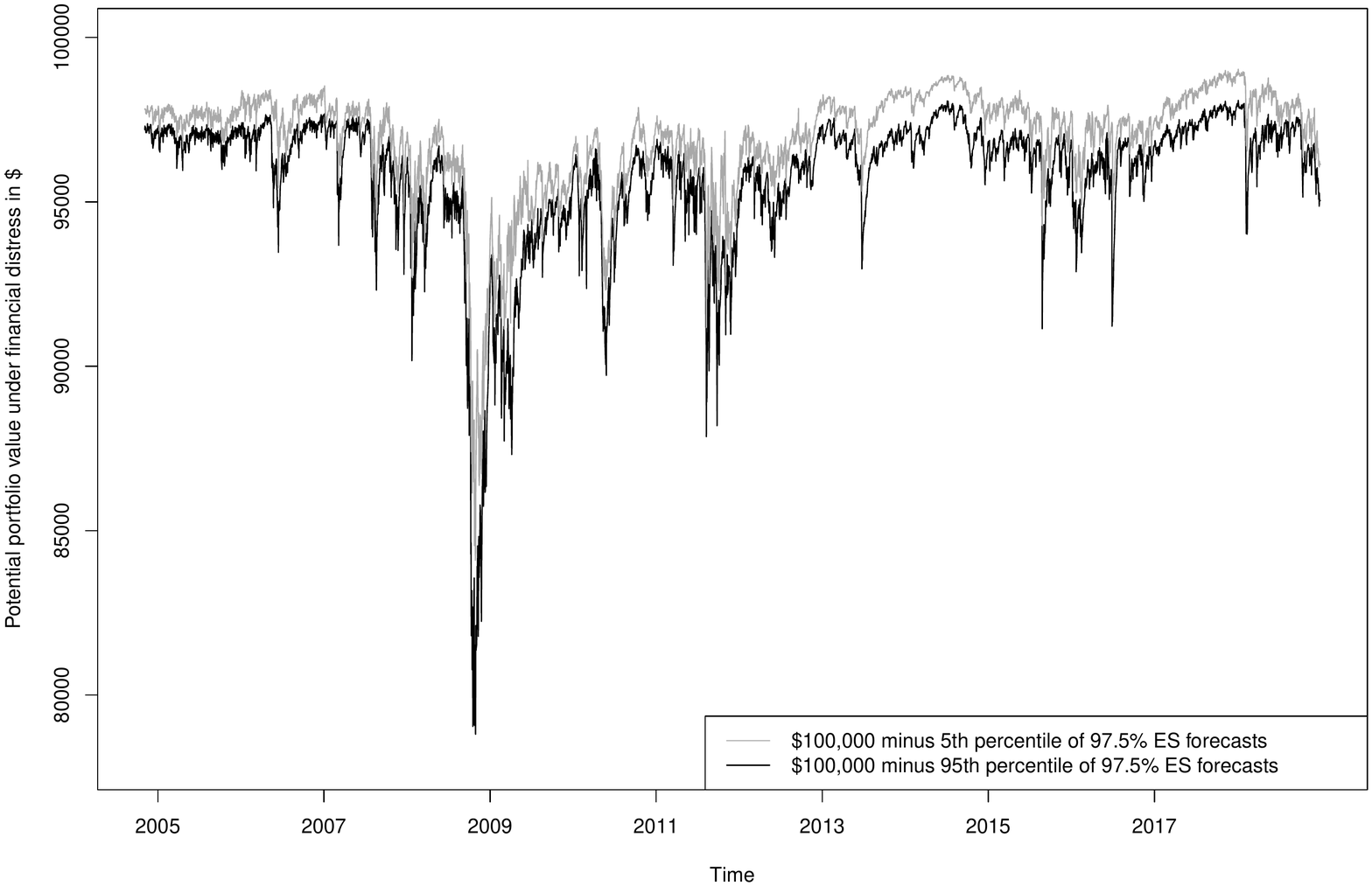}
	\label{fig:minimumPortfolioValueES}
	\end{center}
\end{figure}

\begin{figure}[htbp]
	\caption{Average model risk for alternative model risk measures (model sets with fixed and varying copula only)}
	\scriptsize
	This figure shows the average model risk associated with one day ahead 99\% VaR (Subfigures 1, 3, and 5) and 97.5\% ES (Subfigures 2, 4, and 6) forecasts for a well diversified portfolio per group. \textit{Group 1} (G1) includes all model sets in which a copula function is fixed while varying the marginal distribution. \textit{Group 2} (G2) contains analogously the model sets with fixed marginal distribution and varying copula. Model risk is captured by different measures of one day ahead forecasts by various risk models within a model set. Our baseline measure is the mean absolute deviation (mad). We additionally include the standard deviation (sd) and interquartile range (iqr), see Section \ref{sec:modelRisk} for more details. Values are calculated on a daily basis between November 4, 2004 until December 31, 2018 in percent of the portfolio value based on all models that passed the respective backtest, see Section \ref{sec:backtests} for details.
	\begin{center}
		\begin{subfigure}[t]{0.45\linewidth}
			\centering
			\includegraphics[width=\linewidth]{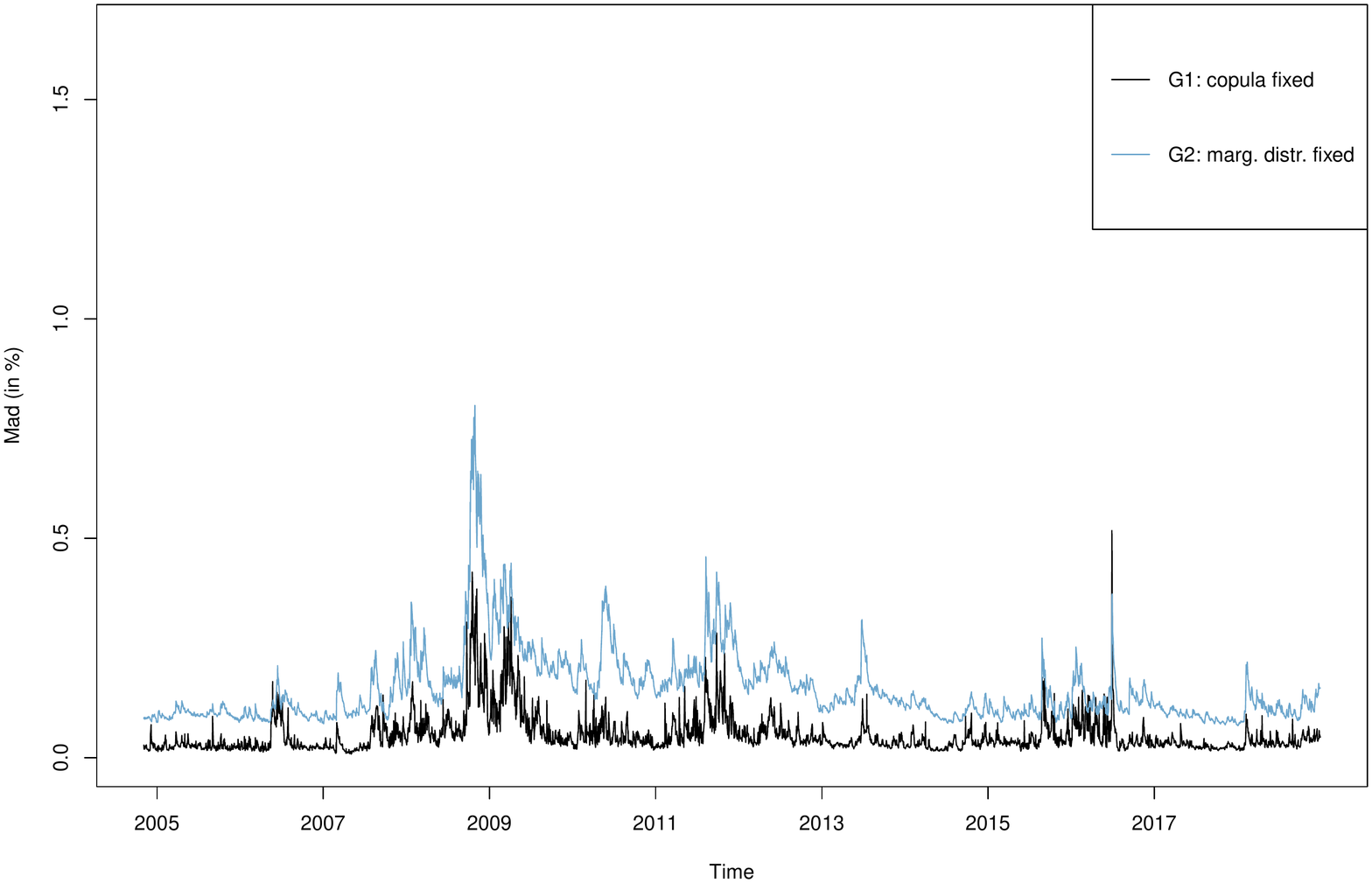}
			\caption{Average model risk (99\% VaR) - mad}\label{fig:Rplot_AnalysisIV_mad1}
		\end{subfigure}
		\begin{subfigure}[t]{0.45\linewidth}
			\centering
			\includegraphics[width=\linewidth]{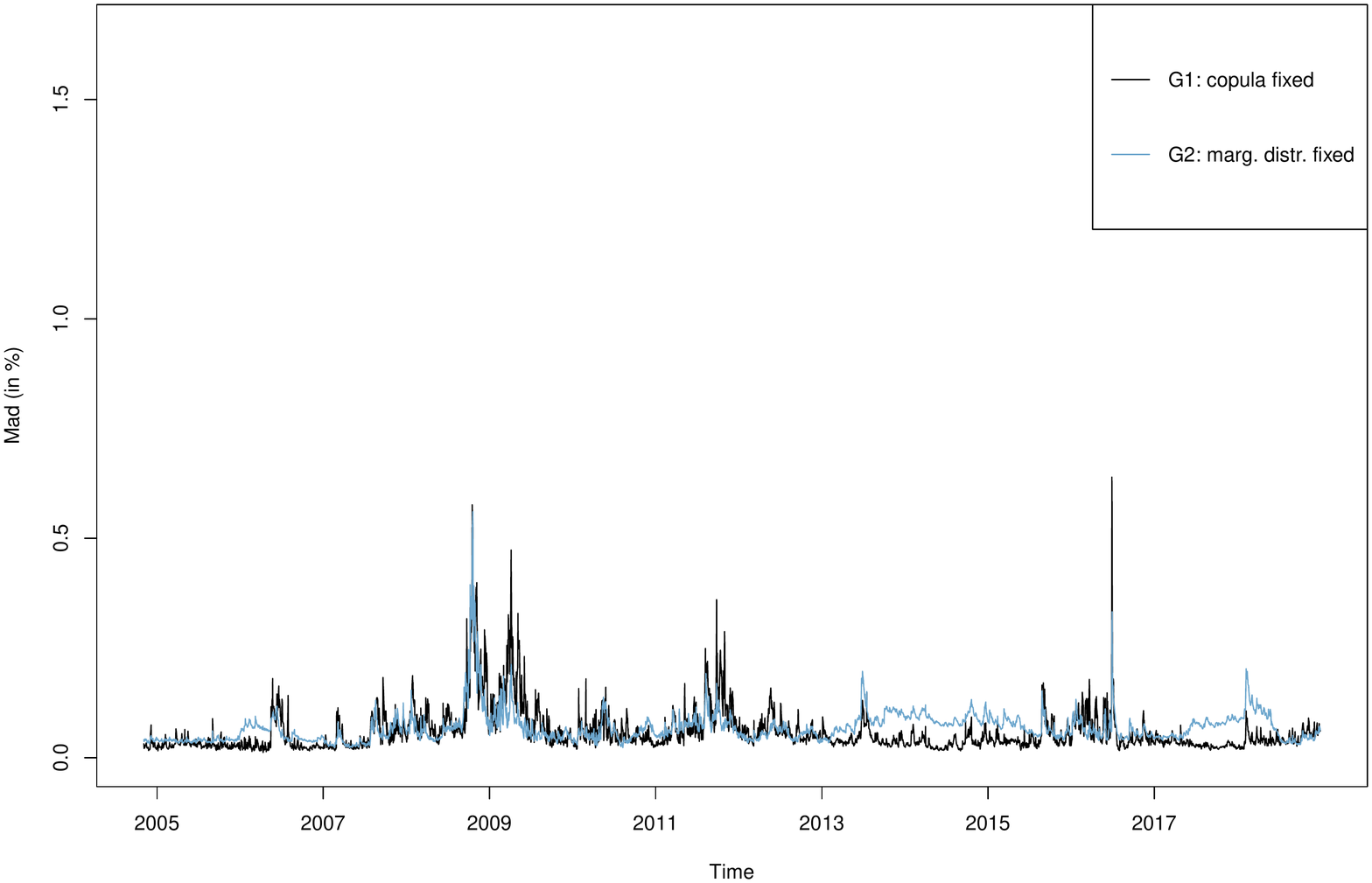}
			\caption{Average model risk (97.5\% ES) - mad}\label{fig:Rplot_AnalysisIV_mad2}
		\end{subfigure}
		\begin{subfigure}[t]{0.45\linewidth}
			\centering
			\includegraphics[width=\linewidth]{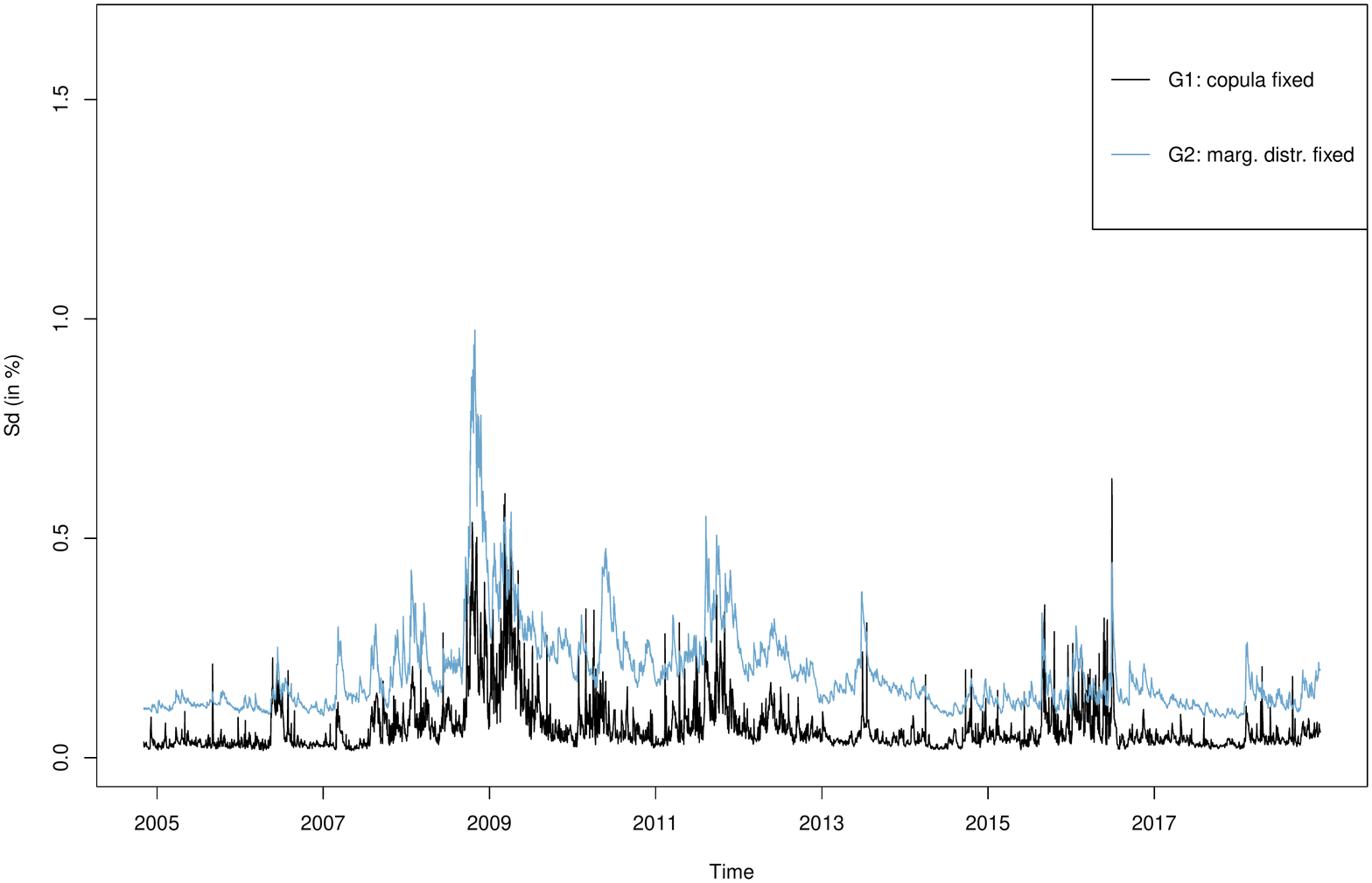}
			\caption{Average model risk (99\% VaR) - sd}\label{fig:Rplot_AnalysisIV_sd1}
		\end{subfigure}
		\begin{subfigure}[t]{0.45\linewidth}
			\centering
			\includegraphics[width=\linewidth]{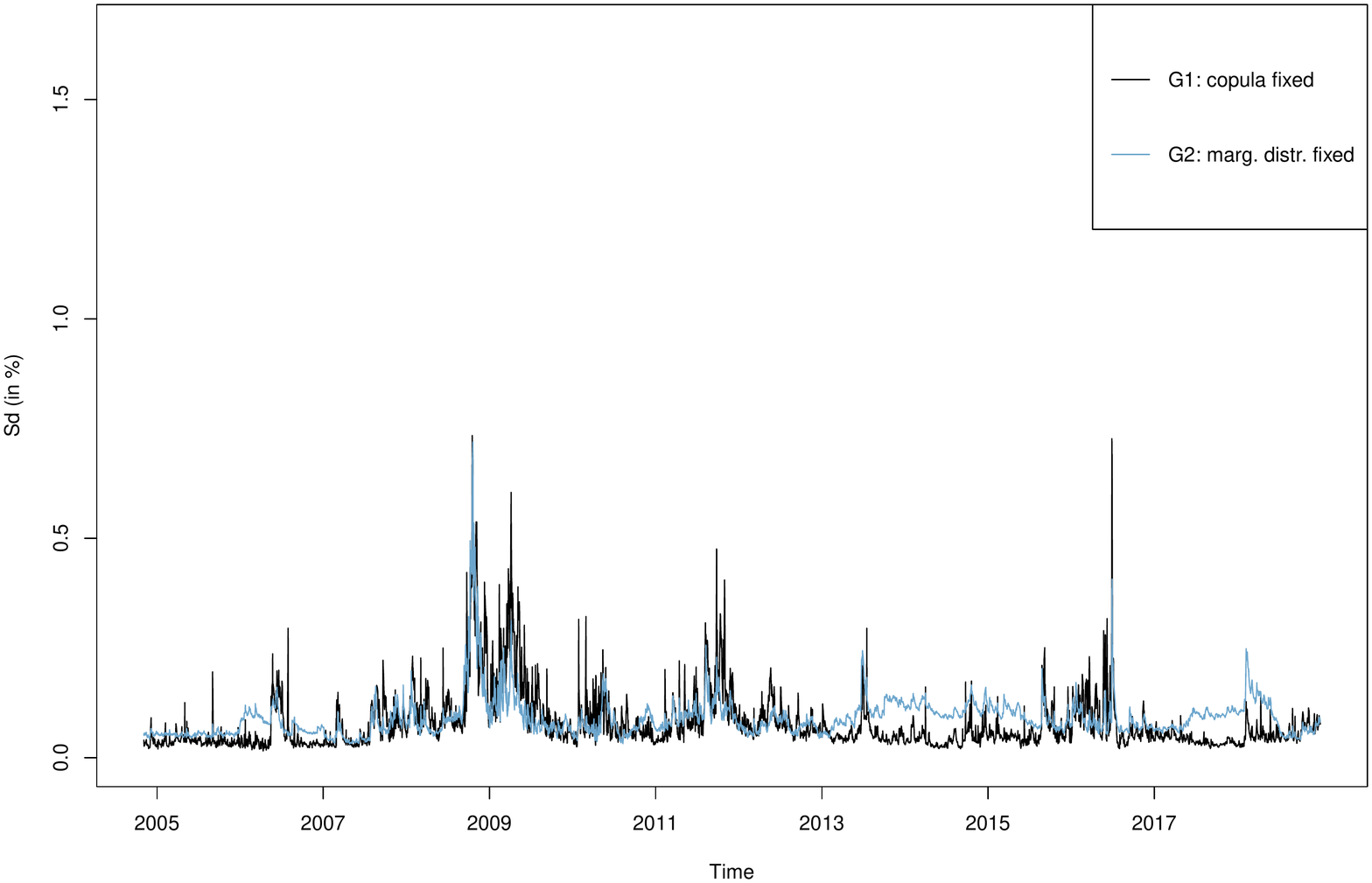}
			\caption{Average model risk (97.5\% ES) - sd}\label{fig:Rplot_AnalysisIV_sd2}
		\end{subfigure}
		\begin{subfigure}[t]{0.45\linewidth}
			\centering
			\includegraphics[width=\linewidth]{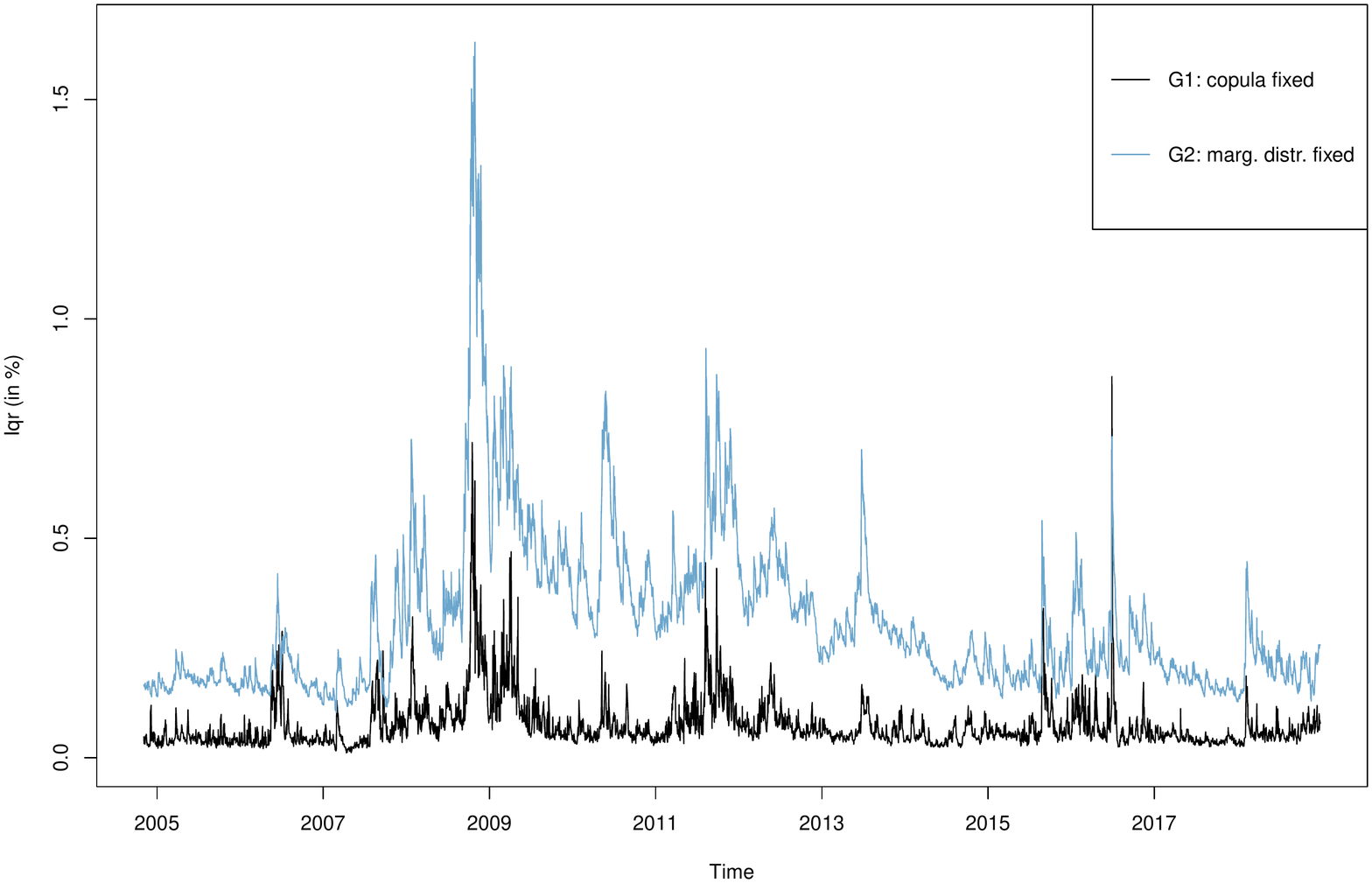}
			\caption{Average model risk (99\% VaR) - iqr}\label{fig:Rplot_AnalysisIV_iqr1}
		\end{subfigure}
		\begin{subfigure}[t]{0.45\linewidth}
			\centering
			\includegraphics[width=\linewidth]{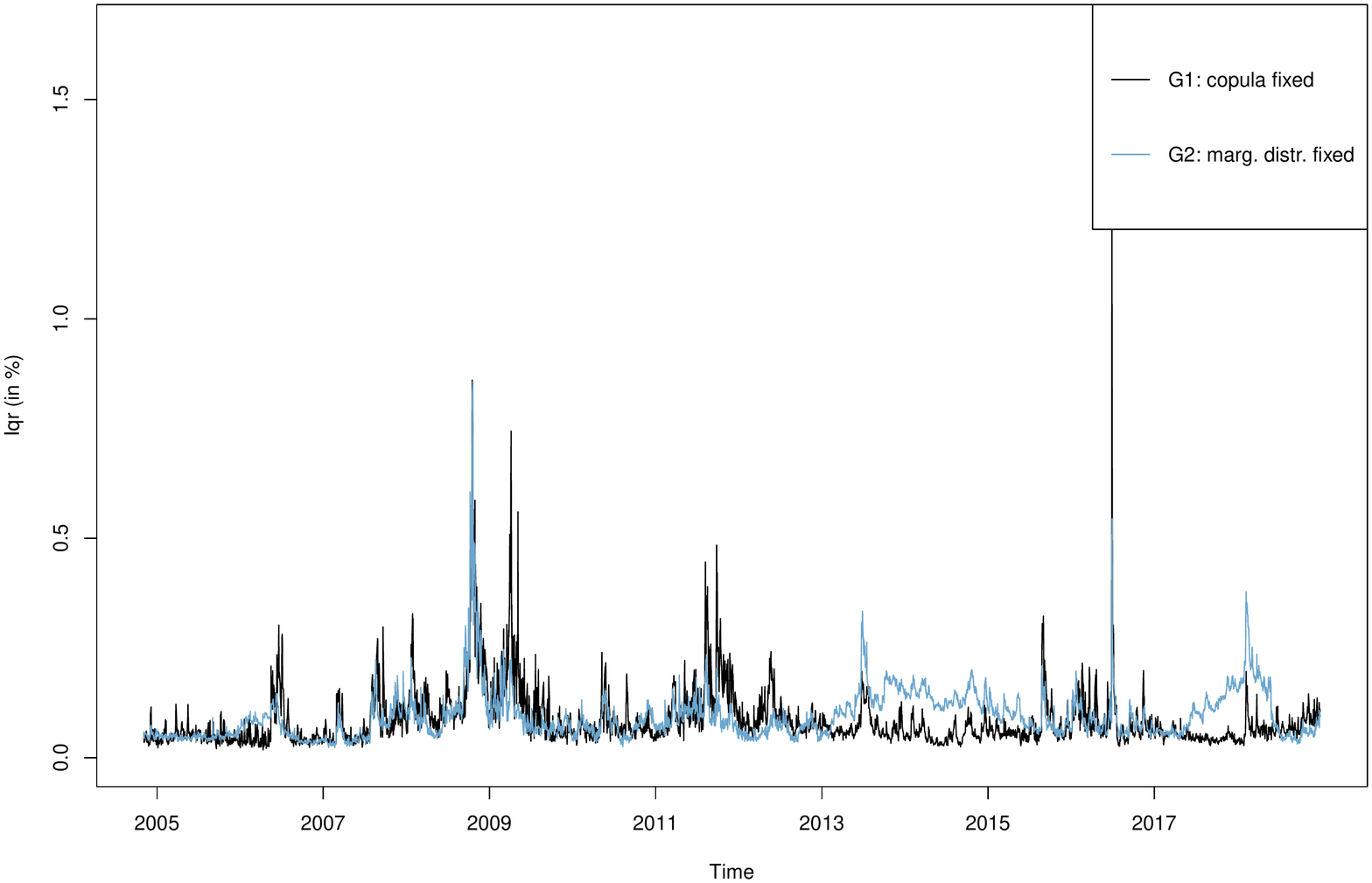}
			\caption{Average model risk (97.5\% ES) - iqr}\label{fig:Rplot_AnalysisIV_iqr2}
		\end{subfigure}
	\end{center}
	\label{fig:Rplot_AnalysisIV}
\end{figure}

\begin{figure}[htbp]
	\caption{Average model risk for all groups under various VaR confidence levels}
	\scriptsize
	This figure shows the average model risk associated with one day ahead VaR forecasts for a well diversified portfolio and a confidence level of 99.9\%, 99\%, 97.5\%, and 95\% (Subfigures 1-4) per group. \textit{Group 1} (G1) includes all model sets in which a copula function is fixed while varying the marginal distribution. \textit{Group 2} (G2) contains analogously the model sets with fixed marginal distribution and varying copula. \textit{Group 3} (G3) consists of all multivariate and \textit{Group 4} (G4) of all univariate models. Model risk is measured in terms of the mean absolute deviation (mad) of one day ahead forecasts by various risk models within a model set. Values are calculated on a daily basis between November 4, 2004 until December 31, 2018 in percent of the portfolio value based on all models that passed the respective backtest, see Section \ref{sec:backtests} for details. 
	\begin{center}
		\begin{subfigure}[t]{0.45\linewidth}
			\centering
			\includegraphics[width=\linewidth]{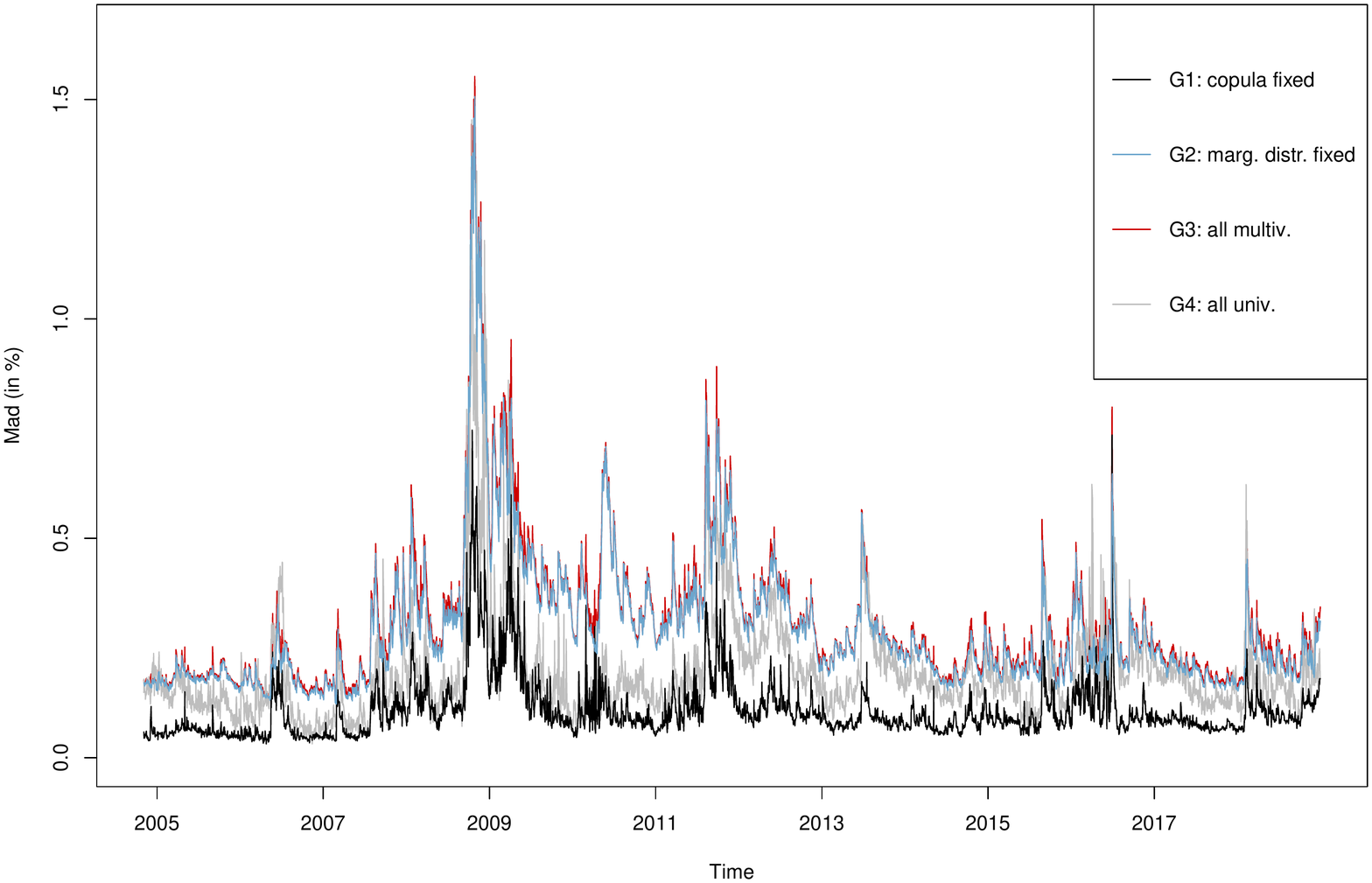}
			\caption{Average model risk (99.9\% VaR)}\label{fig:Rplot_AnalysisI_robust_lev1_1}
		\end{subfigure}
		\begin{subfigure}[t]{0.45\linewidth}
			\centering
			\includegraphics[width=\linewidth]{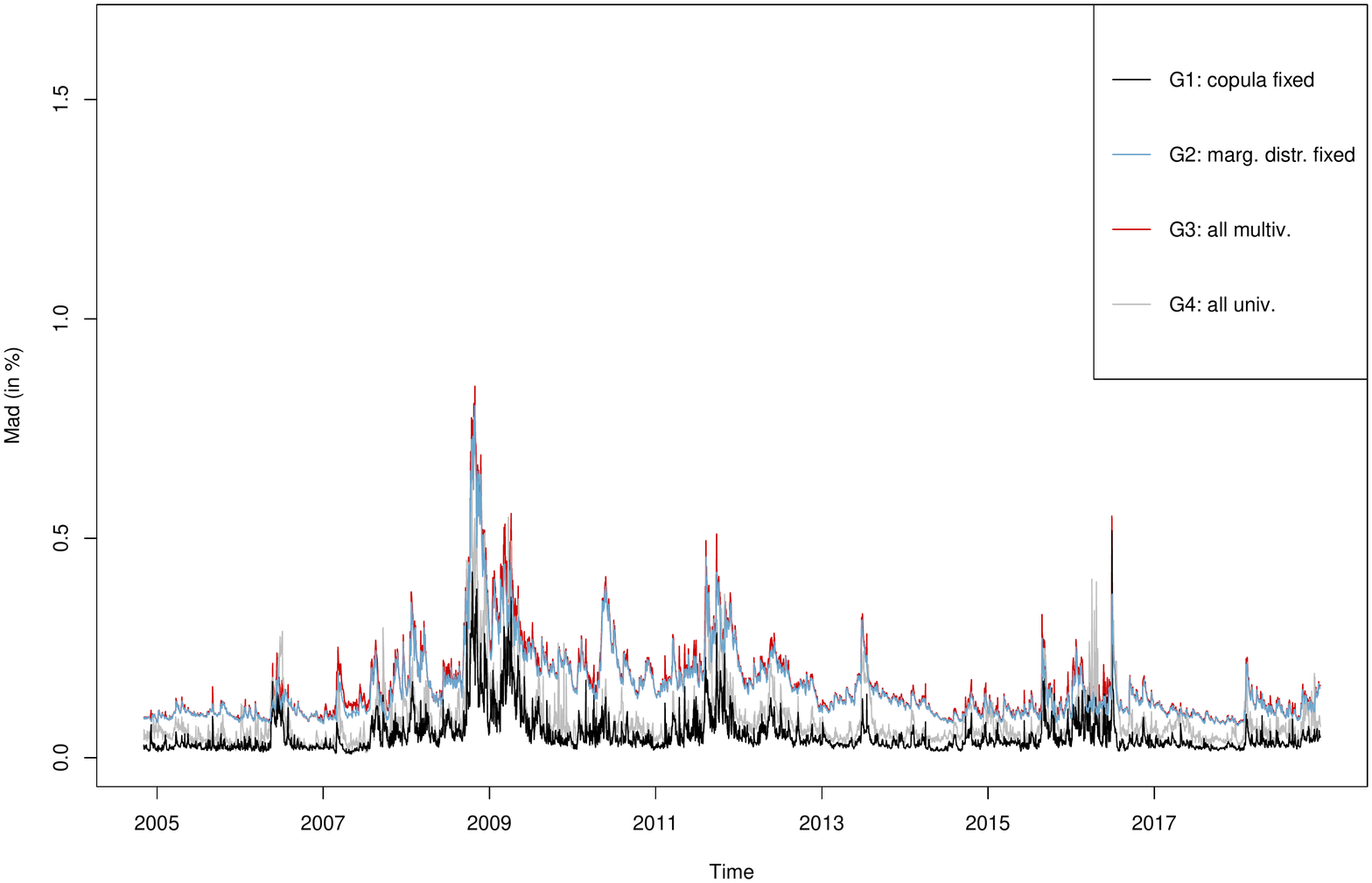}
			\caption{Average model risk (99\% VaR)}\label{fig:Rplot_AnalysisI_robust_lev1_2}
		\end{subfigure}
		\begin{subfigure}[t]{0.45\linewidth}
			\centering
			\includegraphics[width=\linewidth]{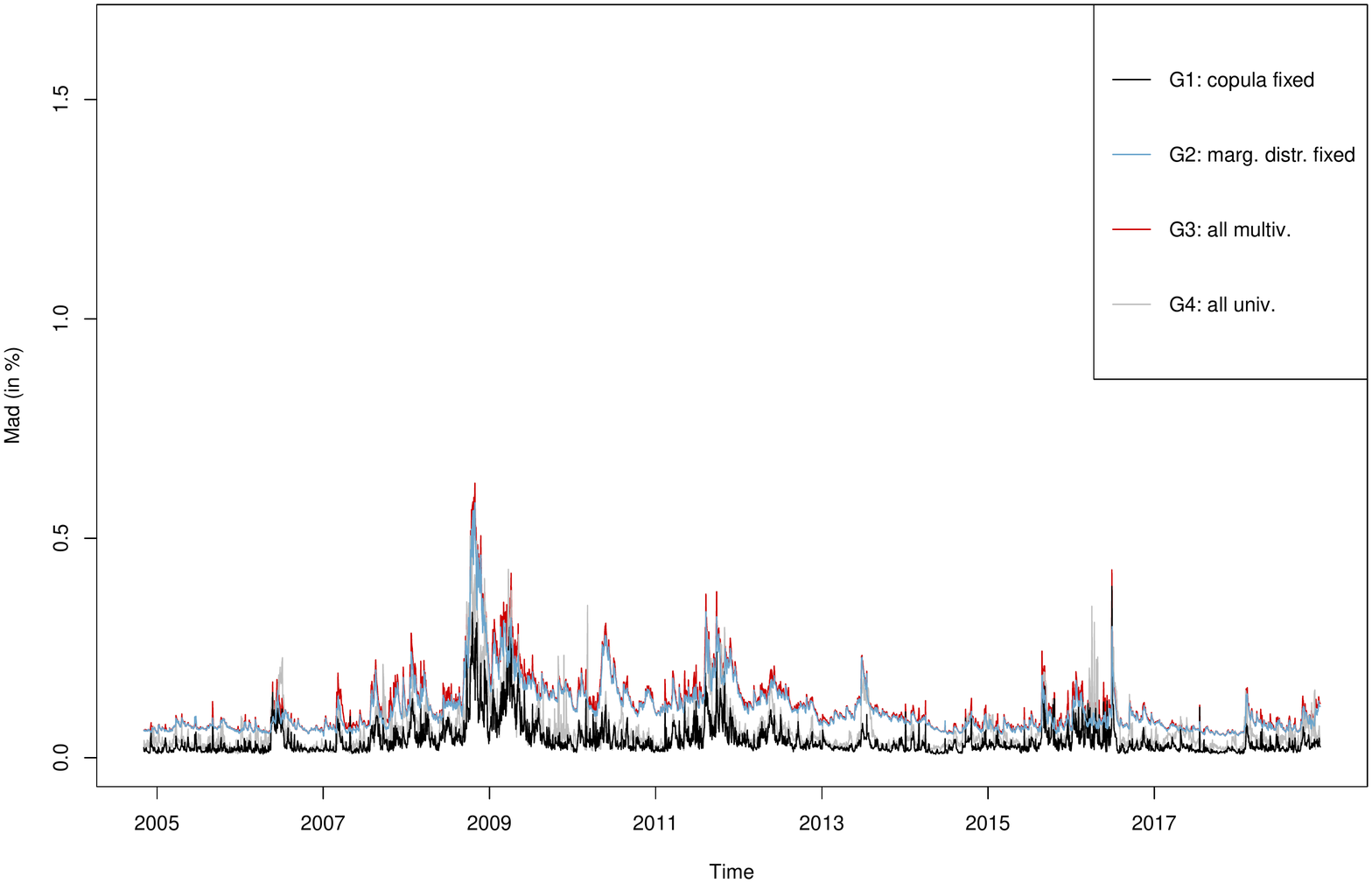}
			\caption{Average model risk (97.5\% VaR)}\label{fig:Rplot_AnalysisI_robust_lev1_3}
		\end{subfigure}
		\begin{subfigure}[t]{0.45\linewidth}
			\centering
			\includegraphics[width=\linewidth]{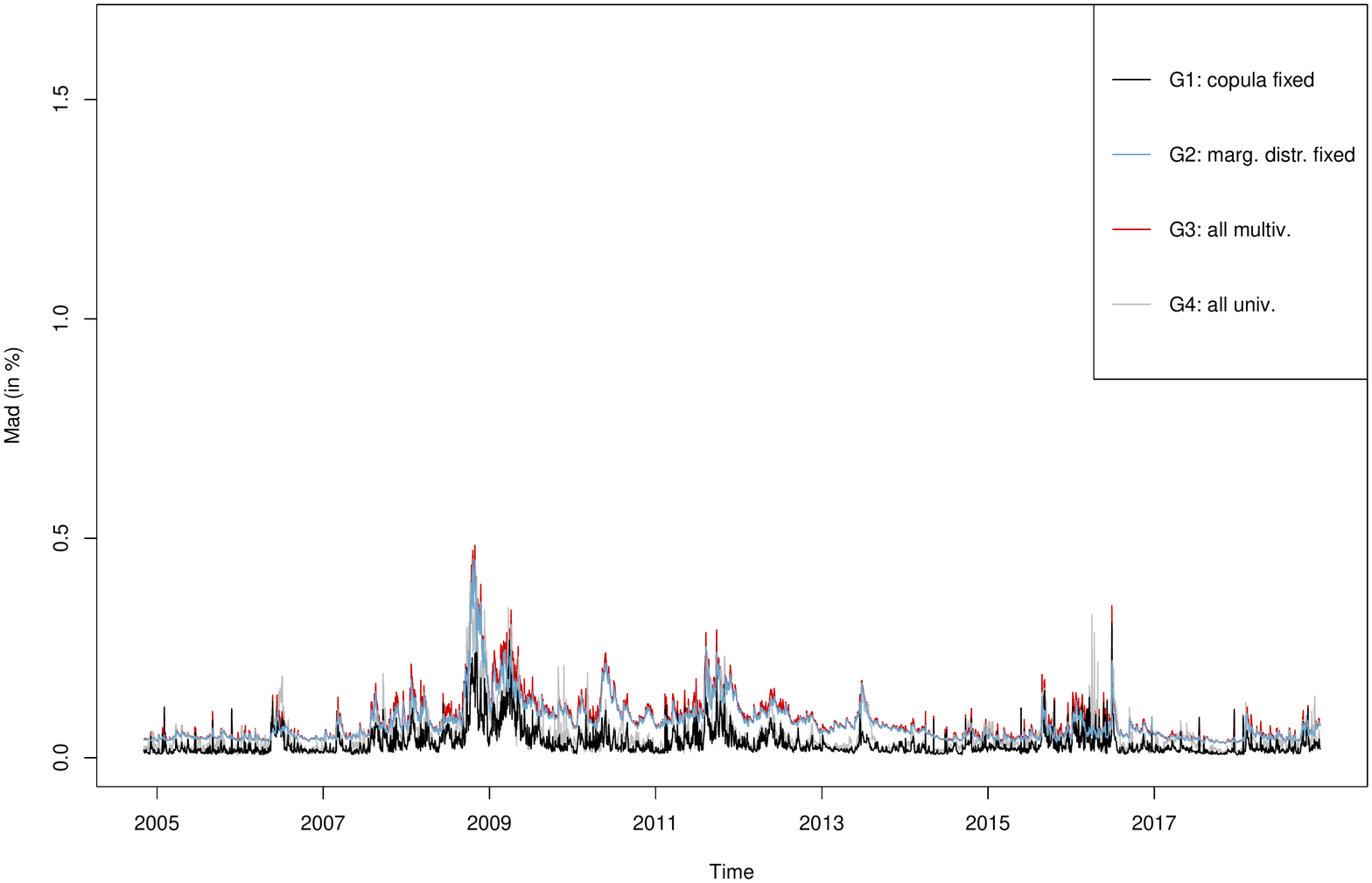}
			\caption{Average model risk (95\% VaR)}\label{fig:Rplot_AnalysisI_robust_lev1_4}
		\end{subfigure}
	\end{center}
	\label{fig:Rplot_AnalysisI_robust_lev1}
\end{figure}

\begin{figure}[htbp]
	\caption{Average model risk for all groups under various ES confidence levels}
	\scriptsize
	This figure shows the average model risk associated with one day ahead ES forecasts for a well diversified portfolio and a confidence level of 99.9\%, 99\%, 97.5\%, and 95\% (Subfigures 1-4) per group. \textit{Group 1} (G1) includes all model sets in which a copula function is fixed while varying the marginal distribution. \textit{Group 2} (G2) contains analogously the model sets with fixed marginal distribution and varying copula. \textit{Group 3} (G3) consists of all multivariate and \textit{Group 4} (G4) of all univariate models. Model risk is measured in terms of the mean absolute deviation (mad) of one day ahead forecasts by various risk models within a model set. Values are calculated on a daily basis between November 4, 2004 until December 31, 2018 in percent of the portfolio value based on all models that passed the respective backtest, see Section \ref{sec:backtests} for details. 
	\begin{center}
		\begin{subfigure}[t]{0.45\linewidth}
			\centering
			\includegraphics[width=\linewidth]{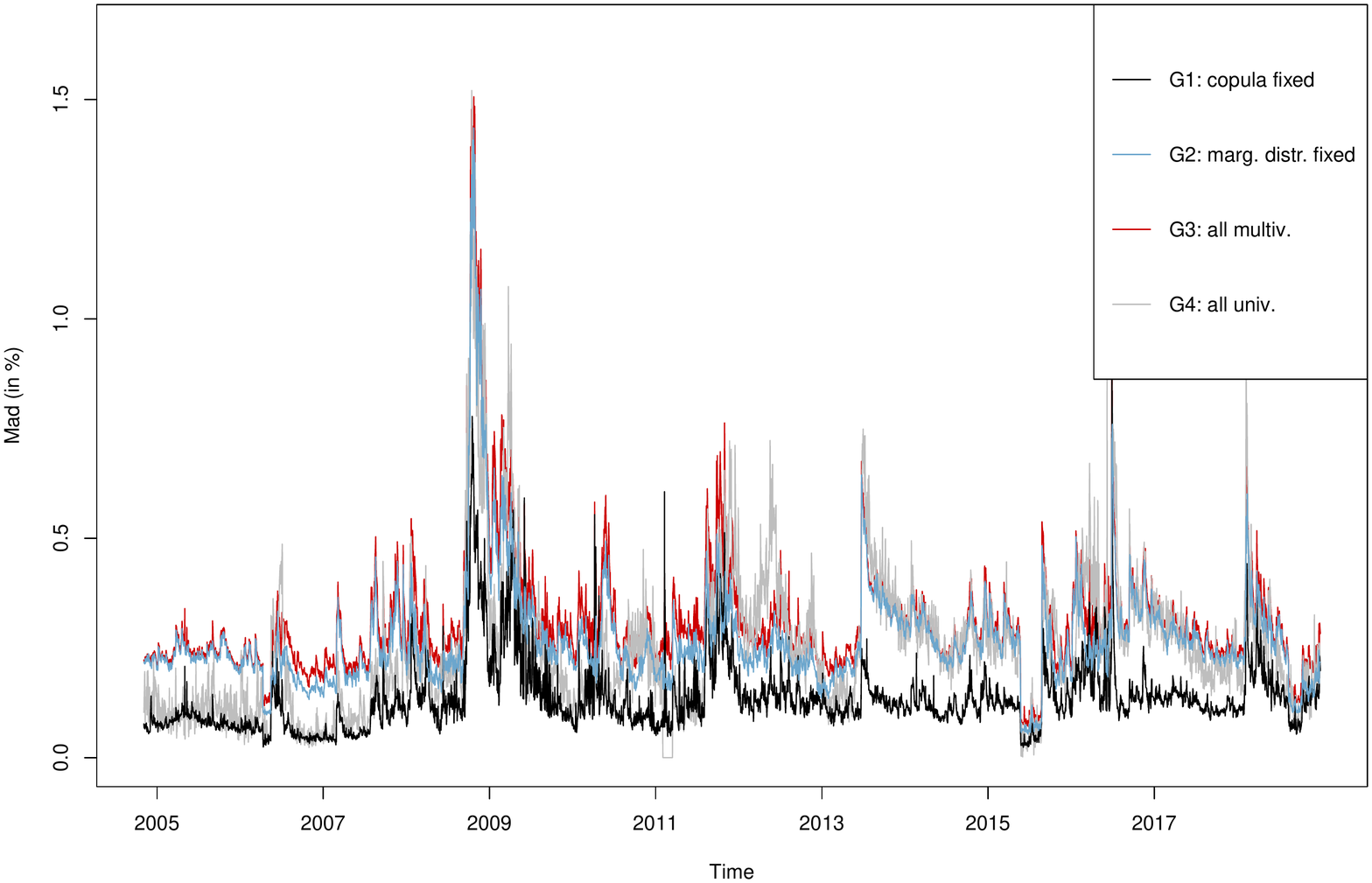}
			\caption{Average model risk (99.9\% ES)}\label{fig:Rplot_AnalysisI_robust_lev2_1}
		\end{subfigure}
		\begin{subfigure}[t]{0.45\linewidth}
			\centering
			\includegraphics[width=\linewidth]{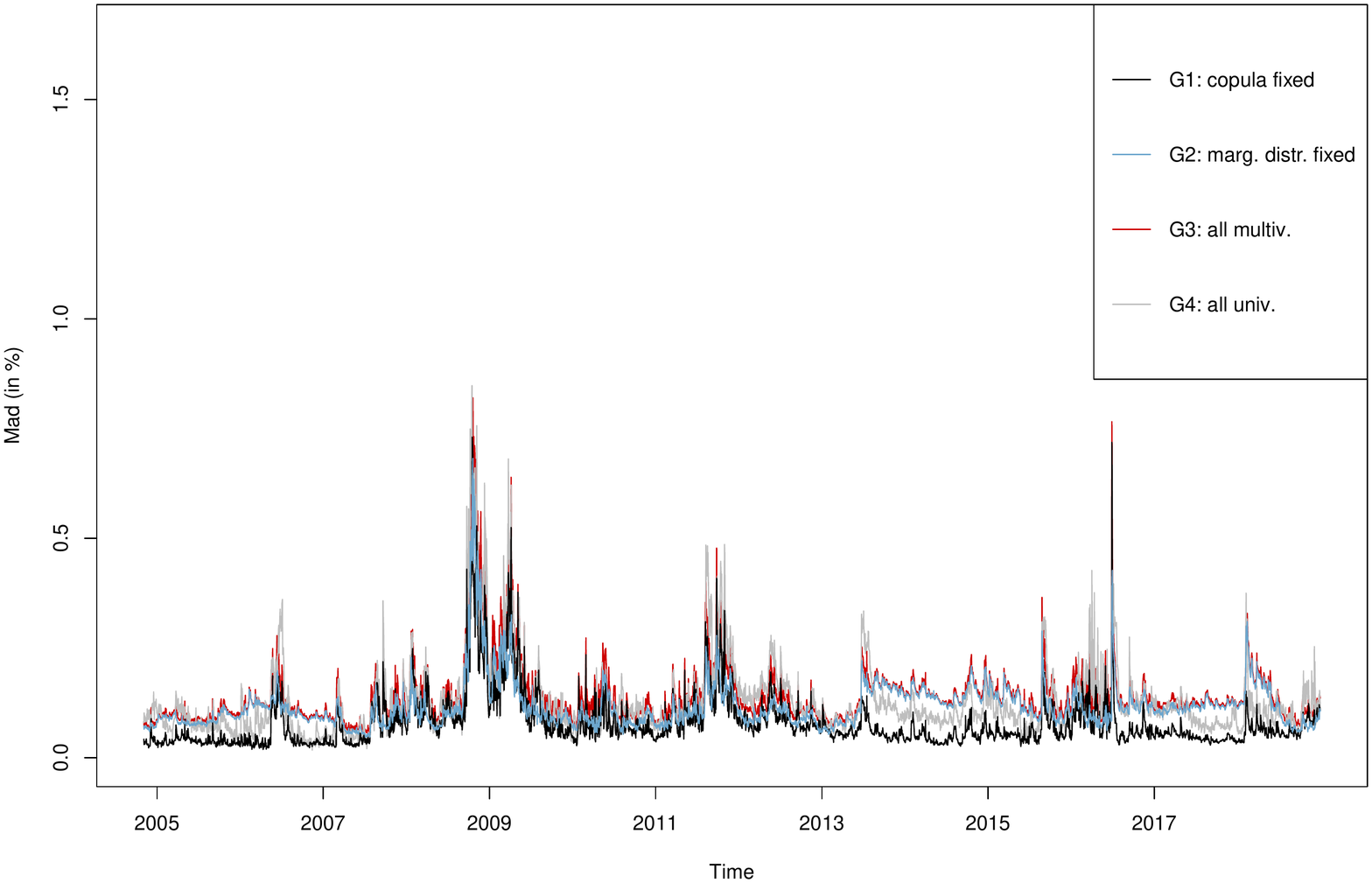}
			\caption{Average model risk (99\% ES)}\label{fig:Rplot_AnalysisI_robust_lev2_2}
		\end{subfigure}
		\begin{subfigure}[t]{0.45\linewidth}
			\centering
			\includegraphics[width=\linewidth]{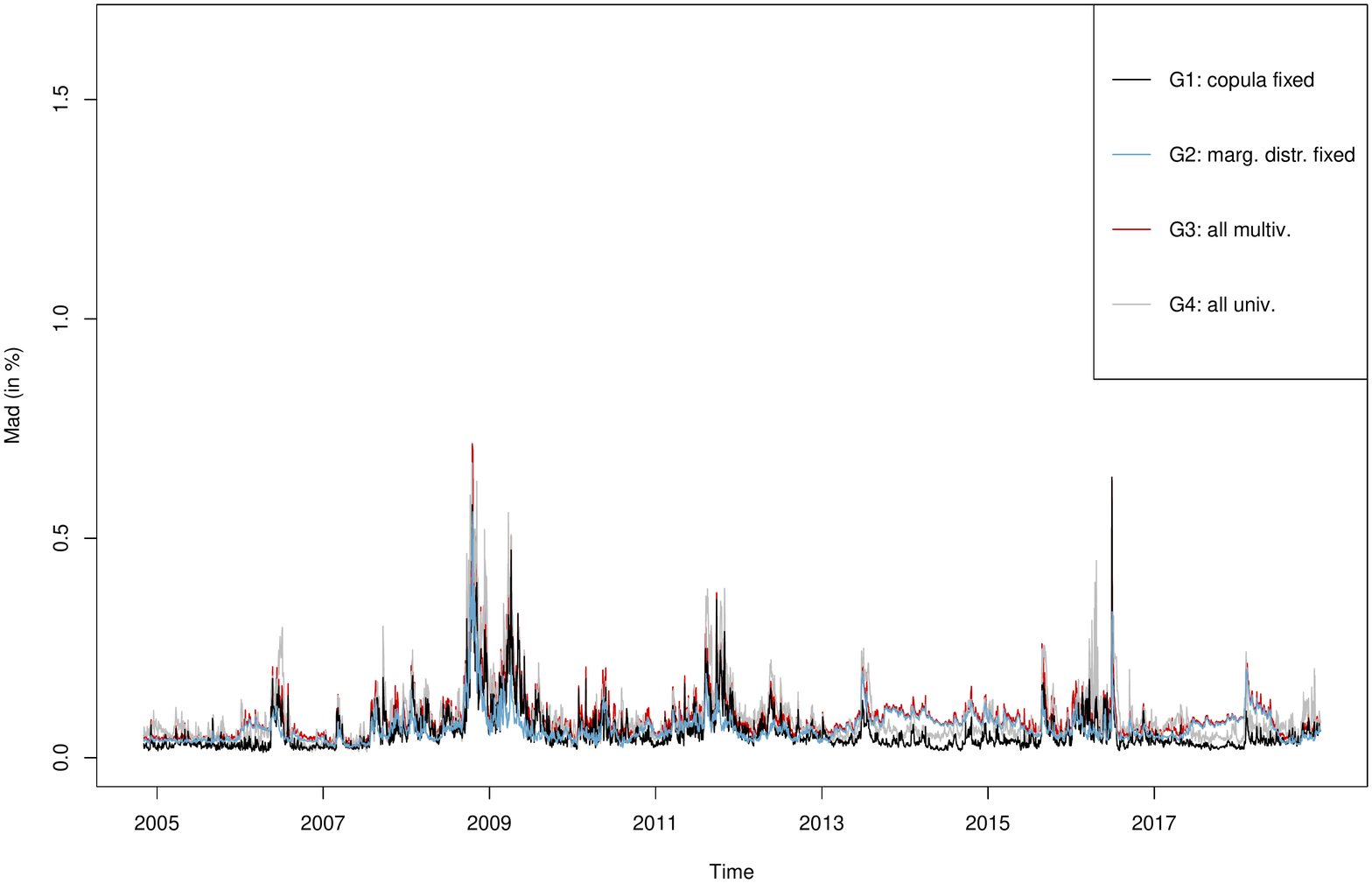}
			\caption{Average model risk (97.5\% ES)}\label{fig:Rplot_AnalysisI_robust_lev2_3}
		\end{subfigure}
		\begin{subfigure}[t]{0.45\linewidth}
			\centering
			\includegraphics[width=\linewidth]{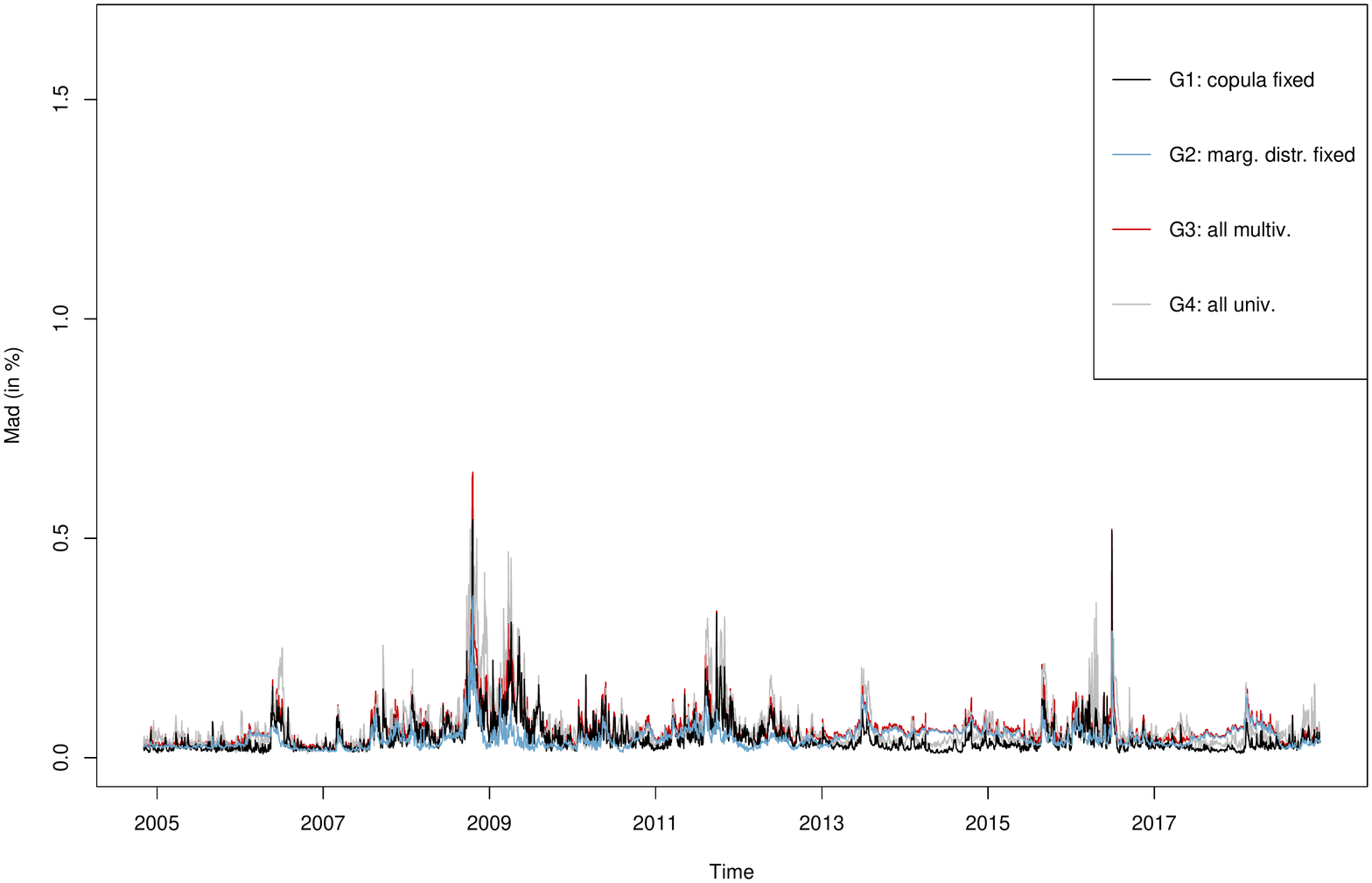}
			\caption{Average model risk (95\% ES)}\label{fig:Rplot_AnalysisI_robust_lev2_4}
		\end{subfigure}
	\end{center}
	\label{fig:Rplot_AnalysisI_robust_lev2}
\end{figure}

\end{document}